\documentclass[a4paper]{memoir}
\usepackage[chapter]{algorithm}
\usepackage{algpseudocode}
\usepackage{amsfonts}
\usepackage{amsmath}
\usepackage{amssymb}
\usepackage{amsthm}
\usepackage[english]{babel}
\usepackage{cancel}
\usepackage{cmll}
\usepackage{illcmolthesis}
\usepackage{listings}
\usepackage{pst-all}
\usepackage{stmaryrd}
\usepackage{tikz}
\usepackage{xspace}

\newcommand{\SP}[1]{#1\xspace}

\newcommand{\N}{\mathbb{N}}
\newcommand{\Np}{\mathbb{N}^+}
\newcommand{\Z}{\mathbb{Z}}
\newcommand{\seq}[1]{\langle #1 \rangle}
\newcommand{\gen}[1]{{<}#1{>}}
\newcommand{\setlen}[1]{\lvert #1 \rvert}
\newcommand{\powerset}{\mathcal{P}}

\newcommand{\BTA}{\SP{\ensuremath{\text{BTA}}}}
\newcommand{\BTAi}{\SP{\ensuremath{\BTA^\infty}}}
\newcommand{\BPPA}{\SP{\ensuremath{\text{BPPA}}}}
\newcommand{\PGA}{\SP{\ensuremath{\text{PGA}}}}
\newcommand{\PGLDg}{\SP{\ensuremath{\text{PGLDg}}}}
\newcommand{\A}{\SP{\ensuremath{A}}}
\newcommand{\B}{\SP{\ensuremath{B}}}
\newcommand{\C}{\SP{\ensuremath{C}}}
\newcommand{\Cp}{\SP{\ensuremath{C'}}}
\newcommand{\Cr}[1]{\SP{\ensuremath{\C_{\leq #1}}}}
\newcommand{\Cg}{\SP{\ensuremath{Cg}}}
\newcommand{\Cgp}{\SP{\ensuremath{Cg'}}}

\newcommand{\Cgr}[1]{\SP{\ensuremath{\Cg_{\leq #1}}}}

\newcommand{\U}{\mathcal{I}}
\newcommand{\Un}[1]{\U^{#1}}
\newcommand{\UA}{\Un{+}}

\newcommand{\UB}{\U_{\B}}
\newcommand{\UnB}[1]{\UB^{#1}}
\newcommand{\UAB}{\UnB{+}}

\newcommand{\UC}{\U_{\C}}
\newcommand{\UnC}[1]{\UC^{#1}}
\newcommand{\UAC}{\UnC{+}}
\newcommand{\FUC}{\UnC{\rightarrow}}
\newcommand{\BUC}{\UnC{\leftarrow}}

\newcommand{\UCp}{\U_{\Cp}}
\newcommand{\UnCp}[1]{\UCp^{#1}}
\newcommand{\UACp}{\UnCp{+}}

\newcommand{\UCr}[1]{\U_{\Cr{#1}}}
\newcommand{\UnCr}[2]{\UCr{#1}^{#2}}
\newcommand{\UACr}[1]{\UnCr{#1}{+}}

\newcommand{\UCg}{\U_{\Cg}}
\newcommand{\UnCg}[1]{\UCg^{#1}}
\newcommand{\UACg}{\UnCg{+}}
\newcommand{\FUCg}{\UnCg{\rightarrow}}
\newcommand{\BUCg}{\UnCg{\leftarrow}}

\newcommand{\UCgp}{\U_{\Cgp}}
\newcommand{\UnCgp}[1]{\UCgp^{#1}}
\newcommand{\UACgp}{\UnCgp{+}}

\newcommand{\UCgr}[1]{\U_{\Cgr{#1}}}
\newcommand{\UnCgr}[2]{\UCgr{#1}^{#2}}
\newcommand{\UACgr}[1]{\UnCgr{#1}{+}}

\newcommand{\aArg}{\_}
\newcommand{\catOp}{\aArg{;}\aArg}

\DeclareMathOperator{\len}{\ell}
\newcommand{\inst}[1]{\sigma_{#1}}

\newcommand{\actions}{\mathcal{A}}

\DeclareMathOperator{\jd}{\delta}
\DeclareMathOperator{\lno}{\lambda}

\newcommand{\TElong}[2]{{\lvert#1, #2\rvert}}
\newcommand{\TE}[2]{{\lvert#2\rvert}^{#1}}
\newcommand{\TEfwd}[1]{\TE{\rightarrow}{#1}}
\newcommand{\TEbwd}[1]{\TE{\leftarrow}{#1}}
\newcommand{\ATElong}[2]{\TElong{#1}{#2}_\A}
\newcommand{\ATE}[2]{\TE{#1}{#2}_\A}
\newcommand{\ATEfwd}[1]{\ATE{\rightarrow}{#1}}
\newcommand{\ATEbwd}[1]{\ATE{\leftarrow}{#1}}
\newcommand{\BTElong}[2]{\TElong{#1}{#2}_\B}
\newcommand{\BTE}[2]{\TE{#1}{#2}_\B}
\newcommand{\BTEfwd}[1]{\BTE{\rightarrow}{#1}}
\newcommand{\BTEbwd}[1]{\BTE{\leftarrow}{#1}}
\newcommand{\CTElong}[2]{\TElong{#1}{#2}_{\C}}
\newcommand{\CTE}[2]{\TE{#1}{#2}_{\C}}
\newcommand{\CTEfwd}[1]{\CTE{\rightarrow}{#1}}
\newcommand{\CTEbwd}[1]{\CTE{\leftarrow}{#1}}

\newcommand{\CpTE}[2]{\TE{#1}{#2}_{\Cp}}

\newcommand{\CgTElong}[2]{\TElong{#1}{#2}_{\Cg}}
\newcommand{\CgTE}[2]{\TE{#1}{#2}_{\Cg}}
\newcommand{\CgTEfwd}[1]{\CgTE{\rightarrow}{#1}}
\newcommand{\CgTEbwd}[1]{\CgTE{\leftarrow}{#1}}
\newcommand{\CgpTElong}[2]{\TElong{#1}{#2}_{\Cgp}}
\newcommand{\CgpTE}[2]{\TE{#1}{#2}_{\Cgp}}

\newcommand{\CgRelTE}[3]{\TE{#2}{#3}_{\Cg,#1}}
\newcommand{\CgRelTEfwd}[2]{\CgRelTE{#1}{\rightarrow}{#2}}
\newcommand{\CgRelTEbwd}[2]{\CgRelTE{#1}{\leftarrow}{#2}}
\newcommand{\PGATE}[1]{\TE{}{#1}_{\PGA}}

\newcommand{\true}{\SP{\texttt{true}}}
\newcommand{\false}{\SP{\texttt{false}}}

\newcommand{\lbl}[1]{\text{\pounds} #1}

\newcommand{\pbi}[1]{#1}
\newcommand{\ppt}[1]{{+}#1}
\newcommand{\pnt}[1]{{-}#1}
\newcommand{\pjmp}[1]{\# #1}
\newcommand{\pterm}{\oc}
\newcommand{\plbl}[1]{\lbl #1}
\newcommand{\pgt}[1]{\#\#\lbl #1}

\newcommand{\primitives}{\mathfrak{I}}

\newcommand{\term}{\oc}
\newcommand{\abrt}{\#}
\newcommand{\fbi}[1]{/#1}
\newcommand{\bbi}[1]{\backslash #1}
\newcommand{\fpt}[1]{{+}/#1}
\newcommand{\bpt}[1]{{+}\backslash #1}
\newcommand{\fnt}[1]{{-}/#1}
\newcommand{\bnt}[1]{{-}\backslash #1}
\newcommand{\fj}[1]{/\# #1}
\newcommand{\bj}[1]{\backslash\# #1}
\newcommand{\flbl}[1]{/\lbl #1}
\newcommand{\blbl}[1]{\backslash\lbl #1}
\newcommand{\fgt}[1]{/\#\#\lbl #1}
\newcommand{\bgt}[1]{\backslash\#\#\lbl #1}

\newcommand{\basics}{\mathfrak{B}}
\newcommand{\positives}{\mathfrak{P}}
\newcommand{\negatives}{\mathfrak{N}}
\newcommand{\jumps}{\mathfrak{J}}
\newcommand{\labels}{\mathfrak{L}}
\newcommand{\gotos}{\mathfrak{G}}
\newcommand{\lago}{\labels\gotos}
\newcommand{\Fbasics}{\basics^\rightarrow}
\newcommand{\Bbasics}{\basics^\leftarrow}
\newcommand{\Fpositives}{\positives^\rightarrow}
\newcommand{\Bpositives}{\positives^\leftarrow}
\newcommand{\Fnegatives}{\negatives^\rightarrow}
\newcommand{\Bnegatives}{\negatives^\leftarrow}
\newcommand{\Fjumps}{\jumps^\rightarrow}
\newcommand{\Bjumps}{\jumps^\leftarrow}
\newcommand{\Flabels}{\labels^\rightarrow}
\newcommand{\Blabels}{\labels^\leftarrow}
\newcommand{\Fgotos}{\gotos^\rightarrow}
\newcommand{\Bgotos}{\gotos^\leftarrow}
\newcommand{\Flago}{\lago^\rightarrow}
\newcommand{\Blago}{\lago^\leftarrow}

\newcommand{\cpppt}[1]{{+}#1}
\newcommand{\cppnt}[1]{{-}#1}

\newcommand{\Term}{\mathsf{S}}
\newcommand{\Inac}{\mathsf{D}}
\newcommand{\ThrT}{\unlhd}
\newcommand{\ThrF}{\unrhd}

\newcommand{\Thr}{\BTA}              
\newcommand{\ThrI}{\BTAi}            
\newcommand{\ThrR}{\BTA^\text{reg}}  
\newcommand{\Pga}{\boldsymbol{P}}
\newcommand{\PgaF}{{\Pga_{\boldsymbol{1}}}}
\newcommand{\PgaS}{\Pga_{\boldsymbol{2}}}

\DeclareMathOperator{\res}{\textsc{res}}

\DeclareMathOperator{\fst}{\textsc{fst}}
\DeclareMathOperator{\snd}{\textsc{snd}}

\DeclareMathOperator{\Fsearch}{\overrightarrow{\textsc{search}}}
\DeclareMathOperator{\Bsearch}{\overleftarrow{\textsc{search}}}

\newcommand{\rel}[1]{\textsc{rel}_{#1}}

\newcommand{\Next}{\textsc{next}}
\newcommand{\Left}{\textsc{left}}
\newcommand{\Right}{\textsc{right}}

\newcommand{\PgaToC}{\textsc{pga2c}}
\newcommand{\PgaSToC}{\textsc{snd2c}}
\newcommand{\CToPga}{\textsc{c2pga}}
\newcommand{\CToCg}{\textsc{c2cg}}
\newcommand{\CrToCg}[1]{\textsc{c2cg}_{#1}}
\newcommand{\CrToCgLong}{\textsc{c2cg}}
\newcommand{\CgToC}{\textsc{cg2c}}

\newcommand{\eq}[1]{=_{#1}}
\newcommand{\acc}[1]{\rightarrow_{#1}}
\newcommand{\corr}[1]{\approx_{#1}}
\newcommand{\te}[1]{\curlyveedownarrow_{#1}}
\newcommand{\gacc}[1]{\curvearrowright_{#1}}
\newcommand{\lgr}[1]{\star_{#1}}

\newcommand{\de}{\thicksim}

\newcommand{\exits}[1]{\mathcal{E}_{#1}}
\newcommand{\reachable}[2]{\mathcal{R}_{#1, #2}}
\newcommand{\unreachable}[2]{\overline{\mathcal{R}_{#1, #2}}}

\newcommand{\rem}[2]{[#2]_{#1}}

\newcommand{\rightof}[1]{R_{#1}}

\newcommand{\node}[1]{\textsc{node}_{#1}}
\newcommand{\tree}[1]{\textsc{tree}_{#1}}

\newcommand{\free}[1]{\textsc{f}_{#1}}

\newcommand{\dual}[1]{\overline{#1}}
\newcommand{\rev}{\textsc{rev}}

\newcommand{\Fdecr}{\textsc{d}^\rightarrow}
\newcommand{\Bdecr}{\textsc{d}^\leftarrow}

\newcommand{\Fjsub}{J^\rightarrow}
\newcommand{\Bjsub}{J^\leftarrow}
\newcommand{\Fgsub}{G^\rightarrow}
\newcommand{\Bgsub}{G^\leftarrow}
\newcommand{\Flsub}{L^\rightarrow}
\newcommand{\Blsub}{L^\leftarrow}

\newcommand{\ConcatInstructions}[1]{\text{\Call{ConcatInstructions}{#1}}}
\newcommand{\Connect}[1]{\text{\Call{Connect}{#1}}}
\newcommand{\RandomSelect}[1]{\text{\Call{RandomSelect}{#1}}}

\newcommand{\code}[1]{\lstinline{#1}}

\algnewcommand\algorithmicto{\textbf{to}}
\algrenewtext{For}[3]%
  {\algorithmicfor\ $#1 \gets #2$ \algorithmicto\ $#3$ \algorithmicdo}

\usetikzlibrary{arrows,calc,fit,trees}

\theoremstyle{plain}
  \newtheorem{prop}{Proposition}[chapter]
  \newtheorem{thm}[prop]{Theorem}
\theoremstyle{definition}
  \newtheorem{defn}[prop]{Definition}

\newcommand{\PropRef}[1]{Proposition~\ref{#1}}
\newcommand{\ThmRef}[1]{Theorem~\ref{#1}}
\newcommand{\DefRef}[1]{Definition~\ref{#1}}
\newcommand{\AlgoRef}[1]{Algorithm~\ref{#1}}

\semiisopage
\checkandfixthelayout
\chapterstyle{madsen}
\maxsecnumdepth{subsection}

\begin{document}
\nouppercaseheads


\thispagestyle{empty}
\vspace*{\fill}\begin{abstract}
  \PGA, short for ProGram Algebra \cite{intro_pga_ta, pga_second_paper},
  describes sequential programs as finite or infinite (repeating) sequences of
  instructions. The semigroup \C of finite instruction sequences
  \cite{inseq_intro} was introduced as an equally expressive alternative to
  \PGA. \PGA instructions are executed from left to right; most \C instructions
  come in a left-to-right as well as a right-to-left flavor. This thesis builds
  on \C by introducing an alternative semigroup \Cg which employs label and
  goto instructions instead of relative jump instructions as control
  structures. \Cg can be translated to \C and vice versa (and is thus equally
  expressive). It is shown that restricting the instruction sets of \C and \Cg
  to contain only finitely many distinct jump, goto or label instructions in
  either or both directions reduces their expressiveness. Instruction sets with
  an infinite number of these instructions in both directions (not necessarily
  all such instructions) do not suffer a loss of expressiveness.

\end{abstract}\vspace*{\fill}
\cleardoublepage
\tableofcontents
\chapter{%
  Introduction
}

Bergstra and Ponse \cite{inseq_intro} introduce an algebra of finite
instruction sequences by presenting a semigroup \C in which programs can be
represented without directional bias: in terms of the next instruction to be
executed, \C has both forward and backward instructions and a \C-expression can
be interpreted starting from any instruction.

\cite{inseq_intro} provides equations for thread extraction, i.e. \C's program
semantics, and defines behavioral equivalence. It considers thread extraction
compatible (anti-)homomorphisms and (anti-)automorphisms. Lastly, it discusses
some expressiveness results.

\C is a recent alternative to \PGA \cite{intro_pga_ta, pga_second_paper},
short for ProGram Algebra. Contrary to \C, \PGA uses infinite instruction
sequences to model infinite behavior. Since both \PGA and \C are tools that aid
in the research on imperative sequential programming, and given that any ``real
world'' programs are always finite, \C appears to be a more realistic approach
to a mathematical representation for sequential programs.

This thesis introduces \PGA and \C and describes their semantics. It then
defines an alternative to \C called \Cg which uses label and goto instructions
as control structures, as opposed to \C's relative jump instructions. Behavior
preserving mappings are defined between \PGA, \C and \Cg, thereby establishing
that they are equally expressive.

The final chapter of this thesis investigates the expressiveness of
subsemigroups of \C and \Cg, particularly those from which a finite or infinite
number of jump or goto instructions has been removed, thereby improving on an
expressiveness result presented in \cite{inseq_intro}.

Lastly, the reader should take note of \Aref{app:mappings_overview}, which
provides a graphical representation of some of the (single-pass) instruction
sequences defined in this thesis and the mappings between them.

\chapter{%
  Preliminaries
}\label{ch:preliminaries}

In this chapter we introduce the concepts on which the remainder of this thesis
builds. In \Sref{sec:bta} basic thread algebra is introduced. This allows us to
describe the semantics of instruction sequences. Next, \Sref{sec:pga} and
\Sref{sec:inseqs} introduce two different takes on the way in which instruction
sequences can be represented: on the one hand there is \PGA which describes
finite or infinite single-pass instruction sequences; on the other hand we can
take the (arguably more natural) stance that all instruction sequences must be
finite while allowing instructions to be executed multiple times. It is the
latter theory which describes instruction sequence semigroups, two concrete
instances of which will be introduced in the following chapters as \C and \Cg.

\section{%
  Basic Thread Algebra
}\label{sec:bta}

Basic thread algebra, \BTA for short, is a means to describe the behavior of
sequential programs upon execution. \BTA takes the position that program
execution consists of a sequence of \emph{basic actions} which are performed
inside some execution environment. It is assumed that a fixed but arbitrary set
of basic actions $\actions$ is specified; this parameter is often kept
implicit. Upon execution of an action the execution environment yields a
boolean reply, the value of which specifies how execution should proceed.

In this section we will briefly introduce basic thread algebra. For more on
this subject we refer to \cite{intro_pga_ta, inseq_intro,
pga_second_paper}\footnote{In \cite{pga_second_paper} \BTA is called \BPPA.}.

\BTA expressions are called \emph{threads}. The set of all threads is denoted
$\Thr$. For any set $\actions$, threads are built using two constants and a
single ternary operator:

\begin{itemize}
  \item The \emph{deadlock} constant $\Inac \colon \Thr$.

  \item The \emph{termination} constant $\Term \colon \Thr$.

  \item The \emph{postconditional composition} operator $\aArg \ThrT \aArg
  \ThrF \aArg \colon \Thr \times \actions \times \Thr \to \Thr$.

\end{itemize}

It follows that each closed \BTA expression performs finitely many actions and
then terminates or becomes inactive (in the case of deadlock).

For $P \in \Thr$ and $a \in \actions$, the thread $P \ThrT a \ThrF P$ is often
more conveniently denoted $a \circ P$. The \emph{action prefix} operator
$\circ$ can be used only if the boolean reply returned after execution of $a$
does not influence further behavior. Action prefix binds stronger than
postconditional composition. Additionally, for all $n \geq 1$ we will define
$a^n \circ P$ to mean the thread which performs $n$ $a$-actions, followed by
the behavior described by the thread $P$. That is, $a^1 \circ P = a \circ P$
and $a^{n + 1} \circ P = a \circ (a^n \circ P)$.

The \emph{approximation operator} $\pi \colon \N \times \Thr \to \Thr$ returns
the behavior of a given thread up to a specified ``depth''\footnote{In this
thesis we will use the convention that $\N$ is the set of all natural numbers,
including $0$. $\Np = \N - \{0\}$. The integers are denoted $\Z$.}, i.e., it
bounds the number of actions performed. For all $P, Q \in \Thr$ and $a \in
\actions$ we define,
\begin{align*}
  \pi(0, P)                     &= \Inac \\
  \pi(n + 1, \Term)             &= \Term \\
  \pi(n + 1, \Inac)             &= \Inac \\
  \pi(n + 1, P \ThrT a \ThrF Q) &= \pi(n, P) \ThrT a \ThrF \pi(n, Q)
\end{align*}
From now on we will write $\pi_n(P)$ instead of $\pi(n, P)$ for brevity. Since
every \BTA thread is finite, it follows that for every $P \in \Thr$ there
exists some $n \in \N$ such that for all $m \in \N$,
\[
  \pi_n(P) = \pi_{n + m}(P) = P.
\]
The inclusion relation on threads in \BTA is the partial ordering generated by
the following two clauses:

\begin{itemize}
  \item For all $P \in \Thr$, $\Inac \sqsubseteq P$.

  \item For all $P, P', Q, Q' \in \Thr$ and $a \in \actions$, if $P \sqsubseteq
  P'$ and $Q \sqsubseteq Q'$ then $P \ThrT a \ThrF Q \sqsubseteq P' \ThrT a
  \ThrF Q'$.

\end{itemize}
  
\BTA has a completion \BTAi which also comprises the infinite threads. \BTAi is
the cpo consisting of all projective sequences. We define,
\[
  \ThrI = \{(P_n)_{n \in \N} \mid
            \forall n \in \N(P_n \in \Thr \land \pi_n(P_{n + 1}) = P_n)\}.
\]
Now $(P_n)_{n \in \N} = (Q_n)_{n \in \N}$ if $P_n = Q_n$ for all $n \in \N$.
Furthermore we overload notation and define,
\begin{align*}
  \Inac
    &= (\Inac, \Inac, \dotsc), \\
  \Term
    &= (\Inac, \Term, \Term, \dotsc), \\
  (P_n)_{n \in \N} \ThrT a \ThrF (Q_n)_{n \in \N}
    &= (R_n)_{n \in \N}, \text{with $\begin{cases}
         R_0 = \Inac, \\
         R_{n + 1} = P_n \ThrT a \ThrF Q_n.
       \end{cases}$}
\end{align*}
This definition also shows how all elements of $\Thr$ have a counterpart in
$\ThrI$. The projective sequence corresponding to a thread $P \in \Thr$ is
$(\pi_n(P))_{n \in \N}$.

The set $\res(P)$ of \emph{residual threads} of $P$ has the following inductive
definition:
\begin{align}\label{eq:res}
  P &\in \res(P), &
  Q \ThrT a \ThrF R \in \res(P) &\implies Q \in \res(P) \land R \in \res(P).
\end{align}
Depending on the execution environment a residual thread may be ``reached'' by
performing zero or more actions.

A thread $P$ is \emph{regular} if $\res(P)$ is finite. Regular threads are also
called \emph{finite state threads}. Every element of $\res(P)$ is a state. We
write $\ThrR \subset \ThrI$ for the set of regular threads.

A \emph{finite linear recursive specification} over $\ThrI$ is a set of
equations
\[
  x_i = t_i
\]
for $i \in I$ with $I$ a finite index set, variables $x_i$ and all $t_i$ terms
of the form $\Term$, $\Inac$ or $x_i \ThrT a \ThrF x_j$ with $j, k \in I$ and
$a \in \actions$. $P \in \ThrR$ iff $P$ is the solution of a finite recursive
specification (see Theorem 1 of \cite{inseq_intro}).

\section{%
  Program Algebra: \PGA
}\label{sec:pga}

A program can be viewed as a single-pass instruction sequence. That is, a
program is a finite or infinite sequence of instructions which is executed from
left to right such that every individual instruction is executed at most
once---it is either executed or skipped. Single-pass instruction sequences are
the main concept underlying \PGA \cite{intro_pga_ta, pga_second_paper}. Given
an (implicit) set $\actions$ of actions, \PGA terms are constructed by
concatenating instructions from the set $\primitives$, defined as,
\[
  \primitives = \bigcup_{a \in \actions} \{\pbi{a}, \ppt{a}, \pnt{a}\}
  \cup \bigcup_{k \in \N} \{\pjmp{k}\} \cup \{\pterm\}.
\]
The instructions in $\primitives$ are called \emph{primitive instructions}.
Let us informally define their behavior (note that $a \in \actions$ and $k \in
\N$):

\begin{itemize}
  \item[$\pbi{a}$] is a \emph{basic instruction}. It instructs the execution
  environment to perform action $a$. The boolean reply returned by the
  environment is disregarded.

  \item[$\ppt{a}$] is a \emph{positive test instruction}. Like $\pbi{a}$, it
  instructs execution of action $a$. However, only if the execution environment
  returns \true will the instruction to its immediate right be executed.
  Otherwise this instruction is skipped and execution proceeds at the next
  instruction.

  \item[$\pnt{a}$] is a \emph{negative test instruction}. This is the dual of
  the positive test instruction, in the sense that it skips the next
  instruction iff the environment returns \true after performing action $a$.

  \item[$\pjmp{k}$] is a \emph{forward jump instruction}. This instruction
  transfers execution to the $k$th instruction to its right (i.e., $k - 1$
  instructions are skipped). Note that $\pjmp{0}$ instructs the indefinite
  repetition of this instruction. Hence the behavior of $\pjmp{0}$ is
  identified with deadlock.

  \item[$\pterm$] is the \emph{termination instruction}. It causes successful
  termination of the program.

\end{itemize}

The set of \PGA terms is denoted $\Pga$. \PGA terms are constructed from
primitive instructions using the binary \emph{concatenation} operator $\catOp$
and the unary \emph{repetition} operator $\aArg^\omega$. That is, $\Pga$ is the
smallest superset of $\primitives$ that is closed under concatenation and
repetition. Thus, for all $X, Y \in \Pga$, also $X;Y \in \Pga$ and $X^\omega
\in \Pga$. Examples of \PGA terms include:
\begin{align}\label{eq:pga_ex}
  \pbi{a}, &&
  \ppt{b};\pjmp{3}, &&
  (\pjmp{3};\pbi{a};\pbi{b})^\omega, &&
  \pnt{c};\pnt{c};(\pnt{a})^\omega.
\end{align}

\subsection{%
  First Canonical Form
}

We define $X^1 = X$ and $X^{n + 1} = X;X^n$, for all $n \in \N$. Using this
notation, \PGA defines the following four axioms for all $X, Y, Z \in \Pga$:
\begin{align}
\tag{{\PGA}1}\label{eq:pga1}
  (X;Y);Z &= X;(Y;Z) \\
\tag{{\PGA}2}\label{eq:pga2}
  (X^n)^\omega &= X^\omega \\
\tag{{\PGA}3}\label{eq:pga3}
  X^\omega;Y &= X^\omega \\
\tag{{\PGA}4}\label{eq:pga4}
  (X;Y)^\omega &= X;(Y;X)^\omega
\end{align}
These four axioms define \emph{instruction sequence congruence}. Instruction
sequence congruent \PGA expressions execute exactly the same instructions and
are thus behaviorally equivalent. In the remainder of this thesis instruction
sequence congruent \PGA terms are identified.

\eqref{eq:pga1} states that concatenation is associative. Using \eqref{eq:pga2}
and \eqref{eq:pga4} we derive that $X^\omega = X;X^\omega$ for all $X \in
\Pga$. Furthermore, using \eqref{eq:pga1}--\eqref{eq:pga4} every \PGA term can
be rewritten to one of the following two forms:

\begin{enumerate}
  \item $X$, where $X$ does not contain the repetition operator, or

  \item $X;Y^\omega$, with $X$ and $Y$ not containing the repetition operator.

\end{enumerate}

Any \PGA term in one of these two forms is said to be in \emph{first canonical
form}. The set $\PgaF \subset \Pga$ contains exactly those \PGA terms which are
in first canonical form. The function $\fst \colon \Pga \to \PgaF$ converts any
given \PGA term to a first canonical form. Let $X_1, X_2, Y_1, Y_2 \in \Pga$,
such that $X_1$ and $X_2$ do not contain repetition. Then $\fst$ can be defined
such that,
\begin{align*}
  \fst(Y_1^\omega) &= \fst(Y_1;Y_1^\omega)         & \fst(X_1) &= X_1 \\
  \fst(Y_1^\omega;Y_2) &= \fst(Y_1^\omega)         & \fst(X_1;X_2^\omega) &= X_1;X_2^\omega \\
  \fst(X_1;Y_1^\omega;Y_2) &= \fst(X_1;Y_1^\omega)
\end{align*}
It is not hard to see that $\fst$ is total and makes use only of
\eqref{eq:pga1}--\eqref{eq:pga4}.

\subsection{%
  Second Canonical Form
}\label{sec:snd}

Another congruence relation defined on \PGA terms is \emph{structural
congruence}. It is defined using the following four axioms which are concerned
with chained jump instructions in \PGA terms in first canonical form:
\begin{align}
\tag{{\PGA}5}\label{eq:pga5}
  \pjmp{n{+}1};u_{1};\dotsc;u_n;\pjmp{0}   &= \pjmp{0};u_1;\dotsc;u_n;\pjmp{0}, \\
\tag{{\PGA}6}\label{eq:pga6}
  \pjmp{n{+}1}; u_1;\dotsc;u_n;\pjmp{m}    &= \pjmp{n{+}m{+}1};u_1;\dotsc;u_n;\pjmp{m}, \\
\tag{{\PGA}7}\label{eq:pga7}
  (\pjmp{k{+}n{+}1};u_1;\dotsc;u_n)^\omega &= (\pjmp{k};u_1;\dotsc;u_n)^\omega,
\end{align}
and,
\begin{multline}\tag{{\PGA}8}\label{eq:pga8}
  \pjmp{n{+}m{+}k{+}2};u_1;\dotsc;u_n;(v_1;\dotsc;v_{m{+}1})^\omega = \\
  \pjmp{n{+}k{+}1};u_1;\dotsc;u_n;(v_1;\dotsc;v_{m{+}1})^\omega.
\end{multline}
Using \eqref{eq:pga1}--\eqref{eq:pga8} every \PGA term in first canonical form
can be rewritten to a structurally congruent \PGA term without chained jump
instructions (this also implies that the jump counter of jump instructions into
and inside the repeating part of a \PGA term is minimal). Such a term is said
to be in \emph{second canonical form}. As with first canonical forms, second
canonical forms are not unique. However, any second canonical form $X;Y^\omega$
can be converted to an equivalent second canonical form $X';Y'^\omega$ where
$X'$ and $Y'$ are minimal. Then $X';Y'^\omega$ \emph{is} unique.

The set $\PgaS \subset \PgaF$ contains exactly those \PGA terms which are in
second canonical form. The function $\snd \colon \Pga \to \PgaS$ converts any
\PGA term to its minimal second canonical form. We do not provide an
implementation here.

\subsection{%
  The Semantics of \PGA
}

Every \PGA term $X \in \Pga$ has uniquely defined behavior, in the form of some
thread $T \in \ThrR$. The \emph{thread extraction operator} $\PGATE{\aArg}
\colon \Pga \to \ThrR$ yields this thread, for every \PGA term. It is defined
as,
\begin{equation}\label{eq:pgate}
  \PGATE{X} =
    \begin{cases}
      a \circ \Inac                              & \text{if $X \in \{\pbi{a}, \ppt{a}, \pnt{a}\}$,} \\
      a \circ \PGATE{Y}                          & \text{if $X = \pbi{a};Y$,} \\
      \PGATE{Y} \ThrT a \ThrF \PGATE{\pjmp{2};Y} & \text{if $X = \ppt{a};Y$,} \\
      \PGATE{\pjmp{2};Y} \ThrT a \ThrF \PGATE{Y} & \text{if $X = \pnt{a};Y$,} \\
      \PGATE{Y}                                  & \text{if $X = \pjmp{1};Y$,} \\
      \PGATE{\pjmp{k{+}1};X}                     & \text{if $X = \pjmp{k{+}2};u;Y$,} \\
      \Inac                                      & \text{if $X \in \{\pjmp{k}, \pjmp{0};Y, \pjmp{k{+}2};u\}$,} \\
      \Term                                      & \text{if $X \in \{\pterm, \pterm;Y\}$,} \\
    \end{cases}
\end{equation}
Note that this definition does not explicitly mention the repetition operator.
Instead it uses the notion that $X$ is ``unfolded'' when needed---by means of
\eqref{eq:pga4} and possibly \eqref{eq:pga2}. Thread extraction on \PGA terms
requires one additional rule:
\begin{equation}\label{eq:pgate_inf_jumps}\parbox{0.8\textwidth}{
  If the equations in \eqref{eq:pgate} can be applied infinitely
  often from left to right without ever yielding an action, then the extracted
  thread is $\Inac$.

}\end{equation}
Observe that \eqref{eq:pgate_inf_jumps} is only relevant for \PGA terms which
contain an infinite sequence of chained jump instructions. As such it is not
applicable to second canonical forms.

\paragraph{%
  Examples
}

Let us apply the thread extraction operator $\PGATE{\aArg}$ to the example \PGA
terms of \eqref{eq:pga_ex}.

\begin{itemize}
  \item The behavior of the term $\pbi{a}$ can be derived in a single step
  according to \eqref{eq:pgate}:
  \[
    \PGATE{\pbi{a}} = a \circ \Inac.
  \]

  \item $\ppt{b};\pjmp{3}$ appears to be a more complicated example, but its
  behavior turns out to be equally simple:
  \[
    \PGATE{\ppt{b};\pjmp{3}}
      = \PGATE{\pjmp{3}} \ThrT b \ThrF \PGATE{\pjmp{2};\pjmp{3}}
      = \Inac \ThrT b \ThrF \Inac
      = b \circ \Inac.
  \]

  \item It turns out that the single pass instruction sequence
  $(\pjmp{3};\pbi{a};\pbi{b})^\omega$ does not perform any action, despite its
  infinite length:
  \[
    \PGATE{(\pjmp{3};\pbi{a};\pbi{b})^\omega}
      = \PGATE{(\pjmp{0};\pbi{a};\pbi{b})^\omega}
      = \PGATE{\pjmp{0};(\pbi{a};\pbi{b};\pjmp{0})^\omega}
      = \Inac.
  \]
  Observe that the first step of this derivation applies \eqref{eq:pga7},
  followed by an application of \eqref{eq:pga4}.

  \item Lastly, $\pnt{c};\pnt{c};(\pnt{a})^\omega$ produces infinite behavior.
  To determine its exact behavior, we start out with a couple of left-to-right
  applications of \eqref{eq:pgate}:
  \begin{align*}
    \PGATE{\pnt{c};\pnt{c};(\pnt{a})^\omega}
      &= \PGATE{\pjmp{2};\pnt{c};(\pnt{a})^\omega} \ThrT c \ThrF \PGATE{\pnt{c};(\pnt{a})^\omega} \\
      &= \PGATE{(\pnt{a})^\omega} \ThrT c \ThrF (\PGATE{\pjmp{2};(\pnt{a})^\omega} \ThrT c \ThrF \PGATE{(\pnt{a})^\omega}) \\
      &= \PGATE{(\pnt{a})^\omega} \ThrT c \ThrF (\PGATE{\pjmp{2};\pnt{a};(\pnt{a})^\omega} \ThrT c \ThrF \PGATE{(\pnt{a})^\omega}) \\
      &= \PGATE{(\pnt{a})^\omega} \ThrT c \ThrF (\PGATE{(\pnt{a})^\omega} \ThrT c \ThrF \PGATE{(\pnt{a})^\omega}) \\
      &= \PGATE{(\pnt{a})^\omega} \ThrT c \ThrF c \circ \PGATE{(\pnt{a})^\omega}. \\
  \end{align*}
  At this stage the behavior of $\pnt{c};\pnt{c};(\pnt{a})^\omega$ has not been
  fully derived, as the thread corresponding to $(\pnt{a})^\omega$ still needs
  to be determined. This thread turns out to be infinite:
  \begin{align*}
    \PGATE{(\pnt{a})^\omega}
    &= \PGATE{\pnt{a};(\pnt{a})^\omega} \\
    &= \PGATE{\pjmp{2};(\pnt{a})^\omega} \ThrT a \ThrF \PGATE{(\pnt{a})^\omega} \\
    &= \PGATE{\pjmp{2};\pnt{a};(\pnt{a})^\omega} \ThrT a \ThrF \PGATE{(\pnt{a})^\omega} \\
    &= \PGATE{(\pnt{a})^\omega} \ThrT a \ThrF \PGATE{(\pnt{a})^\omega} \\
    &= a \circ \PGATE{(\pnt{a})^\omega}.
  \end{align*}
  It follows that $\PGATE{(\pnt{a})^\omega}$ can be described by the recursive
  specification $Q = a \circ Q$. Now, equating
  $\PGATE{\pnt{c};\pnt{c};(\pnt{a})^\omega}$ with $P$, we see that the behavior
  of $\pnt{c};\pnt{c};(\pnt{a})^\omega$ is equals $P_1$, as described by the
  following linear recursive specification:
  \begin{align*}
    P_1 &= P_3 \ThrT c \ThrF P_2, &
    P_2 &= P_3 \ThrT c \ThrF P_3, &
    P_3 &= P_3 \ThrT a \ThrF P_3.
  \end{align*}
  (A shorter notation would be $P = Q \ThrT c \ThrF c \circ Q$, $Q = a \circ
  Q$.)

\end{itemize}

\begin{prop}
  Each thread definable in \PGA is regular, and each regular thread can be
  expressed in \PGA.

\end{prop}

\begin{proof}
  See e.g. Proposition 2 in \cite{intro_pga_ta}. Alternatively, the result
  follows from the following two observations:

  \begin{itemize}
    \item The code semigroup \C introduced in \Cref{ch:c_intro} characterizes
    the regular threads (see \PropRef{prop:c_characterizes_regular_threads}).

    \item There exist total behavior preserving mappings from \PGA to \C and
    vice versa (see \Sref{sec:pga_to_c} and \Sref{sec:c_to_pga}, respectively).
    \qedhere 

  \end{itemize}
\end{proof}

\section{%
  Finite Instruction Sequences and Code Semigroups
}\label{sec:inseqs}

In \PGA each instruction is executed at most once and the repetition operator
$\aArg^\omega$ is used to construct infinite sequences of instructions. The
instruction sequence semigroups introduced in the following chapters, on the
other hand, represent only finite instruction sequences in which instructions
can be executed multiple times in any order. This section introduces some
relevant notions and terminology in preparation of the introduction of concrete
code semigroups in \Cref{ch:c_intro} and \Cref{ch:cg_intro}.

\subsection{%
  Finite Instruction Sequences
}

Consider a non-empty instruction set $\U$ and an associative binary operation
$\catOp$ on $\U$. We will call $\catOp$ the \emph{concatenation operator}.
Instructions can be concatenated, thereby yielding finite \emph{instruction
sequences} (\emph{inseqs}) of arbitrary length. For all $n \in \Np$, let
\begin{align*}
  \Un{1}     &= \U, &
  \Un{n + 1} &= \{X;u \mid X \in \Un{n}, u \in \Un{1}\}.
\end{align*}
Then $\Un{n}$ is the set of instruction sequences of length $n$. We define
\[
  \UA = \bigcup_{n \in \Np} \Un{n}.
\]
$\UA$ contains all finite, non-empty (length greater than zero) sequences of
$\U$-instructions. $X$ is an \emph{$\U$-inseq} iff $X \in \UA$. An $\U$-inseq
will also be called an \emph{$\U$-expression}. We call $\len \colon \UA \to
\N^+$ the \emph{length function}, and it is defined such that $\len(X) = n$ iff
$X \in \Un{n}$.

Concatenation is an associative operation, thus $(X;Y);Z = X;(Y;Z)$ for
arbitrary $X, Y, Z \in \UA$. Parentheses will therefore usually be omitted, and
we write $X;Y;Z$. Note also, that it trivially follows that for arbitrary $n, m
\geq 1$,
\[
  \Un{n + m} = \{X;Y \mid X \in \Un{n}, Y \in \Un{m}\}.
\]
For convenience, we will write $\Un{\leq n}$ for the set of all
$\U$-expressions up to length $n$. Likewise $\Un{\geq n}$ contains all
$\U$-expressions of length $n$ or greater. That is,
\begin{align*}
  \Un{\leq n} &= \{X \in \UA \mid \len(X) \leq n\}, &
  \Un{\geq n} &= \{X \in \UA \mid \len(X) \geq n\}.
\end{align*}
For all $i \in \Np$, we define auxiliary functions $\inst{i} \colon \Un{\geq i}
\to \U$ which return the $i$th instruction in a given $\U$-inseq. That is, if
$X = u_1;u_2;\dotsc;u_n$, then $\inst{i}(X) = u_i$ for all $1 \leq i \leq n$.
We define $i \eq{X} j$ iff $\inst{i}(X) = \inst{j}(X)$. Clearly $\eq{X}$ is an
equivalence relation.

Next, for all $X \in \UA$ and $U \subseteq \U$ we define $U(X) = \{i \mid
\inst{i}(X) \in U\}$. In other words, $U(X)$ contains the positions in the
$\U$-inseq $X$ of instructions contained in $U$.

It will sometimes prove convenient to regard an inseq $X$ as a set whose
elements are the distinct instructions contained in $X$. So for any $X \in \UA$
we write $u \in X$ to indicate that $\inst{i}(X) = u$ for some $i$. $X \cap S$
and $X \cup S$ are defined as one would expect them to be (note that $S$ can be
a set or another inseq).

\paragraph{%
  About Notation
}

Let $X \in \UA$ be an instruction sequence. Throughout this thesis we will
write $X^k$ for $k$ concatenations of $X$. That is,
\begin{align*}
  X^1 &= X, &
  X^{n + 1} = X;X^n.
\end{align*}
What about $X^0$? Our definition of an instruction sequence explicitly excludes
the empty sequence: an $\U$-expression will always contain at least one
instruction. Still, within some contexts it will prove convenient to talk about
$X^k$ for \emph{any} $k \in \N$. Throughout this thesis we will only write
$X^0$ as part of sequences which, as a whole, are guaranteed to be non-empty,
and are as such contained in $\UA$ (i.e., the set of proper instruction
sequences).

\subsection{%
  Code Semigroups
}

Given some instruction set $\U$, every inseq $X \in \UA$ is constructed by
concatenation of a finite number of elements in $\U$. Hence $\U$
\emph{generates} $\UA$, denoted $\gen{\U} = \UA$. $\UA$ is closed under the
associative binary operation $\catOp$ and as such $\UA$ is a semigroup with
respect to $\catOp$. Clearly every instruction set $\U$ gives rise to a
semigroup $(\gen{\U}, \catOp)$. We will call such a semigroup an
\emph{instruction sequence semigroup} or simply \emph{code semigroup}. For an
introduction to semigroup theory we refer to \cite{semigroup_theory}.

\paragraph{%
  About Notation
}

Let $\B$ refer to some code semigroup. Then we write $\UB$ for the instruction
set of $\B$. $\UnB{n}$ denotes the $\UB$-inseqs of length $n$, and $\UAB$
contains all $\UB$-expressions. Hence we write $\B = (\UAB, \catOp)$. When no
confusion can arise, $\UB$-instructions and $\UB$-inseqs may simply be referred
to as $\B$-instructions and $\B$-inseqs ($\B$-expressions), respectively.
Whenever $\B$ is referred to as a set instead of a semigroup, it is identified
with $\UAB$. That is, $\B$ stands for all well-formed $\B$-expressions.
Likewise $\UAB$ may be referred to as a semigroup, in which case it is
identified with $\B$.

\paragraph{%
  Subsemigroups
}

Let $A$ and $B$ be two semigroups with respect to some operator ${\bullet}$,
such that $A \subseteq B$. Then $A$ is a subsemigroup of $B$. Equivalently, if
$(B, \bullet)$ is a semigroup and $A \subseteq B$ such that $a, b \in A$
implies $a \bullet b \in A$, then $A$ is a subsemigroup of $B$. Note that the
intersection of any subsemigroups of $B$ is either empty or itself a
subsemigroup of $B$.\footnote{We will not consider the empty semigroup.}

Given an instruction set $\U$ we can take a subset of these instructions, $\U'
\subseteq \U$. Observe that the semigroup $\gen{\U'}$ is a strict subsemigroup
of $\gen{\U}$. We will define plenty of such subsemigroups later in this
thesis.

\paragraph{%
  Semigroup Homomorphisms
}

Consider two code semigroups $A$ and $B$, and a function $f \colon \UA_A \to
\UA_B$. Then $f$ is a mapping between instruction sequences. A significant part
of this thesis describes mappings between distinct code semigroups. Most of
these mappings are homomorphisms.

In general, a function $f \colon A \to B$ is a homomorphism between semigroups
$(A, \bullet)$ and $(B, \ast)$ iff $f(x \bullet y) = f(x) \ast f(y)$, for all
$x, y \in A$. It is easy to see that $f$ only needs to be defined explicitly on
elements of $A$'s generating set. If $\gen{G} = A$, then for all $a \in A - G$
it is the case that $a = g_0 \bullet g_1 \bullet \dotsb \bullet g_{n + 1}$, for
some $n \in \N$ and $g_0, g_1, \dotsc, g_{n + 1} \in G$, and hence $f(a) =
f(g_0) \ast f(g_1) \ast \dotsb \ast f(g_{n + 1})$ by definition. In the
specific case of code semigroups this implies that a homomorphic function only
needs to be defined explicitly on individual instructions.

\subsection{%
  Instruction Sequence Semantics
}\label{sec:inseq_semantics}

It is the ability to be \emph{executed} that sets instruction sequences apart
from sequences of arbitrary mathematical objects. Execution of an instruction
sequence leads to (possibly unobservable) behavior. Thus, for a sequence of
objects to be called an instruction sequence, it must be ascribed a semantics,
such that its behavior upon execution is defined.

This thesis will use basic thread algebra to that end. This allows us to define
the semantics of the semigroup \C as in \cite{inseq_intro} and provides an easy
way to compare the code semigroups introduced in this thesis to \PGA on a
syntactical as well as a semantic level.

In the tradition of \PGA instructions are viewed as atomic program components:
at any stage during the execution of a program at most one instruction is
``active'' (i.e., being executed).\footnote{One could draw a parallel with the
program counter as found in central processing units (CPUs), which holds the
memory address of the instruction that is currently executed (or the
instruction which is to be executed next, depending on the architecture).} We
will define the behavior of individual instructions based on their position $i$
within an instruction sequence $X$. Execution of an individual instruction may
or may not cause an action to be performed, after which control of execution is
transferred to another position in $X$. Then, given the position of the first
instruction to be executed, the semantics of the instruction sequence as a
whole follows naturally.

The first instruction to be executed is called the \emph{initial} or
\emph{start instruction}. The leftmost and rightmost instruction of an inseq
are obvious candidates to be designated as such, but given a specific
instruction sequence $X$, execution can start at any position within $X$. Thus
for all $X \in \UA$ and $1 \leq i \leq \len(X)$, the pair $(i, X)$ can be
identified with a certain thread, namely the thread which represents the
behavior resulting from the execution of $X$ starting with the $i$th
instruction. Though not strictly necessary, for any invalid instruction
position $i$ (i.e. $i < 1$ or $i > \len(X)$) the pair $(i, X)$ will be
identified with some default thread $D$. Once $D$ has been fixed, every pair
$(i, X) \in \Z \times \UA$ is identified with a certain thread $T$. Throughout
this thesis we will consider only one value for $D$, namely $\Inac$, i.e.
deadlock.

In this way the \emph{thread extraction operator} $\TElong{\aArg}{\aArg} \colon
\Z \times \UA \to \ThrI$ specifies the semantics of a semigroup $\UA$. For
convenience we will usually write $\TE{i}{X}$ instead of $\TElong{i}{X}$, but
this is merely a notational matter. For any $X \in \UA$, the thread describing
the behavior of $X$ if executed starting from the leftmost instruction is
called its \emph{left behavior}, written $\TEfwd{X} = \TE{1}{X}$. Likewise
$\TEbwd{X} = \TE{\len(X)}{X}$ is called the \emph{right behavior} of $X$,
meaning the behavior of $X$ if executed starting from the rightmost
instruction.

Once specific code semigroups have been defined---along with suitable thread
extraction operators---it becomes possible to analyze their expressiveness.
Given equally expressive code semigroups $A$ and $B$ one can define mappings
between them, such that the behavior of any inseq $X$ in the domain is in some
way reflected by the behavior of the corresponding inseq $Y$ to which it is
mapped in the codomain. Similar mappings can also be defined from a semigroup
$A$ onto itself.

\begin{defn}
  Let $A$ and $B$ be two code semigroups on which the thread extraction
  operators $\ATElong{\aArg}{\aArg} \colon \Z \times \UA_A \to \ThrI$ and
  $\BTElong{\aArg}{\aArg} \colon \Z \times \UA_B \to \ThrI$ are defined,
  respectively. Consider arbitrary $X \in \UA_A$ and $Y \in \Pga$ and three
  mappings $f \colon \UA_A \to \UA_B$, $g \colon \UA_A \to \Pga$ and $h \colon
  \Pga \to \UA_A$. Then,

  \begin{itemize}
    \item $f$ is \emph{left behavior preserving} if $\ATEfwd{X} =
    \BTEfwd{f(X)}$.

    \item $f$ is \emph{right behavior preserving} if $\ATEbwd{X} =
    \BTEbwd{f(X)}$.

    \item $f$ is \emph{left-right behavior preserving} if it is both left and
    right behavior preserving.

    \item $f$ is \emph{behavior preserving} if it is left or right behavior
    preserving.

    \item $f$ is \emph{left uniformly behavior preserving} if there exists some
    $b \in \Np$ such that $\ATE{i}{X} = \BTE{b(i - 1) + 1}{f(X)}$ for all $i
    \in \Z$. Observe that every left uniformly behavior preserving mapping is
    left behavior preserving.

    \item $f$ is \emph{right uniformly behavior preserving} if there exists
    some $b \in \Np$ such that $\ATE{i}{X} = \BTE{bi}{f(X)}$ for all $i \in
    \Z$. Observe that every right uniformly behavior preserving mapping is
    right behavior preserving.

    \item $f$ is \emph{left-right uniformly behavior preserving} if it is both
    left and right uniformly behavior preserving.

    \item $f$ is \emph{uniformly behavior preserving} if it is left or right
    uniformly behavior preserving.

    \item $g$ is \emph{behavior preserving} if $\ATEfwd{X} = \PGATE{g(X)}$.

    \item $h$ is \emph{behavior preserving} if $\PGATE{Y} =
    \ATEfwd{h(Y)}$.\footnote{A more general definition would be that $h$ is
    behavior preserving if there exists a function $t \colon \Pga \to \Z$ such
    that $\PGATE{Y} = \ATE{t(Y)}{h(Y)}$, but this definition suffices for our
    purposes.}

  \end{itemize}

  A behavior preserving mapping will also be called a \emph{translation}
  because it preserves the meaning of the original (single pass) instruction
  sequence.

\end{defn}

This concludes the preliminaries. We are now ready to introduce the code
semigroup \C in the next chapter.

\chapter[\C Instruction Sequences]{%
  $\boldsymbol{\C}$ Instruction Sequences
}\label{ch:c_intro}

The previous chapter introduced \PGA as a means to describe programs and \BTA
as a means to describe their behavior. It then introduced an alternative
representation of program objects, namely strictly finite instruction
sequences, as opposed to \PGA's infinite single-pass instruction sequences.
Upon specifying an instruction set $\U$ the set of finite instruction sequences
generated by concatenating elements of $\U$ forms a semigroup. This chapter
introduces one such semigroup and its semantics.

\C was first described in \cite{inseq_intro}. \C is a code semigroup without
directional bias: execution of a \C-inseq can start at the leftmost instruction
(the natural choice for most people in Western society), but may just as well
start at the rightmost instruction. In fact, given some instruction sequence
$X$, any position within $X$ can be designated as starting position.

This chapter is built up as follows: \Sref{sec:c_instr} will introduce \C's
instruction set and provide some basic examples of \C-expressions. It will also
motivate the inclusion in the instruction set of an instruction which upon
execution will cause deadlock. Next, \Sref{sec:c_semantics} formalizes the
semantics of \C-expressions using thread algebra. Based on this,
\Sref{sec:c_accessibility} introduces some accessibility relations on
instruction positions which will be used throughout this thesis. Lastly,
\Sref{sec:c_alternative} briefly discusses a small syntactic and semantic
variation on \C.

\section{%
  The Instruction Set
}\label{sec:c_instr}

Given a set $\actions$ of actions, \C defines basic instructions $\basics$,
positive test instructions $\positives$, negative test instructions
$\negatives$ and relative jumps $\jumps$:
\begin{align*}
  \basics    &= \bigcup_{a \in \actions} \{\fbi{a}, \bbi{a}\}, &
  \jumps     &= \bigcup_{k \in \Np} \{\fj{k}, \bj{k}\}, \\
  \positives &= \bigcup_{a \in \actions} \{\fpt{a}, \bpt{a}\}, &
  \negatives &= \bigcup_{a \in \actions} \{\fnt{a}, \bnt{a}\}.
\end{align*}
$\actions$ is a parameter to \C which is often kept implicit. Additionally, \C
has an abort instruction $\abrt$ and termination instruction $\term$.
Instructions with a backward slash are called \emph{left oriented} or
\emph{backward} instructions; those with a forward slash are called \emph{right
oriented} or \emph{forward} instructions. Instructions with a left (right)
orientation are also said to have a left (right) \emph{directionality}.
Formally, $\C = (\UAC, \catOp)$, with the set of all \C-expressions $\UAC$
generated by \C's instruction set $\UC$, defined as
\[
  \UC = \basics \cup \positives \cup \negatives \cup \jumps \cup \{\abrt, \term\}.
\]
Let $a, b, c \in \actions$. Then examples of \C-expressions are
\begin{align}\label{eq:c_ex}
  \fbi{a}, &&
  \fbi{a};\fpt{a};!;\bj{3}, &&
  \fpt{b};\fnt{c};\bnt{c}, &&
  \bj{2};\bnt{c}.
\end{align}
Each \C-inseq has a semantics. Before we formalize this, it will prove
convenient to informally describe the meaning of some of the instructions:

\begin{itemize}
  \item[$\fbi{a}$] is a \emph{forward basic instruction}. It causes execution
  of the action $a$, after which the instruction to its right is executed, if
  it exists. Otherwise deadlock occurs. Note that the boolean reply resulting
  from $a$'s execution is ignored.

  \item[$\fpt{a}$] is a \emph{forward positive test instruction}. Action $a$ is
  executed. If its boolean reply is \true, then the instruction immediately to
  its right is executed. On \false, however, this instruction is skipped, and
  execution proceeds at the second instruction to its right. If no such
  instruction exists, deadlock follows.

  \item[$\fnt{a}$] is a \emph{forward negative test instruction}. $\fnt{a}$
  mirrors the behavior of $\fpt{a}$, in the sense that the effect of the
  replies \true and \false is reversed.

  \item[$\fj{k}$] is a \emph{forward jump instruction}. It causes execution of
  the instruction $k$ positions to its right, if such instruction exists.
  Otherwise deadlock will follow.

  \item[$\abrt$] is the \emph{abort instruction}. Execution of this instruction
  causes deadlock.

  \item[$\term$] is the \emph{termination instruction}. It causes the program
  to halt successfully.

\end{itemize}

The instructions $\bbi{a}$, $\bpt{a}$, $\bnt{a}$ and $\bj{k}$ are the
\emph{backward} versions of $\fbi{a}$, $\fpt{a}$, $\fnt{a}$ and $\fj{k}$,
respectively, in the sense that they have a right-to-left instead of a
left-to-right orientation. For example, execution of $\bbi{a}$ results in
action $a$, after which the instruction to its \emph{left} is executed (if such
instruction exists).

A jump instruction $\fj{k}$ or $\bj{k}$ has \emph{jump counter} $k$ and
performs a jump of \emph{distance} $k$ instructions. $\fj{k}$ or $\bj{k}$ are
said to be \emph{relative jumps}. The function $\jd \colon \jumps \to \Np$
returns the jump counter of a given jump instruction (e.g. $\jd(\fj{6}) = 6$).

We define $\FUC \subset \UC$ to be the set forward instructions. Likewise,
$\BUC \subset \UC$ denotes the set of backward instructions.
Formally:\footnote{Note, again, that the set $\actions$ of actions is an
implicit parameter for \C (and thereby for $\UC$, $\FUC$ and $\BUC$).}
\begin{align*}
  \FUC &= \bigcup_{a \in \actions} \{\fbi{a}, \fpt{a}, \fnt{a}\}
          \cup \bigcup_{k \in \Np} \{\fj{k}\}, \\
  \BUC &= \bigcup_{a \in \actions} \{\bbi{a}, \bpt{a}, \bnt{a}\}
          \cup \bigcup_{k \in \Np} \{\bj{k}\}.
\end{align*}
The sets $\Fbasics = \basics \cap \FUC$ and $\Bbasics = \basics \cap \BUC$
denote the forward and backward oriented basic instructions, respectively.
Likewise for $\positives$, $\negatives$ and $\jumps$. Note that with the
exception of the abort instruction $\abrt$ and the termination instruction
$\term$, every \C-instruction has a direction, which is either forward or
backward, but not both. That is, $\FUC \cap \BUC = \emptyset$ and $\FUC \cup
\BUC \cup \{\abrt, \term\} = \UC$. We write $u \de v$ if instructions $u$ and
$v$ have the same direction (or no direction). ${\de} \subset \UC \times \UC$
is the \emph{directionality relation}. It is clearly an equivalence relation.

\paragraph{%
  Examples
}

These informal definitions of the meaning of each instruction allow us to
verbally describe the meaning of the example \C-expressions of \eqref{eq:c_ex},
provided that we agree upon which instruction is the first to be executed.
Since this thesis is written in English, which has an obvious left-to-right
bias, we will designate the leftmost instruction to be the initial instruction.
Thus we will informally describe these inseq's left behavior.

\begin{itemize}
  \item $\fbi{a}$: Performs action $a$, after which deadlock occurs.

  \item $\fbi{a};\fpt{a};!;\bj{3}$: Performs action $a$ twice in a row. If the
  second action yields a positive reply, then the program terminates. Otherwise
  it starts all over.

  \item $\fpt{b};\fnt{c};\bnt{c}$: Performs action $b$. If this yields the
  reply \true, then action $c$ will be performed, as specified by the second
  instruction. Here, a positive reply causes deadlock and a negative reply
  causes the third instruction to be executed. If the action $b$ yields \false
  then the third instruction will also be executed. The action $c$ as performed
  by the third instruction causes execution to continue at either the first or
  second instruction, depending on whether it yields is a positive or negative
  reply, respectively.\footnote{Compare the length of this description to that
  of the actual program, and it becomes apparent that natural language is not
  really suited to produce concise descriptions of program behavior. There is
  also the problem of the inherent ambiguity of natural language. Luckily basic
  thread algebra provides a concise and unambiguous alternative!}

  \item $\bj{2};\bnt{c}$: Does not perform any action. Execution of this
  program immediately causes deadlock, since the first instruction jumps
  outside of the inseq.

\end{itemize}

\subsection{%
  The Case for an Explicit Abort Instruction
}\label{sec:c_abrt}

A draft version of the original paper on \C
\cite{inseq_intro_draft_no_abort_instr} provided a definition of the semigroup
\C which differs slightly from the one that was published in \cite{inseq_intro}
(which is introduced in the previous section). Let us refer to the semigroup as
it was introduced in \cite{inseq_intro_draft_no_abort_instr} by the name \Cp.

The instruction set $\UCp$ did not contain an explicit abort instruction. It
did however contain two other instructions which \C lacks: $\fj{0}$ and
$\bj{0}$, both of which signify a jump of distance zero\footnote{The existence
of these instructions was probably inspired by the $\pjmp{0}$ instruction as
found in \PGA.}. That is, $\Cp = (\UACp, \catOp)$, with
\[
  \UCp = \{\fj{0}, \bj{0}\} \cup \UC - \{\abrt\}.
\]
Since $\fj{0}$ and $\bj{0}$ are under all circumstances behaviorally
indistinguishable, \Cp had an extra axiom (aside from the obvious axiom which
states that concatenation is associative) which stated that no distinction is
made between forward and backward jumps of distance $0$:
\[
  \fj{0} = \bj{0}.
\]
A jump of distance $0$ is not really a jump at all and it is rather meaningless
to talk about the direction of such a jump. Semantically both $\fj{0}$ and
$\bj{0}$ signify deadlock. Moreover, the introduction of two distinct but
equivalent instructions allows for the definition of a mapping $f$ on $\UACp$
such that $X = Y$ while $f(X) \neq f(Y)$.

It was therefore argued that \Cp should only contain jumps $\fj{k}$ and
$\bj{k}$ for $k > 0$, together with a single non-directional abort instruction
$\abrt$, thereby eliminating the need for the axiom $\fj{0} = \bj{0}$ while
retaining a single instruction with essentially the same behavior as $\fj{0}$
and $\bj{0}$. This chain of reasoning naturally lead to the definition of an
alternative semigroup, the one introduced in \cite{inseq_intro} and the
previous section under the name \C.\footnote{The introduction of the
instruction $\abrt$ is not really a first. A similar instruction can be found
in \cite{pga_first_paper}, where it is introduced as part of \PGA. It must be
noted though, that \cite{pga_first_paper} ascribes a different semantics to
$\abrt$, namely \emph{meaningless behavior}, than to $\pjmp{0}$, which produces
\emph{divergent behavior}. The latter notion coincides with what is referred to
in this thesis as deadlock ($\Inac$ in basic thread algebra). \BTA does not
provide a constant to represent meaningless behavior. As mentioned in a
footnote in \cite{pga_second_paper}, $\abrt$ was later dropped and should in
hindsight be seen as an abbreviation for $\pjmp{0}$.}

Execution of the abort instruction has the same effect as an attempt to
transfer execution to a non-existing instruction. Since every instruction
sequence is finite, one can take any inseq $X \in \UAC$ and construct a
behaviorally equivalent inseq $X'$ by replacing every abort instruction with a
jump to a position $< 1$ or $> \len(X)$. Hence $\abrt$ does not increase \C's
expressiveness. Still, as we will later see, the abort instruction is a
convenient addition to the instruction set.

For completeness, we define two homomorphisms $f \colon \UACp \to \UAC$ and $g
\colon \UAC \to \UACp$ which make the correspondence between \Cp and \C
explicit. They are defined on individual instructions $u$ as follows:
\begin{align*}
  f(u) &=
    \begin{cases}
      \abrt & \text{if $u \in \{\fj{0}, \bj{0}\}$,} \\
      u     & \text{otherwise,}
    \end{cases} &
  g(u) &=
    \begin{cases}
      \fj{0} & \text{if $u = \abrt$,} \\
      u      & \text{otherwise.}
    \end{cases} &
\end{align*}
Now clearly $f \circ g$ is the identity function on \C-expressions. The axiom
$\fj{0} = \bj{0}$ ensures that likewise $g \circ f$ is an identity function on
\Cp-expressions.

\section{%
  Semantics
}\label{sec:c_semantics}

As discussed in \Sref{sec:inseq_semantics}, \C's semantics are defined using
basic thread algebra. Thus any combination of start position $i$ and inseq $X
\in \UAC$ is assigned some thread $\CTElong{i}{X}$. Writing $\CTE{i}{X}$ for
$\CTE{}{i, X}$, the thread extraction operator $\CTElong{\aArg}{\aArg} \colon
\Z \times \UAC \to \ThrR$ is defined on all $i \in \Z$ and $X \in \UAC$ as,
\begin{equation}\label{eq:cte}
  \CTE{i}{X} =
    \begin{cases}
      \Inac                                       & \text{if $i < 1$ or $i > \len(X)$,}
      \smallskip \\
      a \circ \CTE{i + 1}{X}                      & \text{if $\inst{i}(X) = \fbi{a}$,} \\
      \CTE{i + 1}{X} \ThrT a \ThrF \CTE{i + 2}{X} & \text{if $\inst{i}(X) = \fpt{a}$,} \\
      \CTE{i + 2}{X} \ThrT a \ThrF \CTE{i + 1}{X} & \text{if $\inst{i}(X) = \fnt{a}$,} \\
      \CTE{i + k}{X}                              & \text{if $\inst{i}(X) = \fj{k}$,}
      \smallskip \\
      a \circ \CTE{i - 1}{X}                      & \text{if $\inst{i}(X) = \bbi{a}$,} \\
      \CTE{i - 1}{X} \ThrT a \ThrF \CTE{i - 2}{X} & \text{if $\inst{i}(X) = \bpt{a}$,} \\
      \CTE{i - 2}{X} \ThrT a \ThrF \CTE{i - 1}{X} & \text{if $\inst{i}(X) = \bnt{a}$,} \\
      \CTE{i - k}{X}                              & \text{if $\inst{i}(X) = \bj{k}$,}
      \smallskip \\
      \Inac                                       & \text{if $\inst{i}(X) = \abrt$,} \\
      \Term                                       & \text{if $\inst{i}(X) = \term$.}
    \end{cases}
\end{equation}
In words, the thread $\CTE{i}{X}$ describes the behavior resulting from the
execution of the inseq $X$ starting at the $i$th instruction. Recall that we
defined $\CTEfwd{X} = \CTE{1}{X}$ and $\CTEbwd{X} = \CTE{\len(X)}{X}$ to mean
$X$'s left and right behavior, respectively.

\paragraph{%
  Examples
}

We will apply thread extraction on the instruction sequences of \eqref{eq:c_ex}
to determine their left as well as right behavior.

\begin{itemize}
  \item The \C-expression $\fbi{a}$ consists of a single instruction, and as
  such its left and right behavior are equivalent:
  \begin{align*}
    \CTEfwd{\fbi{a}}
      &= \CTE{1}{\fbi{a}}
       = a \circ \Inac, &
    \CTEbwd{\fbi{a}}
      &= \CTE{\len(\fbi{a})}{\fbi{a}}
       = \CTE{1}{\fbi{a}}
       = a \circ \Inac.
  \end{align*}

  \item Let $X = \fbi{a};\fpt{a};!;\bj{3}$. The left behavior of this
  instruction sequence is infinite, as we have seen in \Sref{sec:c_instr}. This
  is confirmed by several applications of equations in \eqref{eq:cte}:
  \[
    \CTEfwd{X}
      = \CTE{1}{X}
      = a \circ \CTE{2}{X}
      = a \circ (\CTE{3}{X} \ThrT a \ThrF \CTE{4}{X})
      = a \circ (\Term \ThrT a \ThrF \CTE{1}{X}).
  \]
  Observe that $\CTEfwd{X}$ is recursively defined by the equation $P = a \circ
  (\Term \ThrT a \ThrF P)$. As for the right behavior of $X$, we observe that
  $\CTEbwd{X} = \CTE{\len(X)}{X} = \CTE{4}{X} = \CTE{1}{X} = \CTEfwd{X}$.

  \item Let $X = \fpt{b};\fnt{c};\bnt{c}$. Upon trying to extract its behavior,
  we see that
  \begin{align*}
    \CTEfwd{X} &= \CTE{1}{X} \\
      &= \CTE{2}{X} \ThrT b \ThrF \CTE{3}{X} \\
      &= (\CTE{4}{X} \ThrT c \ThrF \CTE{3}{X}) \ThrT b \ThrF (\CTE{1}{X} \ThrT c \ThrF \CTE{2}{X}) \\
      &= (\Inac \ThrT c \ThrF (\CTE{1}{X} \ThrT c \ThrF \CTE{2}{X})) \ThrT b \ThrF (\CTE{1}{X} \ThrT c \ThrF \CTE{2}{X}).
  \end{align*}
  The behavior is clearly infinite, and no single recursive equation can
  describe it. The following linear recursive specification does:
  \begin{align*}
    P_1 &= P_2 \ThrT b \ThrF P_3, &
    P_2 &= P_4 \ThrT c \ThrF P_3, &
    P_3 &= P_1 \ThrT c \ThrF P_2, &
    P_4 &= \Inac.
  \end{align*}
  Now $\CTEfwd{X} = P_1$ and $\CTEbwd{X} = P_3$.

  \item Let $X = \bj{2};\bnt{c}$. Then $\CTEfwd{X} = \Inac$ and $\CTEbwd{X} = c
  \circ \Inac$, because
  \begin{align*}
    \CTEfwd{X} &= \CTE{1}{X} = \CTE{-1}{X} = \Inac \\
    \CTEbwd{X}
      &= \CTE{2}{X}
      = \CTE{0}{X} \ThrT c \ThrF \CTE{1}{X}
      = \Inac \ThrT c \ThrF \CTE{-1}{X}
      = \Inac \ThrT c \ThrF \Inac
      = c \circ \Inac
  \end{align*}

\end{itemize}

\paragraph{%
  Loops Without Activity
}

\C-inseqs may contain chained jump instructions which form a loop. The
equations of \eqref{eq:cte} do not adequately handle this situation, as they do
not assign any specific thread to the execution of such a loop. Hence we
introduce an additional rule for the extraction of behavior from \C instruction
sequences:\footnote{This rule is near identical to the rule
\eqref{eq:pgate_inf_jumps} which assigns a thread to infinite sequences of
chained jump instructions in \PGA.}
\begin{equation}\label{eq:cte_inf_jumps}\parbox{0.8\textwidth}{
  If the equations in \eqref{eq:cte} can be applied infinitely
  often from left to right without ever yielding an action, then the extracted
  thread is $\Inac$.

}\end{equation}
As an example of the application of this rule, consider the \C instruction
sequence $X = \fj{3};\bj{1};\term;\bj{2};\abrt;\bpt{a}$. Its left behavior is
\[
  \CTEfwd{X}
    = \CTE{1}{X}
    = \CTE{4}{X}
    = \CTE{2}{X}
    = \CTE{1}{X}
    = \Inac.
\]
Here we derive that $\CTE{1}{X}$, $\CTE{4}{X}$ and $\CTE{2}{X}$ equal $\Inac$
by means of three left-to-right applications of equations in \eqref{eq:cte}
followed by application of \eqref{eq:cte_inf_jumps}. Indeed, the instructions
at positions $1$, $2$ and $4$ form a closed loop without any non-jump
instructions. This example is also yet another demonstration of the fact that
the left and right behavior of an inseq are in general not equivalent; the
right behavior of $X$ is,
\[
  \CTEbwd{X}
    = \CTE{\len(X)}{X}
    = \CTE{6}{X}
    = \CTE{5}{X} \ThrT a \ThrF \CTE{4}{X}
    = \Inac \ThrT a \ThrF \Inac
    = a \circ \Inac.
\]

\begin{prop}\label{prop:c_characterizes_regular_threads}
  Each thread definable in \C is regular, and each regular thread can be
  expressed in \C.

\end{prop}

\begin{proof}
  Let $X \in \UAC$. Following \eqref{eq:cte} and \eqref{eq:cte_inf_jumps} we
  have that for arbitrary $i \in [1, \len(X)]$ one of the following is the case
  (for some $j, k \in \Z$):
  \begin{align*}
    \CTE{i}{X} &= \Term, &
    \CTE{i}{X} &= \Inac, &
    \CTE{i}{X} &= \CTE{j}{X} \ThrT a \ThrF \CTE{k}{X}.
  \end{align*}
  Let $[i]_X = \{j \in [1, \len(X)] \mid \CTE{i}{X} = \CTE{j}{X}\}$ be an
  equivalence class of positions in $X$ from which identical behavior can be
  extracted. Let $Q$ be the corresponding quotient set of $[1, \len(X)]$. Then
  for all $[i] \in Q$ we define,
  \[
    P_{[i]} = 
      \begin{cases}
        \Term                         & \text{if $\CTE{i}{X} = \Term$,} \\
        \Inac                         & \text{if $\CTE{i}{X} = \Inac$,} \\
        P_{[j]} \ThrT a \ThrF P_{[k]} & \text{if $\CTE{i}{X} = \CTE{j}{X} \ThrT a \ThrF \CTE{k}{X}$.}
      \end{cases}
  \]
  Now for all $i \in [1, \len(X)]$ the thread $\CTE{i}{X}$ equals $P_{[i]}$,
  which is completely specified by the above linear equations and is thus
  regular.

  Conversely, let $T \in \ThrR$ be described by the linear equations $P_0 =
  t_1$, $P_2 = t_2$, \dots, $P_{n - 1} = t_{n - 1}$. Then there exists an $X
  \in \UAC$ with $\len(X) = 3n$ such that $P_i = \CTE{3i + 1}{X}$ and thus
  specifically $\CTEfwd{X} = P_0$. We construct $X$ as follows: If $P_i =
  \Term$ then set $\inst{3i + 1}(X) = \term$. If $P_i = \Inac$ then set
  $\inst{3i + 1}(X) = \abrt$. Otherwise $P_i = P_j \ThrT a \ThrF P_k$, thus we
  set $\inst{3i + 1}(X) = \fpt{a}$.  $\inst{3i + 2}(X)$ and $\inst{3i + 3}(X)$
  are jump instructions to positions $3j + 1$ and $3k + 1$, respectively.
  Positions in $X$ for which no instruction has been specified can be assigned
  an arbitrary instruction.
\end{proof}

\section{%
  The Reachability of Instructions
}\label{sec:c_accessibility}

If the equations in \eqref{eq:cte} are read strictly from left to right, then
they define for a given inseq $X$ and an arbitrary instruction at position $i$
in $X$ which action said instruction performs (if any) and at which program
position(s) $j$ execution may proceed. Let us define this relation between
program positions as follows:

\begin{defn}\label{def:accessibility}
  Let $X \in \UAC$. Then the \emph{accessibility relation} ${\acc{X}} \subset
  \Z^2$ of $X$ is defined as:
  \begin{align*}
    i \acc{X} j \iff &\text{for some $a \in \actions$ and $k \in \Z$, $\CTE{i}{X}$ equals one of} \\
      &\text{$\{\CTE{j}{X}, a \circ \CTE{j}{X}, \CTE{j}{X} \ThrT a \ThrF \CTE{k}{X}, \CTE{k}{X} \ThrT a \ThrF \CTE{j}{X}\}$} \\
      &\text{according to a single left-to-right application of an equation in \eqref{eq:cte}}.
  \end{align*}
  That is, $i \acc{X} j$ iff execution may continue at position $j$ right after
  the instruction at position $i$ has been executed. We then call $i$ the
  \emph{source position} and $\inst{i}(X)$ the \emph{source instruction}.
  Likewise $j$ and $\inst{j}(X)$ are the \emph{target position} and
  \emph{target instruction}, respectively.

  As usual, ${\acc{X}^+}$ denotes the transitive closure of the relation
  ${\acc{X}}$. Likewise ${\acc{X}^*}$ is its reflexive and transitive closure.

\end{defn}

\begin{defn}\label{def:reachable}
  Let $X \in \UAC$. A program position $j$ is \emph{reachable} from position
  $i$ in $X$ if $i \acc{X}^* j$.\footnote{Note that every instruction is
  reachable from itself. This is somewhat unconventional, but convenient for
  our purposes.} The set $\reachable{X}{i} = \{j \mid i \acc{X}^* j\}$ contains
  $i$ and all positions reachable from $i$ in $X$. It's complement
  $\unreachable{X}{i} = \Z - \reachable{X}{i}$ naturally contains those
  positions which are unreachable from $i$. Note that $\reachable{X}{i}$ may
  include ``invalid'' program positions, i.e. positions outside of $X$.

\end{defn}

\begin{defn}\label{def:exit_pos}
  The set $\exits{X} = \{i \in [1, \len(X)] \mid i \acc{X} j, j \notin [1,
  \len(X)]\}$ contains the \emph{exit positions} of $X$. That is, execution of
  an instruction at some position in $\exits{X}$ may cause a position outside
  $X$ to be ``reached''.

\end{defn}

\begin{prop}\label{prop:c_all_instr_reachable_thread}
  Every regular thread can be described by a \C instruction sequence in which
  every instruction is reachable from the start instruction.

\end{prop}

\begin{proof}
  Consider arbitrary $T \in \ThrR$, $X \in \UAC$ and $i \in [1, \len(X)]$ such
  that $\CTE{i}{X} = T$. If $\unreachable{X}{i} \cap [1, \len(X)] = \emptyset$,
  then $T$, $X$ and $i$ meet the requirements. Otherwise, randomly select some
  unreachable position $j \in \unreachable{X}{i} \cap [1, \len(X)]$.

  If the $j$th instruction is removed from $X$, then the jump counter of any
  jump instruction which jumps over position $j$ should be reduced by one, so
  as to ensure that its target instruction remains the same. This is possible
  since said jump counter must be at least $2$. We do not have to be concerned
  with any other instruction which can transfer control of execution to or over
  position $j$; such an instruction must itself not be reachable (because
  position $j$ isn't) and has as such no effect on $X$'s behavior.

  The result of removing the instruction at position $j$ from $X$ is an inseq
  $X'$ such that either $\CTE{i - 1}{X'} = T$ or $\CTE{i}{X'} = T$, depending
  on whether $j < i$ or $j > i$, respectively. This process can be repeated
  until all unreachable instructions are removed.
\end{proof}

\section[A Small Variation on \C]{%
  A Small Variation on $\boldsymbol{\C}$
}\label{sec:c_alternative}

For each $a \in \actions$, \C provides four test instructions: $\fpt{a}$,
$\fnt{a}$, $\bpt{a}$ and $\bnt{a}$. Semantically speaking the first two of
these have immediate counterparts in \PGA: $\ppt{a}$ and $\pnt{a}$. The latter
two are backward versions of the former two, and thus are indirectly based on
(or even inspired by) the \PGA test instructions as well.

\C's lack of directional bias allows for a different semantics for test
instructions, though; one that is instead inspired by the postconditional
composition operator as found in basic thread algebra. Consider the following
two instructions:

\begin{itemize}
  \item[$\cpppt{a}$] is the \emph{positive test instruction}. It performs
  action $a$. If the environment returns \true after completion of action $a$
  the instruction to the left of the current instruction is executed. Otherwise
  the instruction to its right is executed.

  \item[$\cppnt{a}$] is the \emph{negative test instruction}. This instruction
  mirrors the behavior of $\cpppt{a}$, in that it transfers control to the left
  or right if the action $a$ yields \false or \true, respectively.

\end{itemize}

These instructions are syntactically indistinguishable from \PGA's test
instructions, but they differ semantically. We define a code semigroup $\Cp =
(\UACp, \catOp)$ with
\[
  \UCp = (\UC - \positives - \negatives)
         \cup \bigcup_{a \in \actions} \{\cpppt{a}, \cppnt{a}\}.
\]
\Cp's semantics can be formalized by altering the set of equations
\eqref{eq:cte}: the cases related to $\inst{i}(X) \in \{\fpt{a}, \fnt{a},
\bpt{a}, \bnt{a}\}$ are no longer applicable, while two cases to handle
$\inst{i}(X) \in \{\cpppt{a}, \cppnt{a}\}$ need to be added. Thus we define for
all $i \in \Z$ and $X \in \UACp$,
\[
  \CpTE{i}{X} =
    \begin{cases}
      \Inac                                         & \text{if $i < 1$ or $i > \len(X)$,}
      \smallskip \\
      a \circ \CpTE{i + 1}{X}                       & \text{if $\inst{i}(X) = \fbi{a}$,} \\
      a \circ \CpTE{i - 1}{X}                       & \text{if $\inst{i}(X) = \bbi{a}$,}
      \smallskip \\
      \CpTE{i - 1}{X} \ThrT a \ThrF \CpTE{i + 1}{X} & \text{if $\inst{i}(X) = \cpppt{a}$,} \\
      \CpTE{i + 1}{X} \ThrT a \ThrF \CpTE{i - 1}{X} & \text{if $\inst{i}(X) = \cppnt{a}$,}
      \smallskip \\
      \CpTE{i + k}{X}                               & \text{if $\inst{i}(X) = \fj{k}$,} \\
      \CpTE{i - k}{X}                               & \text{if $\inst{i}(X) = \bj{k}$,}
      \smallskip \\
      \Inac                                         & \text{if $\inst{i}(X) = \abrt$,} \\
      \Term                                         & \text{if $\inst{i}(X) = \term$.}
    \end{cases}
\]

\subsection{%
  Behavior Preserving Homomorphisms
}

Now that the behavior of every \Cp-expression has been specified, we can answer
the question whether \Cp is more or less expressive than \C. It turns out that
these code semigroups are equally expressive, because we can define behavior
preserving homomorphisms from \C to \Cp and vice versa.

First, we define a homomorphism $f \colon \UAC \to \UACp$ on individual
instructions as follows:
\begin{align*}
  \fbi{a} &\mapsto \fbi{a};\fj{4};\abrt;\abrt;\bj{4},     &
  \fpt{a} &\mapsto \fj{2};\fj{4};\cpppt{a};\fj{7};\bj{2}, &
  \abrt   &\mapsto \abrt;\abrt;\abrt;\abrt;\abrt,         \\
  \bbi{a} &\mapsto \fj{4};\abrt;\abrt;\bj{4};\bbi{a},     &
  \fnt{a} &\mapsto \fj{2};\fj{4};\cppnt{a};\fj{7};\bj{2}, &
  \term   &\mapsto \term;\abrt;\abrt;\abrt;\term,         \\
  \fj{k}  &\mapsto \fj{5k};\abrt;\abrt;\abrt;\bj{4},      &
  \bpt{a} &\mapsto \fj{2};\bj{2};\cpppt{a};\bj{9};\bj{2}, \\
  \bj{k}  &\mapsto \fj{4};\abrt;\abrt;\abrt;\bj{5k},      &
  \bnt{a} &\mapsto \fj{2};\bj{2};\cppnt{a};\bj{9};\bj{2}.
\end{align*}
Every \C instruction is mapped onto five \Cp instructions. Observe that $f$ is
left-right uniformly behavior preserving. An alternative definition of $f$
could map every \C instruction onto \emph{four} \Cp instructions, at the
expense of being only left \emph{or} right uniformly behavior preserving.

The same holds for the homomorphism $g \colon \UACp \to \UAC$. One can define
left or right uniformly behavior preserving homomorphisms which map every \Cp
instruction onto three \C instructions. Here, however, we define $g$ such that
it is left-right uniformly behavior preserving:
\begin{align*}
  \fbi{a}   &\mapsto \fbi{a};\fj{3};\abrt;\bj{3},  &
  \fj{k}    &\mapsto \fj{4k};\abrt;\abrt;\bj{3},   \\
  \bbi{a}   &\mapsto \fj{3};\abrt;\bj{3};\bbi{a},  &
  \bj{k}    &\mapsto \fj{3};\abrt;\abrt;\bj{4k},   \\
  \cpppt{a} &\mapsto \fpt{a};\bj{2};\fj{2};\bj{3}, &
  \abrt     &\mapsto \abrt;\abrt;\abrt;\abrt,      \\
  \cppnt{a} &\mapsto \fnt{a};\bj{2};\fj{2};\bj{3}, &
  \term     &\mapsto \term;\abrt;\abrt;\term.
\end{align*}

\chapter[\Cg Instruction Sequences]{%
  $\boldsymbol{\Cg}$ Instruction Sequences
}\label{ch:cg_intro}

The semigroup \C introduced in the previous chapter provides two ways to skip
one or more instructions during execution: using a test instruction and using a
jump instruction. In both cases the location of the target instruction (if
present) is at a fixed distance from the source instruction. In other words,
the distance over which control of execution is transferred is static and does
not depend on the context (i.e., the instructions surrounding the instruction
which is currently being executed). As a result, inserting a single instruction
at an arbitrary position in some instruction sequence may completely alter its
semantics.

To alleviate this problem somewhat, we will introduce an alternative means to
transfer control of execution over arbitrary distances within an instruction
sequence. This chapter defines the semigroup \Cg, a close cousin of \C. \Cg
employs \emph{label instructions} to mark specific positions within an
instruction sequence with a natural number (a \emph{label number}). \emph{Goto
instructions} can then specify such a label number as the target of a jump.

\Cg's instruction set is introduced in \Sref{sec:cg_instr}. The semantics of
\Cg-expressions are formalized in \Sref{sec:cg_semantics}. This chapter then
proceeds with \Sref{sec:cg_lnf}, \Sref{sec:cg_label_freeing} and
\Sref{sec:cg_relative_jumps} in which certain properties of label and goto
instructions are analyzed and in which some useful transformations of
\Cg-expressions are defined. Combined, these sections provide us with the tools
required to analyze \Cg and its relation to \C and \PGA in
\Cref{ch:translations}. Finally, \Sref{sec:cg_alternative} briefly discusses an
alternative semantics for goto instructions. After defining behavior preserving
endomorphisms on \Cg to demonstrate that this alternative semantics does not
affect \Cg's expressiveness, we will not consider it any further.

\section{%
  The Instruction Set
}\label{sec:cg_instr}

The semigroup \Cg has basic instructions as well as positive and negative test
instructions, just like \C. \Cg does not have relative jumps $\fj{k}$ and
$\bj{k}$, unlike \C. Instead, it has a set of label instructions $\labels$ and
a set of goto instructions $\gotos$:\footnote{The notation for label and goto
instructions is borrowed from \cite{pga_second_paper, intro_pga_ta}, which
define a language \PGLDg as part of the \PGA language hierarchy. In \PGLDg,
there are label instructions $\plbl{l}$ and goto instructions $\pgt{l}$, for
all $l \in \N$.}
\begin{align*}
  \labels &= \bigcup_{l \in \N} \{\flbl{l}, \blbl{l}\}, &
  \gotos  &= \bigcup_{l \in \N} \{\fgt{l}, \bgt{l}\}.
\end{align*}
Label instructions mark a specific location within an instruction sequence with
a natural number $l$. They come in a forward as well as a backward oriented
flavor, which determines whether the instruction to respectively the right or
left of the label instruction is executed next. Goto instructions too are
marked with a natural number $l$ and jump to the first label $l$ with the same
orientation in the appropriate direction.

Formally, the instruction set $\UCg = \labels \cup \gotos \cup \UC - \jumps$
generates the semigroup $\Cg = (\UACg, \catOp)$. Note that since \Cg has basic
instructions and test instructions, \Cg takes an implicit parameter $\actions$
of actions, just like \C. Examples of \Cg-expressions include:
\begin{align}\label{eq:cg_ex}
  \fpt{a};\abrt, &&
  \fbi{b};\fgt{0};\fbi{a};\flbl{0};!, &&
  \fbi{b};\flbl{3};\fpt{a};\bgt{3}, &&
  \blbl{5};\bnt{c}.
\end{align}
Before formalizing \Cg's semantics, let us first informally describe what the
intended behavior of labels and gotos is.

\begin{itemize}
  \item[$\flbl{l}$] is a \emph{forward label instruction}. Execution of
  $\flbl{l}$ simply causes the instruction to its right to be executed, if it
  exists. Otherwise deadlock occurs.

  \item[$\blbl{l}$] is a \emph{backward label instruction}. It is analogous to
  $\flbl{l}$, except that execution continues with the instruction to its left.

  \item[$\fgt{l}$] is a \emph{forward goto instruction}. Transfers control of
  execution to the nearest $\flbl{l}$ instruction to its right, if such an
  instruction exists. Otherwise deadlock occurs.

  \item[$\bgt{l}$] is a \emph{backward goto instruction}. This instruction will
  cause execution to continue at the nearest $\blbl{l}$ instruction to its
  left. And of course, if such a label does not exist, deadlock will result.

\end{itemize}

For convenience we will write $\lago$ for the set $\labels \cup \gotos$. The
function $\lno \colon \lago \to \N$ returns the label number of a given label
or goto instruction (e.g., $\lno(\flbl{6}) = 6$). As with \C-instructions, we
will define two sets $\FUCg \subset \UCg$ and $\BUCg \subset \UCg$, which
consist of forward and backward \Cg-instructions respectively. That is,
\begin{align*}
  \FUCg &= (\FUC \cap \UCg) \cup \bigcup_{l \in \N} \{\flbl{l}, \fgt{l}\}, \\
  \BUCg &= (\BUC \cap \UCg) \cup \bigcup_{l \in \N} \{\blbl{l}, \bgt{l}\}.
\end{align*}
Clearly $\FUCg \cap \BUCg = \emptyset$ and $\FUCg \cup \BUCg \cup \{\abrt,
\term\} = \UCg$. The sets $\Flabels$, $\Blabels$, $\Fgotos$, $\Bgotos$,
$\Flago$ and $\Blago$ are defined as one would expect them to be. Likewise for
the directionality relation ${\de} \subset \UCg \times \UCg$.

\paragraph{%
  Examples
}

We will formalize \Cg's semantics in \Sref{sec:cg_semantics} below. Still, to
create or improve an intuitive understanding of \Cg-expressions and how they
differ from \C-expressions, let us briefly describe the behavior of the
\Cg-inseqs of \eqref{eq:cg_ex}. As before, we specify that execution starts at
the leftmost position.

\begin{itemize}
  \item $\fpt{a};\abrt$: Performs action $a$, after which deadlock occurs. This
  \Cg-expression is also a valid \C-expression.

  \item $\fbi{b};\fgt{0};\fbi{a};\flbl{0};!$: Performs action $b$ and then
  terminates. Action $a$ is not performed, since the second instruction is a
  goto instruction which causes execution to continue at position $4$.

  \item $\fbi{b};\flbl{3};\fpt{a};\bgt{3}$: Performs action $b$ followed by
  action $a$. Then deadlock results. The action $a$ is \emph{not} repeated,
  regardless of the value returned by the execution environment, because the
  backward goto instruction will not transfer control of execution to the
  forward label instruction: their directionality does not match.

  \item $\blbl{5};\bnt{c}$: Deadlock. After execution of a backward label
  instruction the instruction to its left is executed. Here, no such
  instruction is present.

\end{itemize}

\paragraph{%
  Orphaned Goto Instructions
}

A goto instruction in some \Cg-inseq $X$ which causes deadlock (by lack of a
``matching'' label instruction) will be called \emph{orphaned}. In other words,
given some $X \in \UACg$ and $i \in \gotos(X)$, the $i$th instruction of $X$ is
orphaned iff $i$ is an exit position in $X$.

Note that although some \Cg-expression $X$ may contain labels $\flbl{l}$ and
$\blbl{l}$, this does not preclude the possibility that $X$ contains a goto
instruction $\fgt{l}$ or $\bgt{l}$ which matches neither of these labels (and
is thus orphaned). For example, in the following expression both goto
instructions are orphaned:
\[
  \flbl{0};\bgt{0};\fgt{0};\blbl{0}.
\]
The C programming language \cite{iso_c99, k_and_r} (not to be confused with the
code semigroup \C) allows statements within functions to be marked using
labels. The statement \code{goto lbl;} causes program execution to continue at
the statement marked with label \code{lbl}, provided that \code{lbl} is a label
\emph{within the same function}. The Java programming language \cite{java_spec}
allows the labeling of code blocks. The statement \code{break lbl;} is valid
\emph{only} inside a block labeled \code{lbl}, and indicates that program
execution must be resumed after block \code{lbl}.\footnote{It is actually not
quite as simple as this, because of Java's support for exception handling.
Furthermore, the \code{continue} keyword can also be supplied with an optional
label, but only if said label precedes an iteration statement, not just any
code block. Also note that Java (currently) does not provide a ``regular''
\code{goto} statement, although the language does identify \code{goto} as a
reserved keyword.}

This shows that C and Java, just like the semigroup \Cg, restrict the scope of
\code{label} and \code{goto} statements. The statements \code{goto lbl;} and
\code{break lbl;} may prevent successful compilation of a C or Java program
$X$, even when $X$ contains (multiple) statements labeled with \code{lbl},
because of non-overlapping scopes.

When a C or Java compiler encounters a \code{goto} or \code{break} statement
which references a non-existent or out-of-scope label it may\footnote{Tested
with \emph{gcc 4.3.3} and \emph{javac 1.6.0\_14}.} yield an error claiming that
a certain \emph{label} is \emph{undefined}. Such an error message seems to lay
the ``blame'' for the failure to compile on the non-existence of some label
$l$, rather than on the incorrectly defined \code{goto} (\code{break})
statement. Using the term ``orphaned'' allows us to indicate that some goto
instruction does not have a matching label instruction without blaming any
specific label instruction or label number.

\section{%
  Semantics
}\label{sec:cg_semantics}

As goto instructions transfer control to the nearest label instruction (if
present) in the appropriate direction, their semantics depend on the position
of said label instruction. In order to make this relation precise, we define
two \emph{search functions},
\begin{align*}
  \Fsearch(X, i, S) &= \min(\{j \mid j \geq i, \inst{j}(X) \in S\} \cup \{\len(X) + 1\}), \\
  \Bsearch(X, i, S) &= \max(\{j \mid j \leq i, \inst{j}(X) \in S\} \cup \{0\}).
\end{align*}
$\Fsearch$ performs a forward search in a given inseq $X$, starting at position
$i$, for any instruction in $S$. The first position in $X$ containing one such
instruction is returned. If no instruction from $S$ is found then the first
position outside of $X$, (i.e., $\len(X) + 1$) is returned. $\Bsearch$ behaves
nearly identical, except that it searches from right to left, and returns $0$
if no instruction is found. Both functions have type $\UACg \times \Z \times
\powerset(\UCg) \to \N$, where $\powerset(\UCg)$ denotes the powerset of
$\UCg$.

As with \PGA and \C, we will formally define the semantics of \Cg-expressions
using basic thread algebra. Let $\CgTEfwd{\aArg} \colon \UACg \to \ThrR$ be the
function that yields the behavior of a given \Cg-expression when executed
starting with the leftmost instruction. That is, $\CgTEfwd{\aArg}$ defines its
left behavior. Likewise $\CgTEbwd{\aArg} \colon \UACg \to \ThrR$ yields the
right behavior of a given \Cg-expression. As with \C, we identify $\CgTEfwd{X}$
and $\CgTEbwd{X}$ with $\CgTE{1}{X}$ and $\CgTE{\len(X)}{X}$, respectively, and
define auxiliary functions $\CgTE{i}{\aArg} \colon \UACg \to \ThrR$ for all $i
\in \Z$, such that for all $X \in \UACg$,
\begin{equation}\label{eq:cgte}
  \CgTE{i}{X} =
    \begin{cases}
      \Inac                                         & \text{if $i < 1$ or $i > \len(X)$,}
      \smallskip \\
      a \circ \CgTE{i + 1}{X}                       & \text{if $\inst{i}(X) = \fbi{a}$,} \\
      \CgTE{i + 1}{X} \ThrT a \ThrF \CgTE{i + 2}{X} & \text{if $\inst{i}(X) = \fpt{a}$,} \\
      \CgTE{i + 2}{X} \ThrT a \ThrF \CgTE{i + 1}{X} & \text{if $\inst{i}(X) = \fnt{a}$,} \\
      \CgTE{i + 1}{X}                               & \text{if $\inst{i}(X) = \flbl{l}$,} \\
      \CgTE{\Fsearch(X, i, \{\flbl{l}\})}{X}        & \text{if $\inst{i}(X) = \fgt{l}$,}
      \smallskip \\
      a \circ \CgTE{i - 1}{X}                       & \text{if $\inst{i}(X) = \bbi{a}$,} \\
      \CgTE{i - 1}{X} \ThrT a \ThrF \CgTE{i - 2}{X} & \text{if $\inst{i}(X) = \bpt{a}$,} \\
      \CgTE{i - 2}{X} \ThrT a \ThrF \CgTE{i - 1}{X} & \text{if $\inst{i}(X) = \bnt{a}$,} \\
      \CgTE{i - 1}{X}                               & \text{if $\inst{i}(X) = \blbl{l}$,} \\
      \CgTE{\Bsearch(X, i, \{\blbl{l}\})}{X}        & \text{if $\inst{i}(X) = \bgt{l}$,}
      \smallskip \\
      \Inac                                         & \text{if $\inst{i}(X) = \abrt$,} \\
      \Term                                         & \text{if $\inst{i}(X) = \term$.}
    \end{cases}
\end{equation}
As with \PGA and \C, we equate an infinite sequence of left-to-right
derivations according to \eqref{eq:cgte} which does not yield an action with
deadlock:
\begin{equation}\label{eq:cgte_inf_jumps}\parbox{0.8\textwidth}{
  If the equations in \eqref{eq:cgte} can be applied infinitely
  often from left to right without ever yielding an action, then the extracted
  thread is $\Inac$.

}\end{equation}
This rule is specifically applicable to infinite loops created using label and
goto instructions. For example, $\CgTE{1}{\flbl{1};\blbl{2}} = \Inac$, because
\[
  \CgTE{1}{\flbl{1};\blbl{2}}
    = \CgTE{2}{\flbl{1};\blbl{2}}
    = \CgTE{1}{\flbl{1};\blbl{2}}.
\]
This example allows for an interesting observation: label instructions can act
as control structures even in absence of a matching goto instruction. Another
example is the program $\flbl{5};\bbi{a}$, which left as well as right behavior
is described by the equation $P = a \circ P$. In this sense \Cg's label
instructions are quite unlike labels in C or Java, where labels cannot alter
the flow of control in absence of another statement which references said label
(such as \code{goto}).

\Cg, like \C, characterizes the regular threads (as stated by
\PropRef{prop:c_characterizes_regular_threads}). We will not prove that fact
here; instead we refer to \PropRef{prop:cg_characterizes_regular_threads} in
\Sref{sec:cg_expressiveness}. For completeness we end this chapter with the
left and right behavior of the examples of \Sref{sec:cg_instr}:
\begin{align*}
  \CgTEfwd{\fpt{a};\abrt} &= a \circ \Inac,                            &
  \CgTEbwd{\fpt{a};\abrt} &= \Inac,                                    \\
  \CgTEfwd{\fbi{b};\fgt{0};\fbi{a};\flbl{0};!} &= b \circ \Term,       &
  \CgTEbwd{\fbi{b};\fgt{0};\fbi{a};\flbl{0};!} &= \Term,               \\
  \CgTEfwd{\fbi{b};\flbl{3};\fpt{a};\bgt{3}} &= b \circ a \circ \Inac, &
  \CgTEbwd{\fbi{b};\flbl{3};\fpt{a};\bgt{3}} &= \Inac,                 \\
  \CgTEfwd{\blbl{5};\bnt{c}} &= \Inac,                                 &
  \CgTEbwd{\blbl{5};\bnt{c}} &= c \circ \Inac.
\end{align*}

\subsection{%
  Accessibility and Exit Positions
}

The accessibility relation ${\acc{X}}$ defined on \C-inseqs $X$ by
\DefRef{def:accessibility} is defined analogously on \Cg-expressions. The same
holds for the set $\reachable{X}{i}$ of instruction positions reachable from
position $i$ in $X$ and its complement $\unreachable{X}{i}$ (see
\DefRef{def:reachable}). The set of exit positions $\exits{X}$ of a \Cg-inseq
$X$ is defined as in \DefRef{def:exit_pos}.

Note that for \Cg-expressions the notion of accessibility and reachability is
in a sense more ``artificial'' than for \C-expressions. This is so because for
any orphaned goto instruction on some position $i$ in an inseq $X$ it is the
case that either $i \acc{X} 0$ or $i \acc{X} \len(X) + 1$, due to the
definition of the functions $\Fsearch$ and $\Bsearch$.

We conclude this section with a result analogous to
\PropRef{prop:c_all_instr_reachable_thread}.

\begin{prop}\label{prop:cg_all_instr_reachable_thread}
  Every regular thread can be described by a \Cg instruction sequence in which
  every instruction is reachable from the start instruction.

\end{prop}

\begin{proof}
  Consider arbitrary $T \in \ThrR$, $X \in \UACg$ and $i \in [1, \len(X)]$ such
  that $\CgTE{i}{X} = T$. If $\unreachable{X}{i} \cap [1, \len(X)] =
  \emptyset$, then $T$, $X$ and $i$ meet the requirements. Otherwise, randomly
  select some unreachable position $j \in \unreachable{X}{i} \cap [1,
  \len(X)]$.

  To see why $j$ can be removed from $X$ without problems, we need to make two
  observations. First, any instruction which transfers control of execution to
  position $j$ must itself be unreachable. Second, any instruction which
  transfers control of execution over position $j$ must be a goto instruction;
  the behavior of such instruction will not be affected by the removal of the
  instruction at position $j$ (for $\inst{j}(X)$ cannot be a matching label
  instruction).

  The result of removing the instruction at position $j$ from $X$ is an inseq
  $X'$ such that either $\CgTE{i - 1}{X'} = T$ or $\CgTE{i}{X'} = T$, depending
  on whether $j < i$ or $j > i$, respectively. This process can be repeated
  until all unreachable instructions are removed.
\end{proof}

\section{%
  Normalizing Label Numbers
}\label{sec:cg_lnf}

\Cg-expressions can contain identical goto instructions which, when executed,
cause a jump to distinct positions within the instruction sequence. Likewise,
identical label instructions can occur multiple times within an expression. For
example,
\begin{equation}\label{eq:identical_labels}
  X = \fgt{7};\fbi{a};\flbl{7};\fbi{b};\fgt{7};\fbi{c};\flbl{7}.
\end{equation}
Here, even though $1 \eq{X} 5$, it is easy to see that $\CgTE{1}{X} \neq
\CgTE{5}{X}$. Informally, we may say that the identical instructions in this
expression are not semantically related. In this section we will make the
notion of a semantical relation between label and goto instructions more
precise. This endeavor is motivated by the observation that reasoning about a
\Cg-expression $X$ is greatly simplified if any two label and goto instructions
in $X$ with the same label number and direction are known to be related in
certain ways.

\begin{defn}
  Let $X \in \UACg$. If $i, j \in \lago(X)$, $\inst{i}(X) \de \inst{j}(X)$ and
  $\lno(\inst{i}(X)) = \lno(\inst{j}(X))$, then the label/goto instructions at
  positions $i$ and $j$ have the same label number and direction, and are said
  to \emph{correspond}, written $i \corr{X} j$.

  If $i \in \gotos(X)$, $j \in \labels(X)$ and $i \acc{X} j$, then the goto
  instruction at position $i$ \emph{targets} the label instruction at position
  $j$, written $i \gacc{X} j$.

  If $i, j \in \gotos(X)$, $i \eq{X} j$ and $\exists k(i \acc{X} k \land j
  \acc{X} k)$, then the identical goto instructions at positions $i$ and $j$
  are said to be \emph{target equivalent}, written $i \te{X} j$. Note that
  target equivalent goto instructions can be orphaned. Also, non-target
  equivalent goto instructions need \emph{not} be distinct, as in
  \eqref{eq:identical_labels}.

  Let $\gacc{X}^{-1}$ be the inverse of $\gacc{X}$. We define
  \[
    \lgr{X} = {\te{X}} \cup {\gacc{X}} \cup {\gacc{X}^{-1}} \cup
              \{(i, i) \mid i \in \labels(X)\}.
  \]
  Instructions at positions $i, j \in \lago(X)$ are \emph{related} iff $i
  \lgr{X} j$. $X$ is in \emph{label normal form} (\emph{LNF}) iff $i \corr{X}
  j$ implies $i \lgr{X} j$ for all $i, j \in \lago(X)$. That is, $X$ is in LNF
  if and only if any pair of corresponding instructions is related.

\end{defn}

\begin{prop}
  For all $X \in \UACg$, $\lgr{X}$ is an equivalence relation on $\lago(X)$.

\end{prop}

\begin{proof}
  Let $I = \{(i, i) \mid i \in \labels(X)\}$. $\lgr{X}$ is reflexive since $I
  \subseteq \lgr{X}$ and $i \te{X} i$ for all $i \in \gotos(X)$. $\lgr{X}$ is
  symmetric because $\te{X}$, $(\gacc{X} \cup \gacc{X}^{-1})$ and $I$ are. What
  remains to be proved is that $\lgr{X}$ is transitive. To that end, let $i$,
  $j$ and $k$ be distinct program positions with $i \lgr{X} j$ and $j \lgr{X}
  k$. We distinguish three situations:

  \begin{itemize}
    \item If $i \te{X} j$ then either $j \te{X} k$, in which case $i \te{X} k$,
    or $j \gacc{X} k$, in which case $i \gacc{X} k$.

    \item If $i \gacc{X} j$, then $j \gacc{X}^{-1} k$, and hence $i \te{X} k$.

    \item If $i \gacc{X}^{-1} j$, then $j \te{X} k$, meaning that $k \gacc{X}
    i$ and hence $i \gacc{X}^{-1} k$. (Note that $j \gacc{X} k$ will not be the
    case because that would mean $i = k$, while we defined $i$ and $k$ to be
    distinct positions.) \qedhere 

  \end{itemize}
\end{proof}

\begin{prop}
  Let $X \in \UACg$ be in label normal form. Then the following properties hold
  for all $1 \leq i, j \leq \len(X)$:

  \begin{enumerate}[(a)]
    \item If $i \in \gotos(X)$, $j \in \labels(X)$ and $i \corr{X} j$, then $i
    \gacc{X} j$ (label instructions are targeted by every goto instruction with
    the same label number and directionality).

    \item If $i, j \in \labels(X)$ and $i \eq{X} j$, then $i = j$ (all label
    instructions in $X$ are distinct).

    \item If $i, j \in \gotos(X)$ and $i \eq{X} j$, then $i \te{X} j$
    (identical goto instructions are target equivalent).

  \end{enumerate}
\end{prop}

\begin{proof}
  Let $X \in \UACg$ be in LNF. Note that $i \eq{X} j$ implies $i \corr{X} j$
  for all $i, j \in \lago(X)$. Since $X$ is in LNF, $i \corr{X} j$ implies $i
  \lgr{X} j$. In order, the properties follow from the following identities:

  \begin{align*}
    \lgr{X} \cap (\gotos(X) \times \labels(X))  &= {\gacc{X}}, \\
    \lgr{X} \cap (\labels(X) \times \labels(X)) &= \{(i, i) \mid i \in \labels(X)\}, \\
    \lgr{X} \cap (\gotos(X) \times \gotos(X))   &= {\te{X}}. \qedhere 
  \end{align*}
\end{proof}

\begin{prop}\label{prop:to_lnf}
  For any $X \in \UACg$ there exists an $X' \in \UACg$ such that $X'$ is in
  label normal form and $\CgTE{i}{X} = \CgTE{i}{X'}$ for all $i \in \Z$.

\end{prop}

\begin{proof}
  Let $X \in \UACg$. $\lgr{X}$ is an equivalence relation on $\lago(X)$. Let
  $[i]_{\lgr{X}}$ be the equivalence class of $i$ and let $\lago(X)/\lgr{X}$ be
  the quotient set of $\lago(X)$ by $\lgr{X}$. Let $n = |\lago(X)/\lgr{X}|$ be
  the number of equivalence classes. Now select a bijective mapping $f$ from
  $\lago(X)/\lgr{X}$ onto $[1, n]$, and construct an inseq $X'$ by changing the
  label numbers of each label and goto instruction in $X$ such that
  $\lno(\inst{i}(X')) = f([i]_{\lgr{X}})$ for all $i \in \lago(X)$. Then $X'$
  is in LNF and clearly $\CgTE{i}{X} = \CgTE{i}{X'}$ for all $i \in \Z$.
\end{proof}

\section{%
  Freeing Label Numbers
}\label{sec:cg_label_freeing}

In this section we will briefly describe how certain label numbers can be
removed from a \Cg-inseq. It turns out that defining certain behavior
preserving mappings on \Cg instruction sequences is greatly simplified if one
can assume that no label or goto instruction in the input inseq has a label
number present in some set $L$.

\begin{defn}
  A label number $l$ is \emph{available} in a \Cg-expression $X$ if there is no
  $u \in X$ such that $\lno(u) = l$. That is, no label or goto instruction in
  $X$ has label number $l$. To make a specific label number available, it must
  be \emph{freed}. For each $l \in \N$ we define an endomorphism $\free{l}$
  which frees label number $l$ in a given \Cg-inseq. $\free{l} \colon \UACg \to
  \UACg$ is defined on individual instructions as follows:
  \begin{align}\label{eq:free}
    \free{l}(u) =
      \begin{cases}
        \flbl{l'{+}1} & \text{if $u = \flbl{l'}$ and $l' \geq l$.} \\
        \blbl{l'{+}1} & \text{if $u = \blbl{l'}$ and $l' \geq l$.} \\
        \fgt{l'{+}1}  & \text{if $u = \fgt{l'}$ and $l' \geq l$.} \\
        \bgt{l'{+}1}  & \text{if $u = \bgt{l'}$ and $l' \geq l$.} \\
        u             & \text{otherwise.}
      \end{cases}
  \end{align}
  Some behavior preserving mappings require several label numbers to be
  available. Let $L = \seq{l_1, l_2, \dotsc, l_n}$ be an arbitrary finite
  sequence of natural numbers. Then $\free{L}$ is the endomorphism which frees
  the label numbers in $L$ in order. Formally, $\free{L} = \free{l_n} \circ
  \dotsm \circ \free{l_2} \circ \free{l_1}$.

\end{defn}

\begin{prop}\label{prop:free}
  Let $l \in \N$ and let $L$ be an arbitrary finite sequence of natural
  numbers. Then the endomorphisms $\free{l}$ and $\free{L}$ are left-right
  uniformly behavior preserving. Moreover, if $L$ is monotonically
  nondecreasing, then for every $X \in \UACg$, all label numbers in $L$ are
  available in $\free{L}(X)$.

\end{prop}

\begin{proof}
  $\free{l}$ maps individual instructions onto individual instructions and
  alters only the label number of label and goto instructions with a label
  number $\geq l$. Execution of a label instruction causes the instruction to
  its left or right to be executed, depending on the label's orientation, but
  irrespective of the actual label number. It is not hard to see that likewise
  the position to which goto instructions transfer control of execution is not
  affected by the application of $\free{l}$. Thus $\free{l}$ is left-right
  uniformly behavior preserving. As $\free{L}$ can be decomposed into
  individual applications of functions $\free{l_1}, \dotsc, \free{l_n}$, the
  same holds for $\free{L}$.

  Since $\free{l}$ only increments label numbers $\geq l$, any label number $<
  l$ which is available in some inseq $X$ will also be available in
  $\free{l}(X)$. It follows that if $L$ is monotonically nondecreasing, then
  all $l \in L$ will be available in $\free{L}(X)$.  \end{proof}

\section[\Cg and Relative Jumps]{%
  $\boldsymbol{\Cg}$ and Relative Jumps
}\label{sec:cg_relative_jumps}

\Cg does not have explicit relative jump instructions like \C. Yet in \Cg, too,
some instructions transfer control of execution relative to their own position:
basic instructions, test instruction and label instructions do so. For example,
the label instruction $\flbl{6}$ transfers control to the instruction to its
immediate right, equivalent to a forward relative jump over distance 1.

Section \Sref{sec:cg_alternative} below defines endomorphisms on \Cg in order
to simulate an alternative semantics for goto instructions. These endomorphisms
map single instructions onto a fixed number $b$ of different instructions.
Under those circumstances care must be taken that instructions which perform an
implicit relative jump behave properly: all relative jump distances are
multiplied by $b$.

So how does this work? In this section we will describe how relative jumps over
distances up to some arbitrary value $k$ can be emulated using label and goto
instructions. As a first step, consider the following family of \Cg-inseqs,
defined for every $l \geq 1$ and $k \geq 2$;
\begin{align*}
  \Fdecr_l &= \flbl{l};\fgt{l{-}1} &
  \Left_k  &= \Fdecr_1;\blbl{0};\Fdecr_2;\Fdecr_3;\dotsc;\Fdecr_k \\
  \Bdecr_l &= \bgt{l{-}1};\blbl{l} &
  \Right_k &= \Bdecr_k;\dotsc;\Bdecr_3;\Bdecr_2;\flbl{0};\Bdecr_1
\end{align*}
The \Cg-expressions $\Left_k$ and $\Right_k$ contain alternating label and goto
instructions, and an extra label with label number 0. $\Left_k$ and $\Right_k$
are meant to be used as subsequences of larger instruction sequences. Without
going into the use of $\blbl{0}$ and $\flbl{0}$ for now, observe that $\Left_k$
contains forward label instructions with label numbers $1$ though $k$, each
followed by a forward goto instruction with a label number one less than the
number of the preceding label instruction. The same holds for $\Right_k$,
except that it contains backward label and goto instructions.

Next, for all $k \in \N$, consider the family of functions $\phi_k \colon \UCg
\to \UCg$, defined as
\[
  \phi_k \colon u \mapsto
    \begin{cases}
      \fgt{1} & \text{if $u = \flbl{l}$ and $l \leq k$,} \\
      \bgt{1} & \text{if $u = \blbl{l}$ and $l \leq k$,} \\
      u       & \text{otherwise.}
    \end{cases}
\]
The functions $\phi_k$ map all label instructions with a label number not
greater than $k$ to goto instructions with label number $1$.

We now combine $\Left_k$, $\Right_k$ and $\phi_k$ to create endomorphisms
$\rel{k} \colon \UACg \to \UACg$, for all $k \geq 2$, defined on individual
instructions $u \in \UCg$ such that,
\begin{equation}\label{eq:rel}
  \rel{k} \colon u \mapsto
    \begin{cases}
      \Left_k;\bgt{2};\bgt{1};\phi_k(u);\blbl{0};\Right_k & \text{if $u \in \BUCg$,} \\
      \Left_k;\flbl{0};\phi_k(u);\fgt{1};\fgt{2};\Right_k & \text{otherwise.}
    \end{cases}
\end{equation}
The functions $\rel{k}$ are not quite left or right behavior preserving.
Instead, at some higher level they redefine the semantics of goto instructions
with a label number $l \leq k$, such that their behavior mimics that of a
relative jump over distance $l$. As a special case, $\fgt{0}$ and $\bgt{0}$
signify a jump over distance zero and as such yield deadlock.\footnote{See also
\Sref{sec:c_abrt}.} This alternative semantics can be made explicit by defining
thread extraction operators $\CgRelTE{k}{i}{X}$ which are analogous to
$\CgTE{i}{X}$, except for the fact that the operators $\CgRelTE{k}{i}{X}$ are
defined differently for instances where $i \in \{j \in \gotos(X) \mid
\lno(\inst{j}(X)) \leq k\}$:
\begin{equation}\label{eq:cgrelte}
  \CgRelTE{k}{i}{X} =
    \begin{cases}
      \Inac                                                     & \text{if $i < 1$ or $i > \len(X)$,}
      \smallskip \\
      a \circ \CgRelTE{k}{i + 1}{X}                             & \text{if $\inst{i}(X) = \fbi{a}$,} \\
      \CgRelTE{k}{i + 1}{X} \ThrT a \ThrF \CgRelTE{k}{i + 2}{X} & \text{if $\inst{i}(X) = \fpt{a}$,} \\
      \CgRelTE{k}{i + 2}{X} \ThrT a \ThrF \CgRelTE{k}{i + 1}{X} & \text{if $\inst{i}(X) = \fnt{a}$,} \\
      \CgRelTE{k}{i + 1}{X}                                     & \text{if $\inst{i}(X) = \flbl{l}$,} \\
      \Inac                                                     & \text{if $\inst{i}(X) = \fgt{0}$,} \\
      \CgRelTE{k}{i + l}{X}                                     & \text{if $\inst{i}(X) = \fgt{l}$ and $1 \leq l \leq k$,} \\
      \CgRelTE{k}{\Fsearch(X, i, \{\flbl{l}\})}{X}              & \text{if $\inst{i}(X) = \fgt{l}$ and $l > k$,}
      \smallskip \\
      a \circ \CgRelTE{k}{i - 1}{X}                             & \text{if $\inst{i}(X) = \bbi{a}$,} \\
      \CgRelTE{k}{i - 1}{X} \ThrT a \ThrF \CgRelTE{k}{i - 2}{X} & \text{if $\inst{i}(X) = \bpt{a}$,} \\
      \CgRelTE{k}{i - 2}{X} \ThrT a \ThrF \CgRelTE{k}{i - 1}{X} & \text{if $\inst{i}(X) = \bnt{a}$,} \\
      \CgRelTE{k}{i - 1}{X}                                     & \text{if $\inst{i}(X) = \blbl{l}$,} \\
      \Inac                                                     & \text{if $\inst{i}(X) = \bgt{0}$,} \\
      \CgRelTE{k}{i - l}{X}                                     & \text{if $\inst{i}(X) = \bgt{l}$ and $1 \leq l \leq k$,} \\
      \CgRelTE{k}{\Bsearch(X, i, \{\blbl{l}\})}{X}              & \text{if $\inst{i}(X) = \bgt{l}$ and $l > k$,}
      \smallskip \\
      \Inac                                                     & \text{if $\inst{i}(X) = \abrt$,} \\
      \Term                                                     & \text{if $\inst{i}(X) = \term$.}
    \end{cases}
\end{equation}
As an example, consider the \Cg-inseq $X = \fgt{3};\flbl{3};\fbi{a};\fbi{b}$
and suppose that we want to interpret all goto instructions with a label number
$\leq 7$ as relative jumps. Then,
\[
  \CgRelTE{7}{1}{X}
  = \CgRelTE{7}{4}{X}
  = b \circ \CgRelTE{7}{5}{X}
  = b \circ \Inac.
\]
Observe that the goto instruction on position $1$ transfers control of
execution to position $4$; the label instruction with the matching label number
at position $2$ is bypassed.

Fixing some $k \geq 2$, observe that $\rel{k}$ maps every \Cg-instruction on
$b_k = 4k + 6$ \Cg-instructions. $\rel{k}$ is defined such that the following
equality holds:
\[
  \CgRelTE{k}{i}{X}
    = \CgTE{b_k(i - 1) + 1}{\rel{k}(X)}
    = \CgTE{b_ki}{\rel{k}(X)}.
\]
Specifically,
\begin{align*}
  \CgRelTEfwd{k}{X} &= \CgRelTE{k}{1}{X} = \CgTE{1}{\rel{k}(X)} = \CgTEfwd{\rel{k}(X)}, \\
  \CgRelTEbwd{k}{X} &= \CgRelTE{k}{\len(X)}{X} = \CgTE{\len(\rel{k}(X))}{\rel{k}(X)} = \CgTEbwd{\rel{k}(X)}.
\end{align*}
It follows that the alternative semantics for \Cg as defined by
\eqref{eq:cgrelte} can be simulated using $\rel{k}$ and \Cg's default thread
extraction operator.

\section{%
  Label Instructions as More General Jump Targets
}\label{sec:cg_alternative}

\Cg's goto instructions are defined such that they transfer control to a label
instruction with the same label number \emph{and directionality} in the
appropriate direction (if present). An obvious alternative behavior is for goto
instructions to jump to a label instruction with the same label number in the
appropriate direction, \emph{irrespective of its directionality} (again,
provided such instruction is present). Put more informally: instead of
``accepting'' jumps from a single direction, we may alter \Cg's semantics such
that label instructions accept jumps originating from goto instructions in
either direction. In this section we play with this idea; it turns out that
with respect to expressiveness nothing is gained or lost by using such an
alternative semantics. Therefore we will not consider this idea beyond this
section. As a result, readers may choose to skip this section.

This alternative semantics can be described by a thread extraction operator
$\CgpTElong{\aArg}{\aArg}$ which is nearly identical to the operator
$\CgTElong{\aArg}{\aArg}$ as defined by the set of equations \eqref{eq:cgte}
and rule \eqref{eq:cgte_inf_jumps}, except for the cases involving goto
instructions. Specifically (now using the usual shorthand notation
$\CgpTE{i}{X}$ instead of $\CgpTElong{i}{X}$):
\[
  \CgpTE{i}{X} =
    \begin{cases}
      \Inac                                              & \text{if $i < 1$ or $i > \len(X)$,}
      \smallskip \\
      a \circ \CgpTE{i + 1}{X}                           & \text{if $\inst{i}(X) = \fbi{a}$,} \\
      \CgpTE{i + 1}{X} \ThrT a \ThrF \CgpTE{i + 2}{X}    & \text{if $\inst{i}(X) = \fpt{a}$,} \\
      \CgpTE{i + 2}{X} \ThrT a \ThrF \CgpTE{i + 1}{X}    & \text{if $\inst{i}(X) = \fnt{a}$,} \\
      \CgpTE{i + 1}{X}                                   & \text{if $\inst{i}(X) = \flbl{l}$,} \\
      \CgpTE{\Fsearch(X, i, \{\flbl{l}, \blbl{l}\})}{X}  & \text{if $\inst{i}(X) = \fgt{l}$,}
      \smallskip \\
      a \circ \CgpTE{i - 1}{X}                           & \text{if $\inst{i}(X) = \bbi{a}$,} \\
      \CgpTE{i - 1}{X} \ThrT a \ThrF \CgpTE{i - 2}{X}    & \text{if $\inst{i}(X) = \bpt{a}$,} \\
      \CgpTE{i - 2}{X} \ThrT a \ThrF \CgpTE{i - 1}{X}    & \text{if $\inst{i}(X) = \bnt{a}$,} \\
      \CgpTE{i - 1}{X}                                   & \text{if $\inst{i}(X) = \blbl{l}$,} \\
      \CgpTE{\Bsearch(X, i, \{\flbl{l}, \blbl{l}\})}{X}  & \text{if $\inst{i}(X) = \bgt{l}$,}
      \smallskip \\
      \Inac                                              & \text{if $\inst{i}(X) = \abrt$,} \\
      \Term                                              & \text{if $\inst{i}(X) = \term$.}
    \end{cases}
\]

Observe that $\Fsearch$ and $\Bsearch$ now each search for \emph{two}
instructions, namely $\flbl{l}$, $\blbl{l}$, for some $l \in \N$.

\subsection{%
  Behavior Preserving Homomorphisms
}

It turns out that this alternative semantics does not affect \Cg's
expressiveness. It is straightforward to define a homomorphism $f$ such that
$\CgTE{i}{X} = \CgpTE{i}{f(X)}$ for all $i \in \Z$ and $X \in \UACg$. $f$ is
defined on individual instructions $u \in \UCg$ such that,
\[
  f \colon u \mapsto
    \begin{cases}
      \flbl{2l}     & \text{if $u = \flbl{l}$,} \\
      \blbl{2l{+}1} & \text{if $u = \blbl{l}$,} \\
      \fgt{2l}      & \text{if $u = \fgt{l}$,} \\
      \bgt{2l{+}1}  & \text{if $u = \bgt{l}$,} \\
      u             & \text{otherwise.}
    \end{cases}
\]
Indeed $f$ ensures that any label number $l$ is even for forward label and goto
instructions, while $l$ is odd for backward oriented instructions. As a result,
label instructions in $f(X)$ will in practice ``accept'' jumps from goto
instructions in only one direction, rendering the difference between
$\CgTElong{\aArg}{\aArg}$ and $\CgpTElong{\aArg}{\aArg}$ irrelevant.

Conversely, there exists a homomorphism $g$ such that for all $i \in \Z$ there
exists some $j \in \Z$ such that $\CgpTE{i}{X} = \CgTE{j}{g(X)}$. We define $g
= \phi \circ \rel{2} \circ \free{\seq{0, 1, 2}}$. The functions $\free{\seq{0,
1, 2}}$ and $\rel{2}$ have been defined previously by \eqref{eq:free} and
\eqref{eq:rel}, respectively. The function $\phi$ is a homomorphism, defined on
individual \Cg-instructions $u$ such that,
\[
  \phi \colon u \mapsto
    \begin{cases}
      \fgt{l};\blbl{l};\flbl{l} & \text{if $u = \flbl{l}$ with $l > 2$,} \\
      \blbl{l};\flbl{l};\bgt{l} & \text{if $u = \blbl{l}$ with $l > 2$,} \\
      u                         & \text{otherwise.}
    \end{cases}
\]
The correctness of $g$ hinges on three observations:

\begin{enumerate}
  \item By \PropRef{prop:free}, $\free{\seq{0, 1, 2}}$ is behavior preserving.

  \item The homomorphism $\rel{2}$ alters the semantics of goto instructions
  with label numbers $\leq 2$. These instructions are not present in its input
  because it is passed the output of $\free{\seq{0, 1, 2}}$. As such, $\rel{2}
  \circ \free{\seq{0, 1, 2}}$ is also behavior preserving.

  \item Lastly, $\phi$ does not replace label instructions introduced by
  $\rel{2}$. It \emph{does} replace all other label instructions, such that the
  resulting subsequence of three instructions mimics the behavior of label
  instructions as defined by $\CgpTElong{\aArg}{\aArg}$ if fed to
  $\CgTElong{\aArg}{\aArg}$. Any label replaced by $\phi$ is embedded by
  $\rel{2}$, ensuring that the behavior of other label, basic and test
  instructions is unaffected. This explains the use of $\rel{2}$: it
  accommodates for the implicit relative jumps performed by these instructions.

\end{enumerate}

We conclude with the observation that $g$ is left-right behavior preserving,
but not uniformly so. This is because the number of instructions output by
$\phi$ depends on its input. $g$ can be made left-right uniformly behavior
preserving by using an alternative definition of $\phi$ which always outputs
three instructions:
\[
  \phi \colon u \mapsto
    \begin{cases}
      \fgt{l};\blbl{l};\flbl{l} & \text{if $u = \flbl{l}$ with $l > 2$,} \\
      \blbl{l};\flbl{l};\bgt{l} & \text{if $u = \blbl{l}$ with $l > 2$,} \\
      u;\fgt{1};\fgt{2}         & \text{if $u \in \FUCg \land u \neq \flbl{l}$ for $l > 2$,} \\
      \bgt{2};\bgt{1};u         & \text{otherwise.}
    \end{cases}
\]

\chapter{%
  Translating Instruction Sequences
}\label{ch:translations}

Previous chapters introduced the program algebra \PGA and the code semigroups
\C and \Cg. In this chapter we provide behavior preserving mappings between
these algebras and show some properties of these translations.

Though defined on at a syntactic level, a behavior preserving mapping $f$ makes
explicit certain ways in which (groups of) instructions are related on a
semantic level. If $f \colon A \to B$, then $f$ tells us something about
distinctions and similarities between code semigroups $A$ and $B$. If $f \colon
A \to A$, then $f$ (if it is not the identity function), can be seen as a
reformulation instead of a translation. Additionally, if $f$ is an
(anti-)homomorphism then it provides some additional implicit information about
how $A$ and $B$ are related. Specifically, it shows that an $A$-inseq $X$ can
be translated instruction by instruction, independent of context, and without
taking the length of $X$ as an explicit parameter, to some $B$-inseq $Y$. For
this reason we aim to define homomorphic instead of arbitrary translations
between code semigroups where possible.\footnote{Thinking of $A$ as a high
level programming language and $B$ as a lower level programming language or
even machine code, we can view $f$ as an interpreter or compiler. If $f$ is an
(anti-)homomorphism then parts of an $A$-inseq $X$ can be transformed and
possibly even executed before all of $X$ has been read.}

The translations defined in this chapter will aid us in proving some
expressiveness results in the next chapter. In order, this chapter provides a
translation from \C to \PGA (\Sref{sec:c_to_pga}), from \PGA to \C
(\Sref{sec:pga_to_c}), from \C to \Cg (\Sref{sec:c_to_cg}) and from \Cg to \C
(\Sref{sec:cg_to_c}).

\section[Translating \C to \PGA]{%
  Translating $\boldsymbol{\C}$ to \PGA
}\label{sec:c_to_pga}

In this section we define a behavior preserving mapping $\CToPga \colon \UAC
\to \Pga$. We do so in three steps: the first two steps apply left behavior
preserving mappings to \C itself, thereby converting every \C-inseq $X$ to a
behaviorally equivalent \C-inseq $Y$ which has certain structural properties.
The third step exploits these properties in order to translate every such $Y$
to a behaviorally equivalent \PGA term $Z$. The translation presented here is
based on the behavior preserving mapping from \C onto \PGA as defined in
section 12 of \cite{inseq_intro}.

\begin{enumerate}[1.]
  \item \PGA has basic instructions and test instructions whose semantics are
  identical to \C's forward basic and test instructions. \C's backward basic
  instructions and test instructions have no direct counterpart in \PGA, so we
  wish to eliminate them. Thus we define a left uniformly behavior preserving
  endomorphism $f$ on $\UAC$ which removes these backward instructions. $f$ is
  defined on individual instructions as follows:
  \begin{align*}
    \fbi{a} &\mapsto \fbi{a};\fj{2};\abrt,  &
    \bbi{a} &\mapsto \fbi{a};\bj{4};\abrt,  \\
    \fpt{a} &\mapsto \fpt{a};\fj{2};\fj{4}, &
    \bpt{a} &\mapsto \fpt{a};\bj{4};\bj{8}, \\
    \fnt{a} &\mapsto \fnt{a};\fj{2};\fj{4}, &
    \bnt{a} &\mapsto \fnt{a};\bj{4};\bj{8}, \\
    \\
    \fj{k}  &\mapsto \fj{3k};\abrt;\abrt,  &
    \abrt   &\mapsto \abrt;\abrt;\abrt,    \\
    \bj{k}  &\mapsto \bj{3k};\abrt;\abrt,  &
    \term   &\mapsto \term;\abrt;\abrt.
  \end{align*}

  \item In \cite{inseq_intro} the notion of \emph{\C-programs} is introduced.
  In essence, a \C-program is a \C-inseq which does not contain exit positions.
  I.e, no instruction transfers control of execution outside of the instruction
  sequence; only execution of the termination or abort instruction will cause
  program execution to halt. Every \C-inseq $X$ can be converted to a
  \C-program, simply by prefixing and suffixing sufficiently many abort
  instructions. In order to maintain $X$'s left and right behavior, additional
  jump instructions must be added to its left and right. Let $m \geq 2$ be an
  upper bound on the largest jump counter present in some \C-inseq $X$. Then a
  left-right behaviorally equivalent \C-program $X'$ can be constructed as
  \[
    \fj{m{+}1};(\abrt)^m;X;(\abrt)^m;\bj{m{+}1}.
  \]
  Let $g$ be the left-right behavior preserving mapping which performs the
  above procedure for arbitrary \C-inseqs.

  \item Given $f$ and $g$ as defined in the previous two steps, it is immediate
  that for every \C-inseq $X$ there exists a left behaviorally equivalent
  \C-program $X' = g(f(X))$ which does not contain instructions from the set
  $\Bbasics \cup \Bpositives \cup \Bnegatives$. Let $X' = u_1;\dotsc;u_n$. Then
  the following is a behaviorally equivalent \PGA term:
  \[
    (\phi_n(u_1);\dotsc;\phi_n(u_n))^\omega.
  \]
  For all $n \in \Np$ the function $\phi_n$ is defined as follows (observe that
  due to application of $g$, necessarily $k < n$ and thus $n - k \in \Np$):
  \begin{align*}
    \fbi{a}  &\mapsto \pbi{a},      &
    \fj{k}   &\mapsto \pjmp{k},     &
    \term    &\mapsto \term.        \\
    \fpt{a}  &\mapsto \ppt{a},      &
    \bj{k}   &\mapsto \pjmp{n{-}k}, \\
    \fnt{a}  &\mapsto \pnt{a},      &
    \abrt    &\mapsto \pjmp{0},
  \end{align*}
  Denoting the above procedure by $h$, we have that $\CToPga = h \circ g \circ
  f$.

\end{enumerate}

\section[Translating \PGA to \C]{%
  Translating \PGA to $\boldsymbol{\C}$
}\label{sec:pga_to_c}

Defining a translation $\PgaToC \colon \Pga \to \UAC$ turns out to be be a lot
easier if \PGA terms can be assumed to be in second canonical form. Hence we
start out by defining
\[
  \PgaToC = \PgaSToC \circ \snd.
\]
Recall that $\snd \colon \Pga \to \PgaS$ is the function defined in
\Sref{sec:snd} which converts arbitrary \PGA terms to their structurally (and
behaviorally) equivalent minimal second canonical forms. The mapping $\PgaSToC
\colon \PgaS \to \UAC$ is a behavior preserving mapping defined on second
canonical forms only. Any $X \in \PgaS$ does not contain chained jump
instructions and has one of two forms:

\begin{itemize}
  \item $X$ does not contain repetition and thus $X = u_1;u_2;\dotsc;u_n$ for
  some $n \in \Np$. We define
  \[
    \PgaSToC(u_1;u_2;\dotsc;u_n) = \psi(u_1);\psi(u_2);\dotsc;\psi(u_n).
  \]

  \item $X = Y;Z^\omega$, and $Y$ nor $Z$ contain repetition, meaning that for
  some $n, m \in \Np$, $X = u_1;\dotsc;u_n;(u_{n + 1};\dotsc;u_{n +
  m})^\omega$. Now we define
  \[
    \PgaSToC(u_1;\dotsc;u_n;(u_{n + 1};\dotsc;u_{n +m})^\omega) =
      \psi(u_1);\dotsc;\psi(u_{n + m});(\bj{m})^{\max(2, m - 1)}.
  \]

\end{itemize}

The function $\psi$ is as straightforward as can be:
\begin{align*}
  \pbi{a}  &\mapsto \fbi{a} &
  \ppt{a}  &\mapsto \fpt{a} &
  \pnt{a}  &\mapsto \fnt{a} &
  \term    &\mapsto \term   &
  \pjmp{l} &\mapsto
    \begin{cases}
      \abrt  & \text{if $l = 0$,} \\
      \fj{l} & \text{otherwise.}
    \end{cases}
\end{align*}
$\PgaSToC$ makes extensive use of the assumptions that can be made about its
input (i.e., that it is in second canonical form). Any jump instruction $u_i$
with $i \leq n$ will not jump beyond $u_{n + m}$. Any jump instruction $u_i$
with $i > n$ will not have a jump counter greater than $m - 1$. By appending
$\max(2, m - 1)$ $\bj{m}$ instructions, it is ensured that all jump
instructions which transfer control of execution beyond $\psi(u_{n + m})$
indirectly transfer control to the appropriate instruction. Since $u_{n + m -
1}$ and $u_{n + m}$ can be test instructions, it is important to append at
least two backward jump instructions.

\section[Translating \C to \Cg]{%
  Translating $\boldsymbol{\C}$ to $\boldsymbol{\Cg}$
}\label{sec:c_to_cg}

In this section we focus on translations from \C to \Cg. It turns out that
there does not exist a homomorphism which translates arbitrary \C-expressions
to behaviorally equivalent \Cg-expressions. \ThmRef{thm:no_c_to_cg_hom} below
gives a proof of this fact.

A convenient way to translate \C to \Cg is to start out by categorizing every
\C-expression based on the largest jump counter it contains. We write $\Cr{k}$
for the subsemigroup of \C which consists exactly of those \C-expressions that
do not contain instructions $\fj{k'}$ or $\bj{k'}$ for $k' > k$. Formally,
$\Cr{k} = (\UACr{k}, \catOp)$, with\footnote{Recall that $\jd \colon \jumps \to
\Np$ returns the jump counter of a given jump instruction.}
\begin{equation}\label{eq:ucr}
  \UCr{k} = \UC - \{u \in \jumps \mid \jd(u) > k\}.
\end{equation}
Assume the existence of a family of behavior preserving mappings $\CrToCg{k}
\colon \UACr{k} \to \UACg$ for all $k \in \N$. Writing $\CrToCgLong(k, X)$ for
$\CrToCg{k}(X)$, the behavior preserving mapping $\CToCg \colon \UAC \to \UACg$
can then be defined on all $X \in \UAC$ as,\footnote{Yes, the function name
$\CToCg$ is overloaded here. Its type is either $\N \times \UAC \to \UACg$ or
simply $\UAC \to \UACg$.}
\[
  \CToCg \colon X \mapsto
    \CrToCgLong(\max\{\jd(\inst{i}(X)) \mid i \in \jumps(X)\}, X).
\]
The hypothesized family of functions $\CrToCg{k}$ exists. A straightforward
definition is \eqref{eq:cr_to_cg} in \Sref{sec:cr_to_cg} below. An alternative
homomorphic definition is \eqref{eq:cr_to_cg_hom} in \Sref{sec:c_to_cg_hom}.
Since in both cases $\CrToCg{k}$ is only defined for $k \geq 2$, a slightly
altered definition of $\CToCg \colon \UAC \to \UACg$ is in place:
\begin{equation}\label{eq:c_to_cg}
  \CToCg \colon X \mapsto
    \CrToCgLong(\max(\{\jd(\inst{i}(X)) \mid i \in \jumps(X)\} \cup \{2\}), X).
\end{equation}

\subsection{%
  A Behavior Preserving Mapping from $\boldsymbol{\Cr{k}}$ to
  $\boldsymbol{\Cg}$
}\label{sec:cr_to_cg}

For all $k \geq 2$, we define a function $\CrToCg{k} \colon \UACr{k} \to \UACg$
such that,
\begin{equation}\label{eq:cr_to_cg}
  \CrToCg{k}(u_1;\dotsc;u_n) = \psi_{k,1}(u_1);\dotsc;\psi_{k,n}(u_n).
\end{equation}
In effect $\CrToCg{k}$ replaces the $i$th instruction of its input $X$ with the
output of $\psi_{k,i}(\inst{i}(X))$. The auxiliary functions $\psi_{k,i} \colon
\UC \to \UAC$ are defined as follows:
\[
  \psi_{k,i}(u) =
    \begin{cases}
      \phi_{k,i}(\fbi{a};\fgt{\rem{k + 1}{i{+}1}})                          & \text{if $u = \fbi{a}$,} \\
      \phi_{k,i}(\fpt{a};\fgt{\rem{k + 1}{i{+}1}};\fgt{\rem{k + 1}{i{+}2}}) & \text{if $u = \fpt{a}$,} \\
      \phi_{k,i}(\fnt{a};\fgt{\rem{k + 1}{i{+}1}};\fgt{\rem{k + 1}{i{+}2}}) & \text{if $u = \fnt{a}$,} \\
      \phi_{k,i}(\fbi{a};\bgt{\rem{k + 1}{i{-}1}})                          & \text{if $u = \bbi{a}$,} \\
      \phi_{k,i}(\fpt{a};\bgt{\rem{k + 1}{i{-}1}};\bgt{\rem{k + 1}{i{-}2}}) & \text{if $u = \bpt{a}$,} \\
      \phi_{k,i}(\fnt{a};\bgt{\rem{k + 1}{i{-}1}};\bgt{\rem{k + 1}{i{-}2}}) & \text{if $u = \bnt{a}$,} \\
      \phi_{k,i}(\fgt{\rem{k + 1}{i{+}l}})                                  & \text{if $u = \fj{l}$,} \\
      \phi_{k,i}(\bgt{\rem{k + 1}{i{-}l}})                                  & \text{if $u = \bj{l}$,} \\
      \phi_{k,i}(\abrt)                                                     & \text{if $u = \abrt$.} \\
      \phi_{k,i}(\term)                                                     & \text{if $u = \term$.}
    \end{cases}
\]
In this definition $\rem{k + 1}{n}$ stands for the remainder of $n$ after
division by $k + 1$, i.e. the smallest nonnegative value congruent with $n
\pmod {k + 1}$. Thus $0 \leq \rem{k + 1}{n} \leq k$ for all $n$. For all $i \in
\Np$, $\phi_{k,i} \colon \UACg \to \UACg$ embeds its argument between some
label and goto instructions with label number $\rem{k + 1}{i}$ as follows:
\[
  \phi_{k,i}(U) =
  \fgt{\rem{k + 1}{i}};\blbl{\rem{k + 1}{i}};\flbl{\rem{k + 1}{i}};U;\bgt{\rem{k + 1}{i}}.
\]
Informally, $\phi_{k,i}(U)$ ``guards'' the \Cg instruction sequence $U$ which
replace the \C instruction at position $i$ in the original C-expression using
the labels $\flbl{\rem{k + 1}{i}}$ and $\blbl{\rem{k + 1}{i}}$. In this way a
goto instruction $\fgt{\rem{k + 1}{i + l}}$ or $\bgt{\rem{k + 1}{i + l}}$ in a
\Cg-inseq $U$ which replaces the $i$th \C instruction transfers execution to
the \Cg-inseq $U'$ which replaces the \C instruction at position $i + l$ or $i
- l$, respectively. In this way the transfer of control of execution over a
relative distance in the original \C-inseq is simulated.

Observe that label numbers are repeated (``reused'') with period $k + 1$. This
does not pose a problem because the original \Cr{k}-expression will not contain
relative jumps over a distance greater than $k$. (And since $k \geq 2$, the
implicit relative jumps over distance $1$ or $2$ performed by test instructions
can likewise be simulated.)

The auxiliary functions $\psi_{k,i}$ and their helper functions $\phi_{k,i}$
are defined such that $\CrToCg{k}$ is left-right behavior preserving. Note that
it is possible to omit the rightmost $\bgt{\rem{k + 1}{i}}$ instruction
outputted by each call to $\phi_{k,i}$, but then $\CrToCg{k}$ would no longer
be right behavior preserving.

\subsection{%
  What About a Homomorphic Translation from $\boldsymbol{\C}$ to
  $\boldsymbol{\Cg}$?
}\label{sec:c_to_cg_hom}

The translation $\CToCg \colon \UAC \to \UACg$ defined by \eqref{eq:c_to_cg} is
not homomorphic because it requires knowledge about the largest jump counter
present in its input. It turns out that it is not possible to define a
homomorphic alternative to $\CToCg$.

\begin{thm}\label{thm:no_c_to_cg_hom}
  There does not exist a behavior preserving homomorphism $f \colon \UAC \to
  \UACg$.

\end{thm}

\begin{proof}
  We prove that no homomorphism $f \colon \UAC \to \UACg$ can be left behavior
  preserving. The proof that no such $f$ can be right behavior preserving is
  analogous.

  For all $n \in \Np$ we define the following \C-inseqs:
  \begin{equation}\label{eq:jump_tree}\begin{split}
    \node{n} &= \fpt{a};\fj{3n{-}1};\fj{3n{+}1}, \\
    \tree{n} &= \node{1};\node{2};\dotsc;\node{2^n{-}1}.
  \end{split}\end{equation}
  Observe that $\tree{n}$ contains $2^n$ exit positions (see
  \DefRef{def:exit_pos}), each containing one of the rightmost $2^n$ forward
  jump instructions of $\tree{n}$. Exactly one of these exit positions will be
  reached after $n$ consecutive $a$-tests, provided that execution starts at
  position $1$. Every instruction in $\tree{n}$ is reachable from position $1$.
  \fref{fig:c_binary_tree} provides a graphical representation of $\tree{3}$.

  \begin{figure}
    \centering
    \begin{tikzpicture}
      \draw[->, level distance=1.25cm,
            level 1/.style={sibling distance=5cm},
            level 2/.style={level distance=1cm,sibling distance=2.5cm},
            level 3/.style={sibling distance=1.25cm}]
        node (A) {$1 \colon \node{1}$}
          child { node (B) {$4 \colon \node{2}$}
            child { node (D) {$10 \colon \node{4}$}
              child { node (H) {$\bullet$} }
              child { node (I) {$\bullet$} }
            }
            child { node (E) {$13 \colon \node{5}$}
              child { node (J) {$\bullet$} }
              child { node (K) {$\bullet$} }
            }
          }
          child { node (C) {$7 \colon \node{3}$}
            child { node (F) {$16 \colon \node{6}$}
              child { node (L) {$\bullet$} }
              child { node (M) {$\bullet$} }
            }
            child { node (G) {$19 \colon \node{7}$}
              child { node (N) {$\bullet$} }
              child { node (O) {$\bullet$} }
            }
          };
      \draw
        let \p1 = (B), \p2 = (C) in ($(\p1)!0.5!(\p2) - (0.5cm,-0.128cm)$)
          node[draw,densely dotted,text width=11cm,minimum height=3.25cm]
          {$\tree{3}{:}$};
      \draw
        let \p1 = (D), \p2 = (G) in ($(\p1)!0.5!(\p2) - (0.5cm,1cm)$)
          node[draw,densely dotted,text width=11cm,minimum height=0.75cm]
          {$X$:};
      \begin{scope}[->,dashed]
        \draw (A.east) .. controls +(2cm,-1.25cm) and +(-2cm,1.25cm).. (B.west);
        \draw (B) -- (C);
        \draw (C.east) .. controls +(2cm,-1.25cm) and +(-2cm,1.25cm).. (D.west);
        \draw (D) -- (E);
        \draw (E) -- (F);
        \draw (F) -- (G);
        \draw (G.east) .. controls +(2cm,-1cm) and +(-2cm,1cm).. (H.west);
      \end{scope}
    \end{tikzpicture}
    \caption{Graphical representation of the \C-expression $\tree{3};X$ as
    defined in \eqref{eq:jump_tree}. The dashed arrows show the order in which
    the subexpressions $\node{1}, \dotsc, \node{7}$ are concatenated (the
    prefixes denote their positions in $\tree{3}$). The solid arrows signify
    jumps between which a choice is made based on the boolean reply to the
    $a$-test in the originating node. As depicted here, all instructions at
    exit positions of $\tree{3}$ jump to distinct positions within the inseq
    $X$. This means that $\len(X) \geq 22$.}

    \label{fig:c_binary_tree}
  \end{figure}
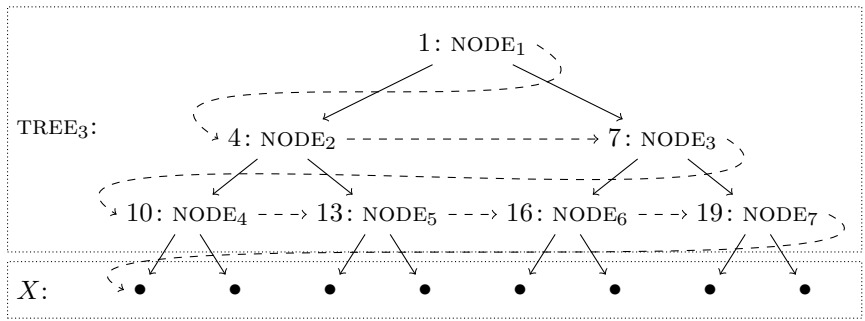

  Towards a contradiction we will now assume that there \emph{does} exist a
  left behavior preserving homomorphism $f$ from the code semigroup \C onto the
  code semigroup \Cg.

  It is easy to see that for any combination of $m \leq 2^n$ exit positions
  $i_1, i_2, \dotsc, i_m$ in $\tree{n}$ there exist some $X \in \UAC$ such that
  all of the following yield distinct behavior:\footnote{In fact, infinitely
  many inseqs $X$ have this property.}
  \[
    \CTE{i_1}{\tree{n};X},
    \CTE{i_2}{\tree{n};X},
    \dotsc,
    \CTE{i_m}{\tree{n};X}.
  \]
  It follows that $f(\tree{n})$ must have at least $2^n - 2$ distinct orphaned
  forward goto instructions, all of which are reachable from the leftmost
  instruction.\footnote{We do not exclude the possibility that either or both
  of the rightmost two instruction positions of $f(\tree{n})$ are exit
  positions containing forward basic instructions, test instructions or label
  instructions. This explains the conservative estimate of $2^n - 2$ instead of
  $2^n$ orphaned forward goto instructions.}

  For all $X \in \UACg$, let $L_X = X \cap \Flabels$ be the set of distinct
  forward label instructions in $X$. Obviously $L_X = L_{X^k}$ for all $k \in
  \Np$.

  Now take some $n, k \in \N$ such that $2^n - 2 > |L_{f(\fbi{a})}|$ and $k
  \geq 3(2^n - 1) + 1$. Then $\CTE{1}{\tree{n};(\fbi{a})^k}$ will perform at
  least $n + 1$ consecutive $a$-actions, irrespective of the boolean replies
  they yield. However, this cannot be the case for
  $\CgTE{1}{f(\tree{n};(\fbi{a})^k)}$. Some of the forward goto instructions in
  $f(\tree{n})$ which are reachable after $n$ $a$-tests cannot have a matching
  label instruction in $f((\fbi{a})^k)$, because the number of distinct forward
  label instructions $|L_{f((\fbi{a})^k)}| = |L_{(f(\fbi{a}))^k}| =
  |L_{f(\fbi{a})}|$ is smaller than the number of distinct forward goto
  instructions (which is at least $2^n - 2 $). Thus we reach a contradiction.
\end{proof}

\paragraph{%
  A Behavior Preserving Homomorphism from $\boldsymbol{\Cr{k}}$ to
  $\boldsymbol{\Cg}$
}

It turns out that the result of \ThmRef{thm:no_c_to_cg_hom} is due to a
surprisingly small lack of information about the context of individual
instructions. Once an upper bound on the size of jump counters in the input
inseq is known, a homomorphism \emph{can} be defined. In other words, there
does exists a homomorphic alternative to the family of behavior preserving
mappings $\CrToCg{k} \colon \UACr{k} \to \UACg$ defined by \eqref{eq:cr_to_cg}
in \Sref{sec:cr_to_cg}. We provide one such alternative definition, by building
on the work of \Sref{sec:cg_relative_jumps}. For all $k \geq 2$ we define,
\begin{equation}\label{eq:cr_to_cg_hom}
  \CrToCg{k} = \rel{k} \circ \phi.
\end{equation}
The homomorphism $\rel{k} \colon \UACg \to \UACg$ is defined by \eqref{eq:rel}
in \Sref{sec:cg_relative_jumps}. Recall that it causes all goto instructions
with label numbers up to and including $k$ to behave as relative jumps. It
should come as no surprise then that the definition of the homomorphism $\phi
\colon \UAC \to \UACg$ is straightforward:
\[
  \phi \colon u \mapsto
    \begin{cases}
      \fgt{k}           & \text{if $u = \fj{k}$,} \\
      \bgt{k}           & \text{if $u = \bj{k}$,} \\
      u                 & \text{otherwise.}
    \end{cases}
\]
Observe that $\CrToCg{k}$ is left-right uniformly behavior preserving. (Like
$\rel{k}$, $\CrToCg{k}$ maps every instruction in the input instruction
sequence to $4k + 6$ instructions in the output.)

\section[Translating \Cg to \C]{%
  Translating $\boldsymbol{\Cg}$ to $\boldsymbol{\C}$
}\label{sec:cg_to_c}

Defining a behavior preserving mapping $\CgToC \colon \UACg \to \UAC$ is rather
straightforward. Label instructions can simply be replaced by relative jumps
over distance $1$ in the appropriate direction. Goto instructions are replaced
by relative jumps to the position of the label instruction which they target,
if any. Orphaned goto instructions can be replaced by an abort instruction or a
jump outside of the instruction sequence. For convenience we will choose to do
the latter.

For all $i \in \Np$ we define functions $\phi_i \colon \UnCg{\leq i} \to \UC$
such that,
\begin{equation}\label{eq:cg_to_c_one_instr}
  \phi_i(X) =
    \begin{cases}
      \fj{j{-}i}  & \text{if $\inst{i}(X) = \fgt{l}$ and $j = \Fsearch(X, i, \{\flbl{l}\}),$} \\
      \bj{i{-}j}  & \text{if $\inst{i}(X) = \bgt{l}$ and $j = \Bsearch(X, i, \{\blbl{l}\}),$} \\
      \fj{1}      & \text{if $\inst{i}(X) = \flbl{l},$} \\
      \bj{1}      & \text{if $\inst{i}(X) = \blbl{l},$} \\
      \inst{i}(X) & \text{otherwise.}
    \end{cases}
\end{equation}
$\phi_i$ determines whether and how the $i$th instruction in a given \Cg-inseq
$X$ should be translated. Only label and goto instructions are replaced,
precisely according to the rules mentioned. Concatenating the results of
appropriate invocations of \eqref{eq:cg_to_c_one_instr}, the mapping $\CgToC
\colon \UACg \to \UAC$ is thus defined:
\begin{equation}\label{eq:cg_to_c}
  \CgToC \colon X \mapsto \phi_1(X);\phi_2(X);\dotsc;\phi_{\len(X)}(X).
\end{equation}
Every label and goto instruction is replaced by a jump instruction which mimics
its transfer of control of execution. Other instructions are unaltered. Thus
$\CgToC$ is left-right uniformly behavior preserving.

\subsection{%
  What About a Homomorphic Translation from $\boldsymbol{\Cg}$ to
  $\boldsymbol{\C}$?
}\label{sec:cg_to_c_hom}

The translation $\CgToC$ defined by \eqref{eq:cg_to_c} is not a homomorphism.
It turns out that this is necessarily so.

\begin{thm}\label{thm:no_cg_to_c_hom}
  There does not exist a behavior preserving homomorphism $f \colon \UACg \to
  \UAC$.

\end{thm}

\begin{proof}
  We prove that no homomorphism $f \colon \UACg \to \UAC$ can be left behavior
  preserving. The proof that no such $f$ can be right behavior preserving is
  analogous.

  For all $n \in \Np$ we define the following \Cg-inseqs:
  \begin{align*}
    \node{n} &= \flbl{n};\fpt{a};\fgt{2n};\fgt{2n{+}1}, \\
    \tree{n} &= \node{1};\node{2};\dotsc;\node{2^n - 1}.
  \end{align*}
  It is not hard to see that $\tree{n}$ contains $2^n$ orphaned goto
  instructions with label numbers $2^n$ through $2^{n + 1} - 1$. For example,
  $\tree{2}$ contains the orphaned goto instructions $\fgt{4}$, $\fgt{5}$,
  $\fgt{6}$ and $\fgt{7}$:
  \begin{align*}
    &\flbl{1};\fpt{a};\fgt{2};\fgt{3}; \\
    &  \qquad\flbl{2};\fpt{a};\fgt{4};\fgt{5}; \\
    &  \qquad\flbl{3};\fpt{a};\fgt{6};\fgt{7}.
  \end{align*}
  If execution of $\tree{n}$ starts at position $1$, then exactly one of the
  orphaned goto instructions will be reached after performing $n$ consecutive
  $a$-actions. Every orphaned goto instruction is reachable.

  Towards a contradiction we will now assume that there \emph{does} exist a
  left behavior preserving homomorphism $f$ from the code semigroup \Cg onto
  the code semigroup \C.

  For all $X \in \UAC$ define $\rightof{X} = \{j - \len(X) \mid i \in [1,
  \len(X)], i \acc{X}^+ j, j > \len(X)\}$. Informally, $\rightof{X}$ contains
  the offsets of ``invalid'' positions to the right of $X$ which are reachable
  from $X$. We fix some $r$ such that $\max(\rightof{f(\fbi{a})}) \leq
  \len(f((\fbi{a})^r))$. Then $\rightof{f(\fbi{a};(\fbi{a})^r)} =
  \rightof{f((\fbi{a})^r)}$, and in fact $\rightof{f((\fbi{a})^k)} =
  \rightof{f((\fbi{a})^r)}$ for all $k \geq r$.

  Next we define $\tree{n,k} = \tree{n};(\fbi{a})^k$ for all $n, k \in \N$, and
  we make two easily verifiable claims:

  \begin{enumerate}[(1)]
    \item For all $n \in \N$, $k, k' \geq 2$ and $X \in \UACg$ the identity
    $\CgTEfwd{\tree{n,k};X} = \CgTEfwd{\tree{n,k'};X}$ holds. To see why this
    is so, observe that all exit positions in $\tree{n,k}$ and $\tree{n,k'}$
    are goto instructions and that $\tree{n,k}$ and $\tree{n,k'}$ do not
    contain backward label instructions. As a result only the last two
    instructions of $\tree{n,k}$ and $\tree{n,k'}$ (which are $\fbi{a}$
    instructions) may be reachable from a position in the ``$X$-part`` of
    $\tree{n,k};X$ and $\tree{n,k'};X$.

    \item For any combination of $m \leq 2^n$ distinct positions of orphaned
    goto instructions $i_1$, $i_2, \dotsc, i_m$ within $\tree{n,k}$ there
    exists an $X \in \UACg$ such that all of the following yield distinct
    behavior:\footnote{Note again that there are in fact infinitely many such
    $X$.}
    \[
      \CgTE{i_1}{\tree{n,k};X}, \CgTE{i_2}{\tree{n,k};X}, \dotsc,
      \CgTE{i_m}{\tree{n,k};X}.
    \]

  \end{enumerate}

  Combining these two claims, we must conclude that $|\rightof{f(\tree{n,k})}|
  \geq 2^n$ for all $n, k \in \N$. Now take some $n$ such that $2^n >
  |\rightof{f((\fbi{a})^r)}|$ and select some $k \geq r$ such that
  $\rightof{f(\tree{n,k})} = \rightof{f(\tree{n};(\fbi{a})^k)} =
  \rightof{f((\fbi{a})^k)} = \rightof{f((\fbi{a})^r)}$. But then
  $|\rightof{f(\tree{n,k})}| = |\rightof{f((\fbi{a})^r)}| < 2^n$.
  Contradiction.
\end{proof}

\paragraph{%
  A Behavior Preserving Homomorphism from $\boldsymbol{\Cgr{k}}$ to
  $\boldsymbol{\C}$
}

Similar to the definition of subsemigroups \Cr{k}, we define subsemigroups
$\Cgr{k} \subset \Cg$ for all $k \in \N$. \Cgr{k} contains precisely those
\Cg-inseqs which do not contain goto instructions with a label number greater
than $k$. That is, we define $\Cgr{n} = (\UACgr{n}, \catOp)$, with
\[
  \UCgr{k} = \UCg - \{u \in \gotos \mid \lno(u) > k\}.
\]
Note that \Cgr{k} places no restriction on label instructions. As such, the
utility of label instructions with a label number greater than $k$ in a
\Cgr{k}-expression is limited.

As per \ThmRef{thm:no_cg_to_c_hom} no total homomorphism from \Cg to \C can be
behavior preserving. However, the family of behavior preserving functions
$\CgToC_k \colon \UACgr{k} \to \UAC$ ($k \in \N$) can be defined such that each
$\CgToC_k$ is a homomorphism. Given arbitrary $k$, we define $\CgToC_k$ on
individual instructions as follows:
\begin{equation}\label{eq:cgr_to_c_hom}
  \CgToC_k(u) = \phi_k(u);\Next_k(u);\Left_k(u);\Right_k(u).
\end{equation}
Here $\phi_k$ is defined as:
\begin{align*}
  \fbi{a}    &\mapsto \fbi{a},        &
    \bbi{a}  &\mapsto \fbi{a},        &
    \flbl{l} &\mapsto \fj{1},         &
    \fgt{l}  &\mapsto \fj{k{+}l{+}4}, \\
  \fpt{a}    &\mapsto \fpt{a},        &
    \bpt{a}  &\mapsto \fpt{a},        &
    \blbl{l} &\mapsto \fj{1},         &
    \bgt{l}  &\mapsto \fj{l{+}3},     \\
  \fnt{a}    &\mapsto \fnt{a},        &
    \bnt{a}  &\mapsto \fnt{a},        &
    \abrt    &\mapsto \abrt,          &
    \term    &\mapsto \term.
\end{align*}
Furthermore, $\Next_k$, $\Left_k$ and $\Right_k$ are defined as follows:
\begin{align*}
  \Next_k &\colon u \mapsto
    \begin{cases}
      \bj{2k{+}6};\bj{4k{+}12} & \text{if $u \in \BUCg$,} \\
      \fj{2k{+}4};\fj{4k{+}8}  & \text{otherwise,}
    \end{cases} \\
  \Left_k &\colon u \mapsto
    \begin{cases}
      (\bj{2k{+}5})^l; \bj{l{+}3}; (\bj{2k{+}5})^{k{-}l} & \text{if $u = \blbl{l}$ and $l \leq k$,} \\
      (\bj{2k{+}5})^{k{+}1}                              & \text{otherwise,}
    \end{cases} \\
  \Right_k &\colon u \mapsto
    \begin{cases}
      (\fj{2k{+}5})^l; \bj{k{+}l{+}4}; (\fj{2k{+}5})^{k{-}l} & \text{if $u = \flbl{l}$ and $l \leq k$,} \\
      (\fj{2k{+}5})^{k{+}1}                                  & \text{otherwise.}
    \end{cases}
\end{align*}
The mapping $\CgToC_k \colon X \mapsto Y$ can be explained using the metaphor
of a ``highway'' that is laid between successive instructions of $X$. The
highway contains a dedicated lane for each goto instruction $\fgt{l}$ and
$\bgt{l}$ for $0 \leq l \leq k$, thus resulting in a highway with $2k + 2$
lanes. The highway is the result of the functions $\Left_k$ and $\Right_k$.
Each $\Cgr{k}$-instruction is mapped onto $2k + 5$ \C-instructions:
\[
  \underbrace{
    u;v;w;
    \overbrace{
      \overbrace{\bj{2k{+}5};\dotsc;\bj{2k{+}5}}^{\text{$k + 1$ ``lanes'' to the left}};
      \overbrace{\fj{2k{+}5};\dotsc;\fj{2k{+}5}}^{\text{$k + 1$ ``lanes'' to the right}}
    }^{\text{label/goto ``highway'' with $2k + 2$ ``lanes''}}
  }_{\text{these $2k + 5$ \C instructions represent a single $\Cgr{k}$ instruction}}
\]
The highway is used solely to mimic the behavior of goto instructions using a
finite number of jumps. The following \C-inseq is yielded by
$\CgToC_k(\fgt{l})$:
\bigskip\[
  \underbrace{
    \rnode{source}{\fj{k{+}l{+}4}};
    \makebox[0pt][l]{$\overbrace{\phantom{
      \fj{2k{+}4};\fj{4k{+}8};(\bj{2k{+}5})^{k + 1};(\fj{2k{+}5})^l
    }}^{\text{$k + l + 3$ instructions}}$}
    \fj{2k{+}4};\fj{4k{+}8}
  }_{\phi_k(\fgt{l})};
  \underbrace{
    (\bj{2k{+}5})^{k + 1}
  }_{\Left_k(\fgt{l})};
  \underbrace{
    (\fj{2k{+}5})^l;
    \overbrace{
      \fj{2k{+}5}
    }^{\text{\rnode{target}{right lane $l$}}};
    (\fj{2k{+}5})^{k - l}
  }_{\Right_k(\fgt{l})}
  \ncbar[nodesep=4pt,angle=90,arm=.25cm,linearc=.15]{**->}{source}{target}
  \lput*{:U}{\text{\scriptsize entering the ``highway''}}
\]
The intention here is that the effect of $\fgt{l}$ is to jump onto the $l$th
highway lane to the right. This lane consists of chained jumps, each of
distance $2k + 5$, until the segment of \C-instructions that is the result of
$\CgToC_k(\flbl{l})$ (note that $l \leq k$, for otherwise $\fgt{l}$ would not
be part of the input). There, a jump instruction off the highway can be found:
\bigskip\[
  \rnode{target}{\fj{1}};
  \makebox[0pt][l]{$\overbrace{\phantom{
    \fj{2k{+}4};\fj{4k{+}8};(\bj{2k{+}5})^{k + 1};(\fj{2k{+}5})^l
  }}^{\text{$k + l + 3$ instructions}}$}
  \underbrace{
    \fj{2k{+}4}
  }_{\substack{\text{to next} \\ \text{$\Cgr{k}$} \\ \text{instruction}}};
  \underbrace{
    \fj{4k{+}8}
    \underbrace{
      (\bj{2k{+}5})^{k + 1}
    }_{\Left_k(\flbl{l})};
    \underbrace{
      (\fj{2k{+}5})^l;
      \overbrace{
        \bj{k{+}l{+}4}
      }^{\text{\rnode{source}{right lane $l$}}};
      (\fj{2k{+}5})^{k - l}
    }_{\Right_k(\flbl{l})}
  }_{\text{$2k + 3$ instructions}}
  \ncbar[nodesep=4pt,angle=90,arm=.25cm,linearc=.15]{<-**}{target}{source}
  \lput*{:U}{\text{\scriptsize leaving the ``highway''}}
\]
$\CgToC_k$ maps each $\Cgr{k}$-instruction in an inseq $X$ onto $2k + 5$
\C-instructions in an inseq $Y$. Thus the \C-instructions corresponding to the
$i$th instruction in $X$ start in $Y$ at position $(i - 1) \cdot (2k + 5) + 1$.

It follows that $\CgTE{i}{X} = \CTE{(i - 1)(2k + 5) + 1}{\CgToC_k(X)}$ for all
$i \in \Z$, $k \leq 2$ and $X \in \UACg$. Thus $\CgToC_k$ is left uniformly
behavior preserving.

\chapter{%
  Some Expressiveness Results
}\label{ch:expressiveness}

As stated in \Sref{sec:c_abrt}, the abort instruction does not enhance \C's
expressiveness as any abort instruction can be replaced by a jump instruction
with a sufficiently large jump counter. In \Sref{sec:c_to_pga} the first of
three steps involving the translation of \C to \PGA involved the elimination of
backward basic/test instructions. These observations naturally lead one to
wonder whether \C contains more redundant instructions. There are at least two
ways to prove that this is indeed the case, both of which will be utilized in
this chapter.

\begin{itemize}
  \item On the one hand one can define a procedure $M$ which, given an
  arbitrary regular thread $T \in \ThrR$, constructs a \C-expression $X$ such
  that $\CTE{i}{X} = T$ for some $i \in [1, \len(X)]$, using only a subset of
  all \C instructions, regardless of $T$. Clearly, any instruction which is not
  utilized by $M$ irrespective of its input is redundant in the sense that it
  does not enhance \C's expressiveness.

  \item On the other hand one can define a function $f$ on $\UAC$ which
  translates any given inseq $X$ to a behaviorally equivalent inseq $Y$, such
  that certain instructions will never be present in $Y$. Again, any such
  instruction can be deemed redundant from the point of view of expressiveness.

\end{itemize}

In our quest to trim \C's instruction set we will inevitably stumble upon
instruction sets which cannot express all threads in $\ThrR$. As we will later
see, there is in fact a hierarchy of expressive power.

Each \C or \Cg instruction $u$ has a \emph{dual} $\dual{u}$: for forward
instructions this is their backward counterpart, and vice versa. The abort and
termination instructions are their own dual. Thus e.g. $\dual{\fbi{a}} =
\bbi{a}$, $\dual{\bnt{b}} = \fnt{b}$ and $\dual{\abrt} = \abrt$. Observe that
the dual operator is an involution: $\dual{\dual{u}} = u$ for all $u \in \UC
\cup \UCg$.

The anti-automorphism $\rev$ reverses a given instruction sequence and converts
all its instructions to their dual. It is defined on \C and well as \Cg
instruction sequences. For example,
\[
  \rev(\fpt{a};\term;\bj{2})
    = \dual{\bj{2}};\dual{\term};\dual{\fpt{a}}
    = \fj{2};\term;\bpt{a}.
\]
Observe that $\rev$ is an involution, because for all $i \in [1, \len(X)]$,
\[
  \inst{i}(X)
    = \dual{\inst{\len(X) - i + 1}(\rev(X))}
    = \dual{\dual{\inst{\len(X) - (\len(X) - i + 1) + 1}(\rev \circ \rev(X))}}
    = \inst{i}(X).
\]
It is not hard to see that $\CTEfwd{X} = \CTEbwd{\rev(X)}$ for arbitrary inseq
$X$. It follows that any code semigroup generated by some set $I \subseteq \UC$
or $I \subseteq \UCg$ is exactly as expressive as the set of its duals
$\{\dual{u} \mid u \in I\}$. Thus $\rev$ tells us something about the
expressiveness of subsemigroups of \C and \Cg.

The remainder of this chapter is organized as follows: in
\Sref{sec:c_expressiveness} we will be concerned with the expressiveness of
several subsemigroups of \C. Specifically, we will show that a reduction of
$\UC$ so that it contains only a finite number of forward or backward jump
instructions (or both) reduces its expressiveness.  In
\Sref{sec:cg_expressiveness} we will combine the results of
\Sref{sec:c_expressiveness} with some of the translations defined in the
previous chapter and use these to make some statements about the expressiveness
of \Cg and some of its subsemigroups.

\section[The Expressiveness of Subsemigroups of \C]{%
  The Expressiveness of Subsemigroups of $\boldsymbol{\C}$
}\label{sec:c_expressiveness}

In \Sref{sec:c_to_pga} it was shown that backward basic instructions and
backward test instructions do not increase \C's expressiveness, by means of a
left behavior preserving endomorphism $f$ on $\UAC$ which does not output any
of these instructions. In other words, the code semigroup generated by the
instruction set $\UC - \Bbasics - \Bpositives - \Bnegatives$ is as expressive
as \C itself. This instruction set is not minimal, however, since the proper
subset $\Fpositives \cup \jumps \cup \{\term\}$ suffices. This is demonstrated
by the left behavior preserving endomorphism $g$, defined on individual
\C-instructions by
\begin{align*}
  \fbi{a} &\mapsto \fpt{a};\fj{2};\fj{1}, &
  \bbi{a} &\mapsto \fpt{a};\bj{4};\bj{5}, \\
  \fpt{a} &\mapsto \fpt{a};\fj{2};\fj{4}, &
  \bpt{a} &\mapsto \fpt{a};\bj{4};\bj{8}, \\
  \fnt{a} &\mapsto \fpt{a};\fj{5};\fj{1}, &
  \bnt{a} &\mapsto \fpt{a};\bj{7};\bj{5}, \\
  \smallskip\\
  \fj{k}  &\mapsto \fj{3k};\term;\term,   &
  \abrt   &\mapsto \fj{1};\bj{1};\term,   \\
  \bj{k}  &\mapsto \bj{3k};\term;\term,   &
  \term   &\mapsto \term;\term;\term.
\end{align*}
The next question which naturally arises is whether the instruction set
$\Fpositives \cup \jumps \cup \{\term\}$ is minimal. For example, can we do
with less than infinitely many jump instructions? And if not, will an infinite
but otherwise arbitrary set of jump instructions suffice? We will now
investigate those questions.

Recall the definition of the subsemigroup \Cr{k} in \Sref{sec:c_to_cg}. As
defined by \eqref{eq:ucr}, \Cr{k}'s instruction set does not contain jump
instructions with a jump counter greater than $k$.

\begin{thm}[Bergstra \& Ponse]\label{thm:b_and_p}
  Let $\setlen{\actions} \geq 2$. There does not exists a value $k \in \Np$
  such that $\Cr{k}$ can express all finite threads.

\end{thm}

See the proof of Theorem 7 in \cite{inseq_intro}; it has been replicated in
\Aref{app:b_and_p_proof}. See the proof of \ThmRef{thm:finite_jumps} below for
a discussion.

\begin{thm}\label{thm:finite_jumps}
  Let $\actions$ be non-empty. There does not exists a value $k \in \Np$ such
  that $\Cr{k}$ can express all finite threads.

\end{thm}

\begin{proof}
  By \ThmRef{thm:b_and_p} we conclude that if $\setlen{\actions} \geq 2$, then
  \Cr{k} cannot express all finite threads. What remains is to be proved is
  that claim also holds if $\setlen{\actions} = 1$. We do this by ``patching''
  the proof by Bergstra \& Ponse. As their proof is rather long we will not
  repeat it here---instead we summarize some key aspects of the proof, point
  out why it requires that $\setlen{\actions} \geq 2$ and then proceed to show
  how this requirement can be eliminated. (Again, the proof is provided
  verbatim in \Aref{app:b_and_p_proof}.)

  The proof uses two key notions:

  \begin{itemize}
    \item Following the definition of residual threads by \eqref{eq:res}, the
    concept of \emph{$n$-residual} threads is defined: $Q$ is a $0$-residual
    thread of $P$ if $P = Q$. $Q$ is an $(n{+}1)$-residual thread of $P$ if $P
    = P_1 \ThrT a \ThrF P_2$ and $Q$ is $n$-residual of either $P_1$ or $P_2$.

    \item Now a thread $P$ has the \emph{$a$-$n$-property} if $\pi_n(P) = a^n
    \circ \Inac$ and $P$ has $2^n - 1$ distinct $n$-residuals with a first
    approximation not equal to $a \circ \Inac$.\footnote{The sentences
    following this definition of the $a$-$n$-property in \cite{inseq_intro}
    make it clear that $P$ is meant to have $2^n$ instead of $2^n - 1$ distinct
    $n$-residuals with a first approximation not equal to $a \circ \Inac$. It
    turns out that this slightly weaker definition of the property does not
    affect the proof in any significant way.} An instruction sequence has the
    $a$-$n$-property if a thread with the $a$-$n$-property can be extracted
    from it.

  \end{itemize}

  The proof by Bergstra and Ponse shows that for every $k \in \N$ there exists
  an $n \in \Np$ such that no \Cr{k}-expression $X$ has the $a$-$n$-property.
  It does so by assuming the contrary and taking the minimal value for $k$ in
  this respect. It is then shown that, given arbitrary $n \in \Np$, one can
  find an $X \in \UACr{k}$ with the $a$-$n$-property for which it is also the
  case that $X \in \UACr{k - 1}$. This contradicts the assumption that $k$ was
  minimal.

  Let $P$ be a thread with the $a$-$n$-property. There are two observations to
  be made. First, if $n > 1$, then the set $\actions$ of actions contains at
  least two elements, for otherwise the requirement that all first
  approximations of the distinct $n$-residuals of $P$ must not equal $a \circ
  \Inac$ cannot be met.

  Second, not only are all of $P$'s $n$-residuals distinct, by extension the
  same holds of all $m$-residuals with $m < n$. Moreover, since all first
  approximations of $n$-residuals of $P$ must not equal $a \circ \Inac$, it
  follows that for any $m$-residual $Q$ and $m'$-residual $R$ with $0 \leq m <
  m' \leq n$ it is necessarily so that $Q \neq R$.

  Summarizing that second observation, we see that every $m$-residual ($m \leq
  n$) of a thread $P$ with the $a$-$n$-property is unique. As a result any
  instruction sequence with the $a$-$n$-property has at least $2^n - 1$
  distinct test instructions with action $a$.

  Analyzing the proof, it turns out that it relies specifically on this second
  observation about threads with the $a$-$n$-property; requiring that threads
  with the $a$-$n$-property ($n > 1$) contain non-$a$ actions is merely a means
  to that end. It turns out that we can define a slightly different class of
  threads with this second property without requiring that $\setlen{\actions}
  \geq 2$: we say that a thread $P$ has the $a$+$n$-property if $\pi_n(P) = a^n
  \circ \Inac$ and $P$ has $2^n$ distinct $n$-residuals, none of which equals
  an $(n{-}m)$-residual of $P$ (for $m > 0$).

  To see why every $m$-residual ($m \leq n$) of a thread $P$ with the
  $a$+$n$-property is unique, assume the contrary: then there are values $m$
  and $m'$ with $m \leq m' \leq n$ such that some $m$-residual $Q$ of $P$
  equals an $m'$-residual $R$ of $P$. But then every $(n - m')$-residual of $R$
  equals some $(n - m')$-residual of $Q$. This yields a contradiction, because
  every $(n - m')$-residual of $R$ is an $n$-residual of $P$, which is by
  definition distinct from any $(n - m')$-residual of $Q$, because $m + (n -
  m') \leq n$. \fref{fig:all_a_n_property_states_unique} attemps to visualize
  this argument using a thread $T$ with the $a$+$5$-property.

  \begin{figure}
    \centering
    \begin{tikzpicture}
      \foreach \h in {0, ..., 5}{
        \pgfmathparse{pow(2, \h) - 1}
        \foreach \c in {0, ..., \pgfmathresult}{
          \pgfmathsetmacro{\xpos}{6.95 * (-1 * (1 - pow(0.5, \h))
                                       + (\c + pow(2, \h)) * pow(0.5, \h - 1))}
          \pgfmathsetmacro{\ypos}{-\h * 0.85}
          \ifthenelse{\h < 5}{
            \draw[shape=circle, inner sep=0mm]
              (\xpos, \ypos) node (\h-\c) {\tiny $\langle a \rangle$};
          }{
            \draw[shape=circle, inner sep=0mm, minimum size=4mm]
              (\xpos, \ypos) node (\h-\c) {\tiny $_{R_{\c}}$};
          }
          \ifthenelse{\h > 0}{
            \pgfmathtruncatemacro{\ph}{\h - 1}
            \pgfmathtruncatemacro{\pc}{floor(\c / 2)}
            \draw[->, shorten >=1mm, shorten <=1mm] (\ph-\pc) -- (\h-\c);
          }{}
        }
      }

      \node[label={120:$T$}] at (0-0) {};
      \foreach \n/\sh/\sc/\zrl/\trl in {1/1/0/120/120, 2/3/5/60/-90}{
        \pgfmathtruncatemacro{\trh}{\sh + 2}
        \begin{scope}[every node/.style={shape=circle, draw, inner sep=1.5mm},
                      every label/.style={draw opacity=0, inner sep=0.5mm}]
          \foreach \h in {\sh, ..., 5}{
            \pgfmathtruncatemacro{\sd}{\sc * pow(2, \h - \sh)}
            \pgfmathparse{\sd + pow(2, \h - \sh) - 1}
            \foreach \c in {\sd, ..., \pgfmathresult}{
              \ifthenelse{\h = \sh}{
                \node[label={\zrl:$P_\n$}] at (\h-\c) {};
              }{\ifthenelse{\h = \trh \and \c = \sd}{
                \node[label={\trl:$Q_\n$}] at (\h-\c) {};
              }{
                \node at (\h-\c) {};
              }}
            }
          }
        \end{scope}
      }
    \end{tikzpicture}
    \caption{Graphical representation of a thread $T$ with the
    $a$+$5$-property. The ``leaves'' $R_n$ in this tree represent
    pairwise distinct $5$-residuals of $T$ which are each also distinct from
    any $m$-residual of $T$ for $m < 5$. This in turns means that all
    $m$-residuals for $m \leq 5$ are pairwise distinct. For if e.g. $P_1$ and
    $P_2$ are not distinct, then $Q_1$ and $Q_2$ are identical as well,
    violating $T$'s $a$+$5$-property. A similar argument holds for any pair of
    $m$-residuals with $m \leq 5$.}

    \label{fig:all_a_n_property_states_unique}
  \end{figure}

  For every $n \in \Np$ a thread $P$ with the $a$+$n$-property can be created,
  such that $P$ performs only $a$ actions. Fix some $n$ and let $g \colon [0,
  2^n - 1] \to \{\true, \false\}^n$ be a bijection, where $\{\true, \false\}^n$
  is the set of all boolean sequences of length $n$. We write $(g(m))_{d + 1}$
  for the $(d{+}1)$th element of $g(m)$. Now we define the family of threads
  $P^l$ for all $1 \leq l < 2^n$ such that:\footnote{In this definition
  relevant values for $d$ and $m$ are in the ranges $[0, n - 1]$ and $[0, 2^n -
  1]$, respectively.}
  \begin{subequations}\label{eq:finite_jumps}\begin{align}
    P^l &=
      \begin{cases}
        P^{2l} \ThrT a \ThrF P^{2l + 1}                 & \text{if $l < 2^{n - 1}$,} \\
        Q_{2l - 2^n}^n \ThrT a \ThrF Q_{2l - 2^n + 1}^n & \text{otherwise,}
      \end{cases} \\
    Q_m^0 &= a \circ \Inac, \\
    Q_m^{d + 1} &=
      \begin{cases}
        Q_m^d \ThrT a \ThrF \Inac & \text{if $(g(m))_{d + 1} = \false$,} \\
        \Inac \ThrT a \ThrF Q_m^d & \text{otherwise.} \\
      \end{cases}
  \end{align}\end{subequations}
  Informally, the thread $P^1$ performs $n$ $a$-actions after which some state
  $Q_m^n$ is reached. Due to the nature of $g$, $Q_m^n \neq Q_{m'}^n$ for
  distinct $m$ and $m'$. For example, for $n = 2$ and taking $g$ such that
  \begin{align*}
    0 &\mapsto \{\false, \false\}, &
    1 &\mapsto \{\false, \true\}, &
    2 &\mapsto \{\true, \false\}, &
    3 &\mapsto \{\true, \true\},
  \end{align*}
  the following equations are defined:
  \begin{align*}
    P^1 &= P^2 \ThrT a \ThrF P^3, &
    P^2 &= Q_0^2 \ThrT a \ThrF Q_1^2, &
    P^3 &= Q_2^2 \ThrT a \ThrF Q_3^2,
  \end{align*}
  and,
  \begin{align*}
    Q_0^2 &= Q_0^1 \ThrT a \ThrF \Inac, &
    Q_1^2 &= Q_1^1 \ThrT a \ThrF \Inac, &
    Q_2^2 &= \Inac \ThrT a \ThrF Q_2^1, &
    Q_3^2 &= \Inac \ThrT a \ThrF Q_3^1, \\
    Q_0^1 &= Q_0^0 \ThrT a \ThrF \Inac, &
    Q_1^1 &= \Inac \ThrT a \ThrF Q_1^0, &
    Q_2^1 &= Q_2^0 \ThrT a \ThrF \Inac, &
    Q_3^1 &= \Inac \ThrT a \ThrF Q_3^0, \\
    Q_0^0 &= a \circ \Inac, &
    Q_1^0 &= a \circ \Inac, &
    Q_2^0 &= a \circ \Inac, &
    Q_3^0 &= a \circ \Inac.
  \end{align*}
  Observe that any thread $Q^n_m$ performs $n + 1$ $a$-actions only if the
  sequence of boolean replies yielded by the first $n$ actions is exactly
  according to $g(m)$. Thus each thread $Q_m^n$ is a unique $n$-residual of
  $P^1$ (recall that $g$ is bijective). Since $\Inac$ is a $1$-residual of
  every thread $Q_m^n$, but not of any thread $P^l$ we conclude that $P^1$
  meets the necessary criteria to have the $a$+$n$-property.

  Replacing any thread with the $a$-$n$-property in the proof of Bergstra \&
  Ponse with a thread with the $a$+$n$-property results in a valid proof which
  requires only that $\setlen{\actions} \neq \emptyset$, as opposed to
  $\setlen{\actions} > 1$. This proves our claim.
\end{proof}

We have now established that arbitrarily many distinct jump instructions are
required to let \C express all finite threads. It turns out that jump
instructions in a single direction suffice.

\begin{prop}\label{prop:one_direction_jumps_only}
  Let $\Fjsub \subseteq \Fjumps$ be an infinite but otherwise arbitrary set of
  forward jump instructions and let the code semigroup \Cp be generated by the
  instruction set $\Fpositives \cup \Fjsub \cup \{\term\}$. Then \Cp can
  express all finite thread but no infinite threads. This also holds if
  $\Fpositives$ is replaced by $\Fnegatives$. If $\Bjsub \subseteq \Bjumps$ is
  an infinite but otherwise arbitrary set of backward jump instructions, then
  the instruction sets $\Bpositives \cup \Bjsub \cup \{\term\}$ and
  $\Bnegatives \cup \Bjsub \cup\{\term\}$ also generate a code semigroup which
  characterizes \BTA.

\end{prop}

\begin{proof}
  As \Cp does not contain backward instructions, it cannot create any kind of
  loop (for all $i, j \in \Z$, if $i \acc{X} j$ according to some $X \in
  \UACp$, then necessarily $i < j$). Every instruction sequence is finite, thus
  so is any thread extracted from a \Cp-inseq $X$. What remains to be shown is
  that all \BTA threads can be described by \Cp.

  Let $P \in \Thr$ be a finite thread. We will inductively construct a \Cp
  instruction sequence $X_P$ such that $\CTEfwd{X_P} = P$. For convenience we
  will define $F = \{\jd(u) \mid u \in \Fjsub\}$ to be the set of jump counters
  of admitted jump instructions.

  If $P = \Term$ then define $X_P = \term$. If $P = \Inac$ then define $X_P =
  \fj{k}$, for some $k \in F$. Otherwise $P = Q \ThrT a \ThrF R$ for some $a
  \in \actions$ and $Q, R \in \Thr$. By induction there are $X_Q, X_R \in
  \UACp$ such that $\CTEfwd{X_Q} = Q$ and $\CTEfwd{X_R} = R$.

  Create an inseq $X_R'$ from $X_R$ by changing the jump counter $k$ of any
  jump instruction at an exit position in $X_R$ to some value $k' \in \{j \in F
  \mid j \geq k + \len(X_Q)\}$. (These are the instructions which upon
  execution cause deadlock).

  Now we define $X_P = \fpt{a};\fj{k};X_R';(\term)^p;X_Q$, where $k \in \{j \in
  F \mid j > \len(X_R')\}$ and $p = k - \len(X_R') - 1$. It is not hard to see
  that indeed $\CTEfwd{X_P} = P$. Note that the termination instructions
  introduced here are solely for the purpose of \emph{padding}. They are not
  reachable from the leftmost instruction.

  A similar construction can be made using negative tests. When using backward
  jump instructions create an inseq $X_P$ such that $\CTEbwd{X_P} = P$.
\end{proof}

Although all finite threads can be expressed using jump instructions in only one
direction, this is not the case for all regular threads. In fact, infinitely
many distinct jump instructions in both directions are necessary.

\begin{defn}
  In an instruction sequence $X = u_1;u_2;\dotsc;u_k \in \UA_A$ an instruction
  $u_j$ is \emph{$i$-$n$-relevant} if there exists an instruction sequence
  $X'$, created from $X$ by changing $u_j$ to some other instruction $u \in
  \U_A$, such that $\pi_n(\ATE{i}{X}) \neq \pi_n(\ATE{i}{X'})$. In other words:
  the $n$th projection of the execution of inseq $X$ starting at position $i$
  depends on $u_j$. Observe that any instruction which is $i$-$n$-relevant is
  also $i$-$(n{+}1)$-relevant.

\end{defn}

\begin{thm}\label{thm:one_direction_jump_restriction_expressiveness}
  Let $\actions$ be non-empty and fix some $k \in \Np$. Let $\UCp$ be the
  largest subset of $\UC$ which does not contain forward (backward) jump
  instructions with a jump counter greater than $k$ (i.e., $\UCp$ contains a
  finite number of forward or backward jump instructions). Then the semigroup
  \Cp generated by $\UCp$ cannot express all regular threads.

\end{thm}

\begin{proof}
  Let $k$ be fixed and select $n$ such that $2^n \geq 2k + 3$. We will assume
  that \Cp restricts forward jump instructions (a similar argument holds if
  backward jump instructions are restricted). Let $g \colon [0, 2^{2n} - 1] \to
  \{\true, \false\}^{2n}$ be a bijection, where $\{\true, \false\}^{2n}$ is the
  set of all boolean sequences of length $2n$. We write $(g(m))_{d + 1}$ for
  the $(d{+}1)$th element of $g(m)$. Now we define the family of threads $P^l$
  for all $1 \leq l < 2^{2n}$ such that:\footnote{In this definition relevant
  values for $d$ and $m$ are in the ranges $[0, 2n - 1]$ and $[0, 2^{2n} - 1]$,
  respectively.}
  \begin{subequations}\label{eq:one_direction_jump_restriction}\begin{align}
    P^l &=
      \begin{cases}
        P^{2l} \ThrT a \ThrF P^{2l + 1}                             & \text{if $l < 2^{2n - 1}$,} \\
        Q_{2l - 2^{2n}}^{2n} \ThrT a \ThrF Q_{2l - 2^{2n} + 1}^{2n} & \text{otherwise,}
      \end{cases} \\
    Q_m^0 &= \Inac, \\
    Q_m^{d + 1} &=
      \begin{cases}
        Q_m^d \ThrT a \ThrF P^{2^n + \rem{2^n}{m}} & \text{if $(g(m))_{d + 1} = \false$,} \\
        P^{2^n + \rem{2^n}{m}} \ThrT a \ThrF Q_m^d & \text{otherwise.}
      \end{cases}
  \end{align}\end{subequations}

  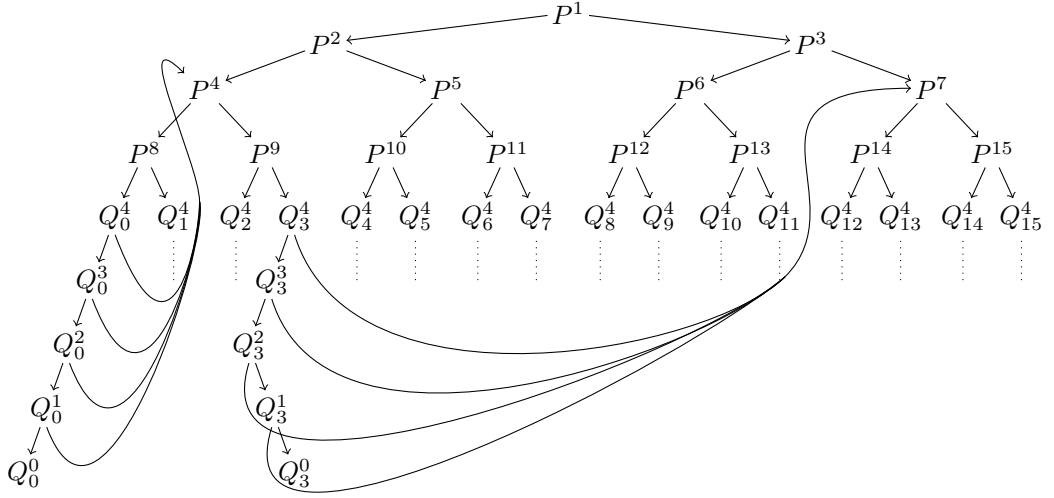
\begin{figure}
    \centering
    \begin{tikzpicture}
      \newcommand{\todec}[1]{\pgfmathbasetodec{\dec}{#1}{2}\dec}

      \draw[->, level distance=1cm, inner sep=0.05cm,
            level 1/.style={level distance=0.4cm, sibling distance=6.38cm},
            level 2/.style={level distance=0.6cm, sibling distance=3.18cm},
            level 3/.style={level distance=0.85cm, sibling distance=1.59cm},
            level 4/.style={sibling distance=0.77cm}]
        node (P1) {$P^1$}
        child foreach \a in {0, 1} {
          node (P1\a) {$P^{\todec{1\a}}$}
          child foreach \b in {0, 1} {
            node (P1\a\b) {$P^{\todec{1\a\b}}$}
            child foreach \c in {0, 1} {
              node (P1\a\b\c) {$P^{\todec{1\a\b\c}}$}
              child foreach \d in {0, 1} {
                node (Q4-\a\b\c\d) {$Q^4_{\todec{\a\b\c\d}}$}
              }
            }
          }
        };

      \begin{scope}
        \draw[->, inner sep=0.05cm, level distance=0.85cm, sibling distance=0.6cm,
              level 1/.style={shorten <=2mm},
              level 2/.style={shorten <=0mm}]
          node at (Q4-0000) {}
          child { node (Q3-0000) {$Q^3_0$}
            child { node (Q2-0000) {$Q^2_0$}
              child { node (Q1-0000) {$Q^1_0$}
                child { node (Q0-0000) {$Q^0_0$} }
                child[draw opacity=0] { node (Q1-0000-D) {} }
              }
              child[draw opacity=0] { node (Q2-0000-D) {} }
            }
            child[draw opacity=0] { node (Q3-0000-D) {} }
          }
          child[draw opacity=0] { node (Q4-0000-D) {} };

        \foreach \n/\x in {4/-4.85, 3/-4.8, 2/-4.75, 1/-4.7}{
          \draw[-] (Q\n-0000.south) .. controls +(0.75, -2) and
                                                (\x , -3) .. (-4.85, -2.5);
        }
        \draw[->] (-4.85, -2.5) .. controls ($ (-4.85, -3) !2! (-4.85, -2.5) $)
                                   and      +(-0.75, 0.75) .. (P100.north west);
      \end{scope}

      \begin{scope}
        \draw[->, inner sep=0.05cm, level distance=0.85cm, sibling distance=0.6cm,
              level 1/.style={shorten <=2mm},
              level 2/.style={shorten <=0mm}]
          node at (Q4-0011) {}
          child { node (Q3-0011) {$Q^3_3$}
            child { node (Q2-0011) {$Q^2_3$}
              child[draw opacity=0] { node (Q2-0011-D) {} }
              child { node (Q1-0011) {$Q^1_3$}
                child[draw opacity=0] { node (Q1-0011-D) {} }
                child { node (Q0-0011) {$Q^0_3$} }
              }
            }
            child[draw opacity=0] { node (Q3-0011-D) {} }
          }
          child[draw opacity=0] { node (Q4-0011-D) {} };

        \foreach \n/\xd/\xa in {4/0.75/1.85, 3/0.75/1.8, 2/-0.75/1.75, 1/-0.75/1.7}{
          \draw[-] (Q\n-0011.south) .. controls +(\xd, -2.45) and
                                                (\xa , -4.5) .. (2.85, -3.5);
        }
        \draw[->] (2.85, -3.5) .. controls ($ (1.85, -4.5) !2! (2.85, -3.5) $)
                                  and      +(-2.5, 0) .. (P111.west);
      \end{scope}

      \pgfmathsetbasenumberlength{4}
      \foreach \s/\e in {1/2, 4/15}{
        \foreach \l in {\s, ..., \e}{
          \pgfmathdectobase{\bin}{\l}{2}
          \draw[-, dotted, shorten <=2mm, level distance=1cm]
            node at (Q4-\bin) {} child { node (Q4-\bin-hidden) {} };
        }
      }
    \end{tikzpicture}
    \caption{Graphical representation of the thread described by $P^1$ as defined
    by \eqref{eq:one_direction_jump_restriction}, for $n = 2$. Observe that the
    threads $P^4$, $P^5$, $P^6$ and $P^7$ (i.e. the threads $P^{2^n}$ through
    $P^{2^{n + 1} - 1}$) are $n$-residuals of of $P^1$. Likewise each thread
    thread $Q^4_m = Q^{2n}_m$ is a $2n$-residual of $P^1$. Each thread $Q^4_m$ is
    distinct, and each of $P^1$'s $n$-residuals is a residual thread of each
    thread $Q^4_m$. Expanded are threads $Q^4_0$ and $Q^4_3$ which are defined
    according to $g(0) = \{\false, \false, \false, \false\}$ and $g(3) =
    \{\false, \false, \true, \true\}$, respectively. Note that $Q^0_m = \Inac$
    for all $m \in [0, 15]$, thus in particular $Q^0_0 = Q^0_3 = \Inac$.}

    \label{fig:one_direction_jump_restriction_proof_thread}
  \end{figure}

  \fref{fig:one_direction_jump_restriction_proof_thread} presents a graphical
  representation of thread $P^1$ for $n = 2$. Observe the similarities of this
  set of equations to those presented in \eqref{eq:finite_jumps}. Recall from
  \Sref{sec:cr_to_cg} that $\rem{2^n}{m}$ is the remainder of $m$ after
  division by $2^n$. Informally, the thread $P^1$ performs $2n$ $a$-actions
  after which some state $Q_m^{2n}$ is reached. Distinct sequences of boolean
  replies to these actions result in distinct values for $m$ ($0 \leq m <
  2^{2n}$). Due to the nature of $g$, $Q_m^{2n} \neq Q_{m'}^{2n}$ for distinct
  $m$ and $m'$. (To see why, observe that the $2n$-residual $\Inac$ of
  $Q_m^{2n}$ can be reached starting in state $Q_m^{2n}$ only if the replies to
  the first $2n$ $a$-actions are precisely according to $g(m)$---and $g$ is a
  bijection). Thus each thread $Q_m^{2n}$ is a unique $2n$-residual of $P^1$.
  Since $\Inac$ is a $2n$-residual of every thread $Q_m^{2n}$, but not of any
  thread $P^l$ we conclude that $P^1$ meets the necessary criteria to have the
  $a$+$2n$-property.

  Towards a contradiction assume that there exists a \Cp-expression $X$ such
  that $\CTE{i}{X} = P^1$ for some $i \in [1, \len(X)]$. We define $f(l) =
  \min\{i \mid \CTE{i}{X} = P^l\}$ to be the function which returns the
  leftmost position in $X$ from which the thread $P^l$ can be extracted.
  Without loss of generality we will assume that all instructions in $X$ are
  reachable from position $i$, for if not, then by
  \PropRef{prop:c_all_instr_reachable_thread} we can create an instruction
  sequence $X'$ for which this \emph{does} hold. The largest jump counter of
  any forward jump instruction in $X'$ would be less than or equal to the
  largest forward jump distance in $X$.

  For distinct $l, l' < 2^{2n}$ it is the case that $P^l \neq P^{l'}$ (because
  $P^1$ has the $a$+$2n$-property) and thus necessarily $f(l) \neq f(l')$. The
  $n$-residuals of $P^1$ are the threads $P^l$ for $l \in [2^n, 2^{n + 1} -
  1]$. The integers in this range are totally ordered by the function
  $f$:\footnote{The ordering on $[2^n, 2^{n + 1} - 1]$ imposed by $f$ does not
  need to be the natural ordering of these integers!}
  \[
    l_0, l_1, \dotsc, l_{2^n - 1}.
  \]
  No instruction in $X$ is both $f(l_i)$-$n$-relevant and $f(l_j)$-$n$-relevant
  for distinct $i$ and $j$, because every thread $P^{l_i}$ is an $n$-residual
  of $P^1$, and $P^1$ has the $a$+$2n$-property. Moreover, the $n$-residuals of
  any thread $P^{l_i}$ are the threads $Q_{i2^n + m}^{2n}$, for $0 \leq m <
  2^n$. The thread $P^{l_m}$ in turn is an $1$-residual (and a $2, 3, \dotsc,
  2n$-residual) of the thread $Q_{i2^n + m}^{2n}$. Thus every thread $P^{l_j}$
  is a residual thread of every thread $P^{l_i}$.

  Recall that $2^n \geq 2k + 3$ and that \Cp does not contain forward jump
  instructions over a distance greater than $k$. Thus for some $i < k + 1$ all
  $f(l_i)$-$n$-relevant instructions are left of position $f(l_{k + 1})$. For
  if not, then there are $k + 1$ distinct positions $< f(l_{k + 1})$ containing
  jump instructions which target $k + 1$ distinct positions $> f(l_{k + 1})$.
  This is not possible because of the restriction on forward jump counters.

  Fix said $i$, and note that there are at least $k + 1$ instructions which are
  $f(l_i)$-$(n{+}1)$-relevant to the right of $f(l_{k + 1})$: namely $f(l_{k +
  2}), f(l_{k + 3}), \dotsc, f(l_{2k + 2})$. This leads to a contradiction,
  since this, too, is not possible because of the restriction on jump counters.
\end{proof}

Now that it has been established that an upper bound on the value of jump
counters limits expressiveness, even if only in a single direction, the
question naturally arises whether \emph{any} two infinite collections of
forward and backward jump instructions suffice to express all regular threads.
We prove that this is indeed the case.

\begin{thm}\label{thm:two_directions_infinitely_many_jumps_enough}
  Let $\Fjsub \subseteq \Fjumps$ and $\Bjsub \subseteq \Bjumps$ be two infinite
  but otherwise arbitrary sets of jump instructions and let the code semigroup
  \Cp be generated by the set $\Fpositives \cup \Fjsub \cup \Bjsub \cup
  \{\term\}$. Then all regular threads can be expressed by \Cp. This also holds
  if $\Fpositives$ is replaced by $\Fnegatives$, $\Bpositives$ or
  $\Bnegatives$.

\end{thm}

\begin{proof}
  Fix some infinite $\Fjsub \subseteq \Fjumps$ and $\Bjsub \subseteq \Bjumps$
  and select arbitrary $T \in \ThrR$ with states $P_0, P_1, \dotsc, P_{n - 1}$.
  Then the result of the procedure \Call{ConstructInseq}{$T$, $\{\jd(u) \mid u
  \in \Fjsub\}$, $\{\jd(u) \mid u \in \Bjsub\}$} as outlined in
  \AlgoRef{alg:c_inseq_gen_algo} is a \Cp-inseq $X$ such that $\CTEfwd{X} = T$.

  \begin{algorithm}
    \caption{\C-expression construction using a restricted set of jump counters}
    \label{alg:c_inseq_gen_algo}

    \begin{algorithmic}[1]
      \Require A regular thread $T$ with states $P_0, P_1, \dotsc, P_{n - 1}$ and infinite sets $F, B \subseteq \N$.
      \Ensure A \C-inseq $X$ with $\CTEfwd{X} = P_0$, $\{\jd(u) \mid u \in \Fjumps(X)\} \subset F$, $\{\jd(u) \mid u \in \Bjumps(X)\} \subset B$.

      \Statex
      \Procedure{ConstructInseq}{$T$, $F$, $B$}
        \State $s \gets \RandomSelect{$\{j \in F \mid j \geq 4\}$}$
        \State $z \gets n \cdot s \cdot (s - 1)$               \Comment{Largest (rightmost) instruction position}
        \State $I \gets \emptyset$                             \Comment{Set of $(position, instruction)$ tuples}
        \For{i}{0}{n - 1}                                      
          \For{r}{0}{s - 1}                                    
            \State $c \gets (i \cdot s + r) \cdot (s - 1) + 1$ 
            \If{$P_i = \Term$}
              \State $I \gets I \cup \{(c, \term)\}$
            \ElsIf{$P_i = \Inac$}
              \State $d \gets \RandomSelect{$\{j \in B \mid j \geq c\}$}$
              \State $I \gets I \cup \{(c, \bj{d})\}$        \Comment{Jump outside program: deadlock}
            \ElsIf{$P_i = P_j \ThrT a \ThrF P_k$}
              \State $I \gets I \cup \{(c, \fpt{a})\}$
              \State $I \gets I \cup \Connect{$c + 1$, $j \cdot s \cdot (s - 1) + 1$, $z$, $s$, $F$, $B$}$
              \State $z \gets \max\{p \mid \exists u[(p, u) \in I]\}$
              \State $I \gets I \cup \Connect{$c + 2$, $k \cdot s \cdot (s - 1) + 1$, $z$, $s$, $F$, $B$}$
              \State $z \gets \max\{p \mid \exists u[(p, u) \in I]\}$
            \EndIf
          \EndFor
        \EndFor
        \State \Return $\ConcatInstructions{$I \cup \{(p, \term) \mid 0 < p < z, \lnot\exists u[(p, u) \in I]\}$}$
      \EndProcedure
      \Statex
      \Procedure{Connect}{$i$, $j$, $z$, $s$, $F$, $B$}
        \State $r \gets i + \RandomSelect{$\{k \in F \mid i + k > z\}$}$
        \State $l \gets r - \RandomSelect{$\{k \in B \mid r - k \leq j\}$}$
        \State $p \gets \lfloor(j - l)/s\rfloor$
        \State $p \gets p + j - (l + p \cdot s)$
        \State \Return $\{(i, \fj{r{-}i})\} \cup \{(r + k \cdot s, \fj{s}) \mid 0 \leq k < p\} \cup \{(r + p \cdot s, \bj{r{-}l})\} $
      \EndProcedure
    \end{algorithmic}
  \end{algorithm}

  Suppose we want to transfer control of execution in an inseq $X$ from
  position $i$ to position $j$. Obviously, $\Fjsub \cup \Bjsub$ may not contain
  the jump instruction required to jump immediately from $i$ to $j$. In fact,
  it may be so that no sequence of jump instructions permitted by $\Fjsub \cup
  \Bjsub$ can transfer control of execution from position $i$ to $j$. For
  example, if only even jump counters are available, then control of execution
  cannot be transferred from $i$ to $j$ if $i - j$ is odd.

  \AlgoRef{alg:c_inseq_gen_algo} solves this issue by producing an instruction
  sequence $X$ in which functionally equivalent subsequences of instructions
  are repeated $s$ times at evenly spaced intervals of length $s - 1$. The
  value of $s$ is selected from the set of permissible forward jump counters
  $\Fjsub$, with the sole restriction that $s \geq 4$. Thus, for any $P_k$
  there are at least $s$ positions $j_0, j_1, \dotsc, j_{s - 1}$ (with $j_{m +
  1} = j_m + s - 1$) in $X$ from which $P_k$ can be extracted and for any
  position $i$ in $X$ there is at least one such position $j_m$ such that $i =
  j_m \pmod s$.

  Now the general procedure to ``connect'' a position $i$ to one such $j_m$ in
  $X$ using a sequence of permissible jump instructions is to extend $X$ with a
  sequence of jump instructions to the right of $X$, as follows. First, select
  a sufficiently large forward jump instruction $f$ which, if placed at
  position $i$, jumps outside of $X$ to some position $r$. Second, select a
  sufficiently large backward jump instruction $b$ which, if placed at position
  $r$, jumps to a position $l \leq j_0$. Now observe that, instead of placing
  $b$ at position $r$, we can add a sequence of chained $\fj{s}$ instructions,
  starting at position $r$ and extending to the right, such that they transfer
  control of execution to some position $r' > r$. $r'$ can be selected such
  that if the backward instruction $b$ were placed there, it would jump to a
  position $l'$ between $j_0 - (s - 1)$ and $j_0$. By adding another $j_0 - l'$
  chained $\fj{s}$ instructions starting at position $r'$, control of execution
  will be transferred to a position $r'' > r'$ from which the instruction $b$
  will target exactly one of the positions $j_m$. Specifically, $m = j_0 - l'$.
  The procedure described here is performed by \Connect{$i$, $j_1$, $\len(X)$,
  $s$, $\{\jd(u) \mid u \in \Fjsub\}$, $\{\jd(u) \mid u \in \Bjsub\}$}, which
  returns the required jump instructions and the positions where they should be
  placed.

  The procedure \Call{ConstructInseq}{$T$, $\{\jd(u) \mid u \in \Fjsub\}$,
  $\{\jd(u) \mid u \in \Bjsub\}$} selects a suitable value $s$ and ensures that
  for every thread $P_i$ there are $s$ positions $j_0, j_1, \dotsc, j_{s - 1}$
  from which $P_i$ can be extracted. At each of these positions it places a
  suitable instruction: $\term$ if $P_i = \Term$, $\abrt$ if $P_i = \Inac$ and
  $\fpt{a}$ if $P_i = P_{i'} \ThrT a \ThrF P_{i''}$. In the latter case
  \Connect{\dots} is used to ensure that indeed either of $P_{i'}$ and
  $P_{i''}$ will be reached after exectution of action $a$.
\end{proof}

\section[The Expressiveness of Subsemigroups of \Cg]{%
  The Expressiveness of Subsemigroups of $\boldsymbol{\Cg}$
}\label{sec:cg_expressiveness}

Equipped with the translations of \Cref{ch:translations} and the theorems of
\Sref{sec:cg_expressiveness}, we are now ready to make statements about the
expressiveness of \Cg and some of its subsemigroups.

\begin{prop}\label{prop:cg_characterizes_regular_threads}
  Each thread definable in \Cg is regular, and each regular thread can be
  expressed in \Cg.

\end{prop}

\begin{proof}
  This follows immediately from the fact that $\CToCg$ and $\CgToC$ are
  behavior preserving and total. Since \C characterizes the regular threads
  (see \PropRef{prop:c_characterizes_regular_threads}), so does \Cg.
\end{proof}

\begin{thm}\label{thm:finite_labels_gotos}
  Let $\actions$ be non-empty. There does not exists a value $k \in \Np$ such
  that $\Cgr{k}$ can express all finite threads.

\end{thm}

\begin{proof}
  Upon analyzing the family of translations $\CgToC_k$ as defined in
  \Sref{sec:cg_to_c_hom}, we see that they map $\Cgr{k}$-expressions to
  behaviorally equivalent $\Cr{4k + 12}$-expressions.

  Thus if $\Cgr{k}$ can express all finite threads, then so can $\Cr{4k + 12}$.
  But by \ThmRef{thm:finite_jumps} this is impossible.
\end{proof}

\begin{prop}
  Let $\Fgsub \subseteq \Fgotos$ be an infinite but otherwise arbitrary set of
  forward goto instructions and let $\Flsub \subseteq \Flabels$ constitute the
  set of label instructions which match the goto instructions in $\Fgsub$. Then
  the code semigroup \Cgp generated by the instruction set $\Fpositives \cup
  \Fgsub \cup \Flsub \cup \{\term\}$ can express all finite threads but no
  infinite threads. This also holds if $\Fpositives$ is replaced by
  $\Fnegatives$. If the infinite sets $\Bgsub \subseteq \Bgotos$ and $\Blsub
  \subseteq \Blabels$ are defined analogously, then the instruction sets
  $\Bpositives \cup \Bgsub \cup \Blsub \cup \{\term\}$ and $\Bnegatives \cup
  \Bgsub \cup \Blsub \cup \{\term\}$ also generate a code semigroup capable of
  expressing all finite threads.

\end{prop}

\begin{proof}
  As in the proof of \PropRef{prop:one_direction_jumps_only} we observe that
  $\Cgp$ does not contain backward instructions. Thus it can only express
  finite threads, as loops (a requirement for infinite behavior) cannot be
  constructed in \Cgp. Now we need to show all \BTA threads can be expressed by
  \Cgp.

  We will inductively define a \Cgp instruction sequence $X_P$ for every $P \in
  \Thr$ such that $\CgTEfwd{X_P} = P$. Let $F = \{\lno(u) \mid u \in \Fgsub\}$
  be the set of label numbers of available goto instructions.

  If $P = \Term$ then $X_P = \term$. If $P = \Inac$ then set $X_P = \fgt{l}$,
  where $l$ is an arbitrary element of $F$. Otherwise $P = Q \ThrT a \ThrF R$
  and there are $X_Q, X_R \in \UACgp$ such that $\CgTEfwd{X_Q} = Q$ and
  $\CgTEfwd{X_R} = R$. Select some label number $l \in F$ such that it is not
  present in $X_Q$ or $X_R$. Then $X_P = \fpt{a};\fgt{l};X_R;\flbl{l};X_Q$.

  A similar construction can be made using negative tests. When using backward
  goto instructions create an inseq $X_P$ such that $\CgTEbwd{X_P} = P$.
\end{proof}

\begin{thm}\label{thm:one_direction_label_goto_restriction_expressiveness}
  Let $\actions$ be non-empty and fix some value $k \in \Np$. Let $\UCgp$ be
  the largest subset of $\UCg$ which does not contain forward (backward) goto
  instructions with a label number $k$ or greater (i.e., $\UCgp$ contains a
  finite number of forward or backward goto instructions). Then the semigroup
  \Cgp generated by $\UCgp$ cannot express all regular threads.

\end{thm}

\begin{proof}
  The proof is analogous to that of
  \ThmRef{thm:one_direction_jump_restriction_expressiveness}. Again select $n$
  such that $2^n \geq 2k + 3$ and consider the thread $P_1$ as defined by
  \eqref{eq:one_direction_jump_restriction}. As before the function $f(l) =
  \min\{i \mid \CTE{i}{X} = P^l\}$ induces a total ordering on the range $[2^n,
  2^{n + 1} - 1]$, say $l_0, l_1, \dotsc l_{2^n - 1}$. Observe that for some $i
  < k + 1$ all $f(l_i)$-$n$-relevant instructions are left of position $f(l_{k
  + 1})$, for otherwise there must be $k + 1$ distinct goto instructions on
  positions $< f(l_{k + 1})$ which target $k + 1$ distinct label instructions
  on positions $> f(l_{k + 1})$; impossible, as $\UCgp$ contains only $k$
  distinct forward goto instructions.

  Fixing said $i$ we note that there are at least $k + 1$ positions which are
  $f(l_i)$-$(n{+}1)$-relevant to the right of $f(l_{k + 1})$: this too is
  impossible, for the same reason. Contradiction.
\end{proof}

\chapter{Discussion}
\label{ch:discussion}

This thesis can be divided into four parts: the introduction of \C and the
theory behind it, the introduction of \Cg as an alternative to \C, the
definition of translations between these, and several results about the
expressiveness of \C and \Cg.

We have proved that \C and \Cg are equally expressive by means of the total
mappings $\CToCg$ and $\CgToC$. We have also proved that such translations are
only possible if the maximum jump counter (or label number) in the input inseq
is known. As a result $\CToCg$ and $\CgToC$ cannot be homomorphic.

We then went on to prove that any subsemigroup of \C (\Cg) needs to contain
infinitely many jump instructions (matching label and goto instructions) in
order to express all finite threads (\ThmRef{thm:finite_jumps},
\ThmRef{thm:finite_labels_gotos}). In order to express all regular threads it
is even necessary that such a semigroup contains infinitely many jump
instructions (label/goto instructions) \emph{in both directions}
(\ThmRef{thm:one_direction_jump_restriction_expressiveness},
\ThmRef{thm:one_direction_label_goto_restriction_expressiveness}). The upshot
is that any such infinite collection of jump instructions (label/goto
instructions) suffices
(\ThmRef{thm:two_directions_infinitely_many_jumps_enough}, the corresponding
result for \Cg is trivial).

\section{Further Work}

The translations between \C and \Cg in \Cref{ch:translations} use label and
goto instructions to mimic the behavior of jump instructions and vice versa.
There are some open questions about the nature of these translations: it is not
known whether alternative behavior preserving mappings can be defined which
employ less jump instructions or label/goto instructions. More precisely,

\begin{itemize}
  \item Given an arbitrary value $k \in \N$, what is the smallest value $k' \in
  \N$ for which there exists a behavior preserving mapping $f \colon \UACgr{k}
  \to \UACr{k'}$? (By definition of equation \eqref{eq:cgr_to_c_hom} in
  \Sref{sec:cg_to_c_hom} we already know that $k' \leq 4k + 12$.)

  \item As demonstrated by the translations defined in \Sref{sec:c_to_cg},
  there exist behavior preserving mappings $f \colon \UACr{k} \to \UACgr{k}$
  for all $k > 2$. Is there any value $k \in \N$ such that for some $k' < k$
  the mapping $f \colon \UACr{k} \to \UACgr{k'}$ is behavior preserving?

\end{itemize}

\section{Acknowledgements}

First and foremost I want to thank my supervisor, Alban Ponse, for his guidance
and most of all patience; the writing of this thesis took much longer than it
should have. I thank Kyndylan Nienhuis for asking some smart questions about
the semantics of \Cg, which led to the inclusion of
\Sref{sec:cg_alternative}.

I thank my family and especially Vera Matei for their support during the
writing of this thesis.

\appendix
\chapter{%
 Overview of Defined Translations
}\label{app:mappings_overview}

\begin{figure}
  \centering
  \begin{tikzpicture}[
    draw=black,fill=black!50,fill opacity=0.25,
    arrows=o->,thick,
    every node/.style={black,opacity=1}]

    \filldraw (0cm,0cm) ellipse (4cm and 2cm) node[left=3.25cm]
              {$\vphantom{\UACr{k}}\UAC$};
    \filldraw (0.3cm,0cm) ellipse (3.6cm and 1.85cm) node[left=2cm]
              {$\UACr{4k + 12}$};
    \filldraw (1.05cm,0cm) ellipse (2.75cm and 1.65cm) node[left=1.3cm]
              {$\UACr{k}$};
    \filldraw (4.5cm,0cm) ellipse (2.5cm and 2cm) node[above=1.35cm]
              {$\UACg$};
    \filldraw (4.5cm,-0.25cm) ellipse (2.25cm and 1.65cm) node[above=1cm]
              {$\UACgr{k}$};
    \filldraw (-1.5cm,-4.5cm) ellipse (2.5cm and 1.5cm) node[right=1.65cm]
              {$\Pga$};
    \filldraw (-1.5cm,-4.5cm) ellipse (1.5cm and 1.25cm) node[left=0.6cm]
              {$\PgaF$};
    \filldraw (-1.5cm,-4.5cm) ellipse (0.5cm and 1.0cm) node
              {$\PgaS$};

    \draw[dotted] (0.8cm,-0.75cm)
      .. controls +(down:1.5cm) and +(down:1.5cm) .. node[below] {$\CToCg_k$}
      (4.5cm,-1.25cm);

    \draw[dotted] (5cm,-1.25cm)
      .. controls +(down:2.25cm) and +(down:1.5cm) .. node[below] {$\CgToC_k$}
      (-0.8cm,-1.55cm);

    \draw (4cm,1.6cm)
      .. controls +(up:1cm) and +(up:1.25cm) .. node[below] {$\CgToC$}
      (-1.75cm,1.65cm);

    \draw (-3.1cm,0.95cm)
      .. controls +(up:2.5cm) and +(up:1.5cm) .. node[above] {$\CToCg$}
      (5cm,1.7cm);

    \draw (-3.4cm,-5.2cm)
      .. controls +(up:0.5cm) and +(left:0.25cm) .. node[above,sloped,pos=0.45] {$\fst$}
      (-2.6cm,-4.2cm);

    \draw (0.2cm,-5.2cm)
      .. controls +(up:0.25cm) and +(right:0.75cm) ..  node[above,sloped,pos=0.65] {$\snd$}
      (-1.4cm,-4.1cm);

    \draw (-3.5cm,-4cm)
      .. controls +(left:1cm) and +(down:0.5cm) .. node[above,sloped] {$\PgaToC$}
      (-3cm,-1cm);

    \draw (-1.5cm,-3.75cm)
      .. controls +(up:1cm) and +(down:0.5cm) .. node[above,sloped,pos=0.4] {$\PgaSToC$}
      (-3cm,-1cm);
  \end{tikzpicture}
  \caption{Overview of semigroups and single-pass instruction sequences and
  certain behavior preserving mappings defined between then, as introduced in
  this thesis. Dotted arrows represent homomorphisms. There is also a
  non-homomorphic version of $\CToCg_k$ (\Sref{sec:cr_to_cg}).}

  \label{fig:projections}
\end{figure}
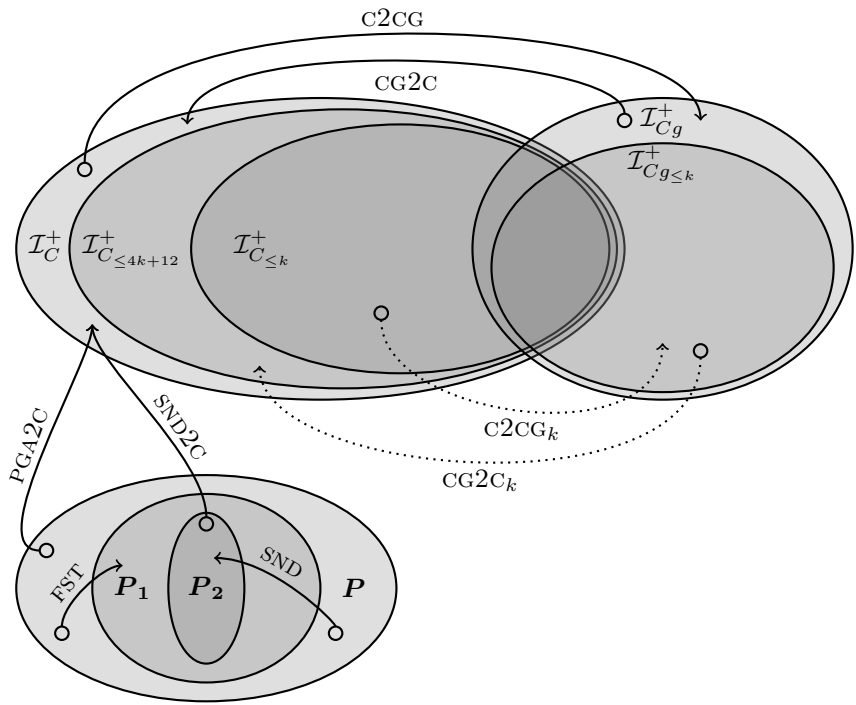

\fref{fig:projections} provides a graphical representation of the most
important sets of (single pass) instruction sequences introduced in this
thesis. Recall that the set $\UAC$ contains all \C-expressions. For arbitrary
$k \in \N$, $\UACr{k}$ is the largest subset of $\UAC$ which does not contain
\C-inseqs with relative jumps over a distance greater than $k$. Similarly,
$\UACg$ contains all \Cg-expressions, and $\UACgr{k}$ contains those inseqs
without goto instructions with a label number greater than $k$. All \PGA terms
are contained in $\Pga$; the set $\PgaF$ is the largest set which is restricted
to single pass instruction sequences in first canonical form. $\PgaS$ contains
\PGA's second canonical forms.

\chapter{%
  Proof by Bergstra \& Ponse
}\label{app:b_and_p_proof}

The proof of \ThmRef{thm:b_and_p} is presented in Section 9 of
\cite{inseq_intro}. As the proof of \ThmRef{thm:finite_jumps} builds upon this
result, Section 9 of \cite{inseq_intro} is reproduced here verbatim, with
kind permission of the authors. Three minor changes have been applied: a
section reference has been updated to point to an equivalent section in this
thesis, a footnote has been added and the last paragraph has been left out, as
it is merely an introduction to Section 10 of that publication.

Observe that \cite{inseq_intro} uses notation which in some places differs
slightly from notation introduced in this thesis.

\begin{quote}
  \newtheorem{theorem}{Theorem}
  \newtheorem{lemma}{Lemma}
  \newtheorem{claim}{Claim}
  \newcommand{\length}{\ell}
  \newcommand{\instr}[1]{\ensuremath{/#1}}
  \newcommand{\pinstr}[1]{\ensuremath{{+}/#1}}
  \newcommand{\rlinstr}[1]{\ensuremath{\backslash#1}}
  \newcommand{\prlinstr}[1]{\ensuremath{{+}\backslash#1}}
  \newcommand{\jum}[1]{\ensuremath{/\##1}}
  \newcommand{\bjum}[1]{\ensuremath{\backslash\##1}}
  \newcommand{\ter}{\ensuremath{\:!\:}}
  \newcommand{\abo}{\ensuremath{\#\:}}
  \newcommand{\extr}[1]{\ensuremath{|#1|}}
  \newcommand{\di}{\mathsf{D}}
  \newcommand{\st}{\mathsf{S}}
  \newcommand{\Nat}{{\mathbb N}}
  \newcommand{\Nplus}{{\mathbb N}^+}
  \newcommand{\tr}{\ensuremath{{\mathtt{true}}}}
  \newcommand{\fa}{\ensuremath{{\mathtt{false}}}}
  \newcommand{\pcc}[3]{\ensuremath{#2 \unlhd #1 \unrhd #3}}

  \section{Expressiveness and reduced instruction sets}

  In this section we further consider $C$'s
  instructions in the perspective of expressiveness.
  We show that
  setting a bound on the size of
  jump counters in $C$ does have consequences
  with respect to expressiveness:
  let
  \[C_k\]
  be defined by allowing only jump instructions
  with counter value $k$ or less.

  \bigskip

  We first introduce some auxiliary notions:
  following the definition of residual threads
  in Section~\ref{sec:bta}, we say that
  thread $Q$ is a \emph{$0$-residual}
  of thread $P$ if $P=Q$, and an
  \emph{$n+1$-residual} of $P$ if for some $a\in A$,
  $P=\pcc a{P_1}{P_2}$ and $Q$ is an $n$-residual of $P_1$
  or of $P_2$.
  Note that a finite thread (in \BTA) only
  has $n$-residuals for finitely many $n$,
  while for the thread
  $P$ defined by $P=a\circ P$ it holds that $P$ is an
  $n$-residual of
  itself for each $n\in\Nat$.

  Let $a\in A$ be fixed and $n\in\Nplus$.
  Thread $P$ has the \emph{$a$-$n$-property} if
  $\pi_n(P)=a^n\circ \di$ and $P$ has $2^n-1$ (different)
  $n$-residuals which all have a first approximation not equal
  to $a\circ\di$.\footnote{It appears that the authors meant to use $2^n$ instead
  of $2^n - 1$ in this sentence, though this does not affect the proof in any
  serious way. ---Stephan} So, if a thread $P$ has the
  {$a$-$n$-property}, then
  $n$ consecutive $a$-actions can be executed and each
  sequence of $n$ replies leads
  to a unique $n$-residual. Moreover, none of these
  residual threads starts with an $a$-action (by the
  requirement on their first approximation).
  We note that for each $n\in\Nplus$ we can find a finite
  thread with the $a$-$n$-property. In the next section we
  return to this point.

  A piece of code $X$ has the {$a$-$n$-property}
  if for some $i$,
  $\extr X_i$ has this property.
  It is not hard to see that in this case
  $X$ contains at least $2^n-1$ different $a$-tests.
  As an example, consider
  \[X=\ter;\rlinstr b;\prlinstr a;\pinstr a;\bjum 2;
  \pinstr a;\jum2; \instr c;\abo\]
  Clearly, $X$ has the $a$-2-property because
  $\extr X_4$ has this property: its 2-residuals are
  $b\circ\st$, $\st$, $\di$ and $c\circ\di$, so each
  thread is not equal to one of the others and does
  not start with an $a$-action.

  \bigskip

  Note that if a piece of code $X$ has the
  {$a$-$(n+k)$-property}, then it also has the $a$-$n$-property.
  In the example above, $X$ has the
  $a$-1-property because $\extr X_3$ has this property
  (and $\extr X_6$ too).

  \begin{lemma}
  \label{a-n-prop}
  For each $k\in\Nat$ there exists $n\in\Nplus$
  such that no $X\in C_k$
  has the $a$-$n$-property.
  \end{lemma}

  \begin{proof}
  Suppose the contrary and let $k$ be minimal in this respect.
  Assume for each $n\in\Nplus$,
  $Y_n\in C_k$ has the $a$-$n$-property.

  Let $B=\{\tr,\fa\}$. For $\alpha,\beta\in B^*$
  we write
  \[\alpha\preceq\beta\]
  if $\alpha$ is a prefix of $\beta$, and we write
  $\alpha\prec\beta$ or $\beta\succ\alpha$
  if $\alpha\preceq\beta$ and $\alpha\neq\beta$.
  Furthermore,
  let
  \[B^{\leq n}=\bigcup_{i=0}^n B^i,\]
  thus $B^{\leq n}$ contains all $B^*$-sequences
  $\alpha$ with
  $\length(\alpha)\leq n$ (there are
  $2^{n+1}-1$ such sequences).

  Let $g:\Nat\rightarrow \Nat$ be such that
  $\extr{Y_n}_{g(n)}$ has the $a$-$n$-property.
  Define
  \[f_n: B^{\leq n}\rightarrow\Nplus\]
  by $f_n(\alpha)=m$ if the instruction
  reached in $Y_n$ when execution started at position
  $g(n)$ after the replies to $a$ according to
  $\alpha$
  has position $m$.
  Clearly, $f_n$ is an injective function.

  In the following claim we show that under the supposition
  made in this proof a certain form of squeezing holds:
  if $k'$ is sufficiently large, then for all $n>0$
  there exist
  $\alpha,\beta,\gamma\in B^{k'}$ with
  $f_{k'+n}(\alpha)<f_{k'+n}(\beta)<f_{k'+n}(\gamma)$
  with the property that
  $f_{k'+n}(\alpha)<f_{k'+n}(\beta')<f_{k'+n}(\gamma)$
  for each extension $\beta'$ of $\beta$ within
  $B^{\leq k'+n}$.
  This claim is proved by showing that not having this
  property implies that ``too many'' such extensions
  $\beta'$ exist.
  Using this claim
  it is not hard to contradict the minimality of $k$.

  \begin{claim}
  \label{claimB}
  Let $k'$ satisfy $2^{k'}\geq2k+3$. Then for all $n>0$
  there exist $\alpha,\beta,\gamma\in B^{k'}$
  with \[f_{k'+n}(\alpha)<f_{k'+n}(\beta)<f_{k'+n}(\gamma)\]
  such that for
  each extension $\beta'\succeq\beta$ in
  $B^{\leq k'+n}$,
  \[f_{k'+n}(\alpha)<f_{k'+n}(\beta')<f_{k'+n}(\gamma).\]
  \end{claim}

  \begin{proof}[Proof of Claim~\ref{claimB}.]
  Let $k'$ satisfy $2^{k'}\geq2k+3$.
  Towards a contradiction, suppose the stated
  claim is not true for some $n>0$.
  The sequences in $B^{k'}$
  are totally ordered by $f_{k'+n}$, say
  \[f_{k'+n}(\alpha_1)<f_{k'+n}(\alpha_2)< \ldots
  <f_{k'+n}(\alpha_{2^{k'}}).\]
  Consider the following list of sequences:
  \begin{align*}
  \alpha_1,\underbrace{\alpha_2,\dots,\alpha_{2k+2}},
  \alpha_{2k+3}\\
  \text{choices for }\beta\hspace{10mm}
  \end{align*}
  By supposition there is
  for each choice $\beta
  \in\{\alpha_2,\ldots,\alpha_{2k+2}\}$
  an
  extension $\beta'\succ\beta$ in $B^{\leq {k'+n}}$
  with
  \[\text{either}\quad
  f_{k'+n}(\beta')<f_{k'+n}(\alpha_1),\quad\text{or}
  \quad
  f_{k'+n}(\beta')>f_{k'+n}(\alpha_{2k+3}).
   \]
  Because there are
  $2k+1$ choices for $\beta$, assume
  that at least $k+1$
  elements $\beta\in\{\alpha_2,
  \ldots,\alpha_{2k+2}\}$
  have an extension $\beta'$ with
  \[f_{k'+n}(\beta')<f_{k'+n}(\alpha_1)\]
  (the assumption $f_{k'+n}(\beta')>
  f_{k'+n}(\alpha_{2k+3})$ for at
  least $k+1$ elements $\beta$ with extension $\beta'$
  leads to a similar argument).
  Then we obtain a contradiction
  with respect to $f_{k'+n}$:
  for each of the sequences $\beta$ in the subset just
  selected
  and its extension $\beta'$,
  \[f_{k'+n}(\beta')<f_{k'+n}(\alpha_1)<f_{k'+n}(\beta),\]
  and there are at least
  $k+1$ different such  pairs $\beta,\beta'$
  (recall $f_{k'+n}$ is injective). But
  this is not possible with jumps of at
  most $k$ because the $f_{k'+n}$ values of
  each of these pairs
  define a path in $Y_{k'+n}$
  that never has a gap
  that exceeds $k$ and
  that passes position $f_{k'+n}(\alpha_1)$, while
  different paths never share a position.
  This finishes the proof of Claim~\ref{claimB}.
  \end{proof}

  Take according to
  Claim~\ref{claimB} an appropriate
  value $k'$, some value $n>0$ and
  $\alpha,\beta,\gamma\in B^{k'}$.
  Consider ${Y_{k'+n}}$
  and mark the positions that are used for
  the computations according to $\alpha$ and $\gamma$:
  these computations both start in position $g({k'+n})$
  and end in
  $f_{k'+n}(\alpha)$ and $f_{k'+n}(\gamma)$, respectively.
  Note that the set of marked positions never has a gap
  that exceeds $k$.

  Now consider a computation that starts from instruction
  $f_{k'+n}(\beta)$ in $Y_{k'+n}$, a position in
  between $f_{k'+n}(\alpha)$
  and $f_{k'+n}(\gamma)$. By Claim~\ref{claimB}, the first
  $n$ $a$-instructions have positions in between
  $f_{k'+n}(\alpha)$
  and $f_{k'+n}(\gamma)$ and none of these are marked.
  Leaving out all marked positions and adjusting the
  associated jumps yields a piece of code, say $Y$, with
  smaller jumps, thus in $C_{k-1}$, that has the
  $a$-$n$-property.
  Because $n$ was chosen arbitrarily, this contradicts
  the initial supposition that $k$ was minimal.
  \end{proof}

  \begin{theorem}
  \label{thm:2}
  For any $k\in\Nplus$, not all threads
  in $\BTA$ can be expressed in $C_k$. This
  is also the case if
  thread extraction may start at arbitrary
  positions.
  \end{theorem}

  \begin{proof}
  Fix some value $k$. Then, by
  Lemma~\ref{a-n-prop} we can find a value $n$
  such that no $X\in C_k$ has the $a$-$n$-property.
  But we can define a finite thread that has this property.
  \end{proof}

\end{quote}

\bibliographystyle{amsalpha}
\bibliography{sources}

\end{document}